\documentclass[%
 reprint,
 amsmath,amssymb,
 aps,
floatfix,
aps, nofootinbib, superscriptaddress,
]{revtex4-2}

\usepackage[dvipdfmx]{graphicx}\usepackage{dcolumn}\usepackage{bm}\usepackage[braket, qm]{qcircuit}
\usepackage[caption=false,labelformat=simple]{subfig}

\usepackage{float}
\usepackage{amsmath}
\usepackage{amsthm}
\usepackage{algorithm,algpseudocode}
\usepackage{color}
\usepackage{here}
\usepackage[title]{appendix}
\usepackage[normalem]{ulem}
\usepackage{xcolor}

\theoremstyle{definition}
\newtheorem{dfn}{Definition}[section]

\newtheorem{rem}[dfn]{Remark}

\newif\ifoverleaf
\overleaftrue

\ifoverleaf
\usepackage[whole]{bxcjkjatype}
\else
\usepackage{bxcjkjatype}
\fi
\usepackage{comment}

\newif\ifhidedone
\hidedonetrue

\usepackage{soul}
\newif\ifcomment
\commenttrue

\ifcomment
\newcommand{\shumpei}[1]{\textcolor[RGB]{254,0,144}{\textbf{Shumpei: }#1}}
\newcommand{\nakajima}[1]{ \textcolor[RGB]{144,0,254}{\textbf{Nakajima: }#1}}
\newcommand{\tran}[1]{\textcolor[RGB]{60,104,234}{\textbf{Tran: }#1}}
\else
\newcommand{\shumpei}[1]{\ignorespaces} \newcommand{\nakajima}[1]{\ignorespaces} \newcommand{\tran}[1]{\ignorespaces} \fi

\ifhidedone
\newcommand{\done}[1]{\ignorespaces}
\else
\newcommand{\done}[1]{#1}
\fi

\newcommand{\diff}[1]{\textcolor{red}{#1}}

\DeclareFontEncoding{FMS}{}{}
\DeclareFontSubstitution{FMS}{futm}{m}{n}
\DeclareFontEncoding{FMX}{}{}
\DeclareFontSubstitution{FMX}{futm}{m}{n}
\DeclareSymbolFont{fouriersymbols}{FMS}{futm}{m}{n}
\DeclareSymbolFont{fourierlargesymbols}{FMX}{futm}{m}{n}
\DeclareMathDelimiter{\VERT}{\mathord}{fouriersymbols}{152}{fourierlargesymbols}{147}

\newif\ifrevise
\revisefalse

\ifrevise
\DeclareRobustCommand{\erase}{\bgroup\markoverwith{\textcolor{blue}{\rule[.5ex]{2pt}{0.4pt}}}\ULon}
\newcommand{\add}[1]{\textcolor{blue}{#1}}
\else
\newcommand{\erase}[1]{}
\newcommand{\add}[1]{#1}
\fi

\begin{document}
\title{Extending echo state property for quantum reservoir computing
}
\author{Shumpei Kobayashi}
 \affiliation{Department of Creative Informatics, The University of Tokyo, Japan}
 \author{Quoc Hoan Tran} \thanks{Present address for Q. H. Tran: Quantum Laboratory, Fujitsu Research, Fujitsu Limited}
 \affiliation{Next Generation Artificial Intelligence Research Center (AI Center), The University of Tokyo, Japan}
\author{Kohei Nakajima}
 \affiliation{Department of Creative Informatics, The University of Tokyo, Japan}
 \affiliation{Next Generation Artificial Intelligence Research Center (AI Center), The University of Tokyo, Japan}
 \affiliation{Department of Mechano-Informatics, The University of Tokyo, Japan}

\date{\today}

\begin{abstract}
The echo state property (ESP) represents a fundamental concept in the reservoir computing (RC) framework that ensures output-only training of reservoir networks by being agnostic to the initial states and far past inputs. However, the traditional definition of ESP does not describe possible non-stationary systems in which statistical properties evolve. To address this issue, we introduce two new categories of ESP: \textit{non-stationary} ESP, designed for potentially non-stationary systems, and \textit{subspace} and \textit{subset} ESP, designed for systems whose subsystems have ESP. Following the definitions, we numerically demonstrate the correspondence between non-stationary ESP in the quantum reservoir computer (QRC) framework with typical Hamiltonian dynamics and input encoding methods using non-linear autoregressive moving-average (NARMA) tasks. We also confirm the correspondence by computing linear/non-linear memory capacities that quantify input-dependent components within reservoir states. Our study presents a new understanding of the practical design of QRC and other possibly non-stationary RC systems in which non-stationary systems and subsystems are exploited.

\end{abstract}
\maketitle

\section{Introduction}

Physical reservoir computing (PRC) \cite{TANAKA2019100, Nakajima_2020, Nakajima2021}, which utilizes non-linear natural dynamics of physical substrate for temporal information processing, has garnered much attention. It is seen as a way to mitigate the massive computational resource needs of sophisticated machine learning methods, such as deep learning. However, not all physical systems are effective as reservoir substrates due to potential initial-state sensitivity in their natural dynamics, such as in chaotic systems. One precondition for excluding such systems is the echo state property (ESP), which requires the initial-state dependency to diminish over time \cite{Jaeger2001ESP, yildiz2012re}.

The current state of quantum computing is based on noisy intermediate-scale quantum (NISQ) \cite{Preskill_2018} technology, which represents non-fault-tolerant and small to medium-sized quantum computer environments. In the NISQ  era, non-universal quantum computation schemes gained much attention because of their near-term feasibility on physical devices. Such computational procedures include, for instance, variational quantum computation (VQC) \cite{McClean_2016, Mitarai_2018} and quantum reservoir computing (QRC) \cite{Fujii_2017, Ghosh_2019}. VQC and QRC apply to one-shot and autoregressive quantum machine learning algorithms, which have also become a general prospective application of quantum computation.

QRC can also be understood as a specific type of PRC that uses a quantum system as its physical reservoir. It has recently been shown to be capable of implementing temporal quantum tomography \cite{Tran2021}, predicting large-scale spatiotemporal chaos \cite{tran2020higherorder}, and emulating functions requiring both classical and quantum inputs simultaneously with a single quantum reservoir \cite{Tran2023}. Other works on QRC include proposals of QRC in various physical apparatus \cite{negoro2018machine, Ghosh_2019_2, Chen_2020, Nokkala_2021, Govia_2021, Spagnolo_2022}, with some performing actual physical experiments \cite{negoro2018machine, Chen_2020, Spagnolo_2022}, and theoretical analyses \cite{Mart_nez_Pe_a_2023}. Specifically, some works \cite{Chen_2020, Suzuki_2022,sannia2022dissipation, Kubota_2023,fry2023optimizing} focus on the dissipative nature of the natural quantum system to find a relationship between the existence of dissipation and the trainability of QRC. Kubota et al. \cite{Kubota_2023} analyzed the behavior of QRC driven by natural noise in quantum processing units. Additionally, the work in \cite{Mart_nez_Pe_a_2023} expands on this research direction by describing ESP from the perspective of a time-independent filter and dynamical systems, which the author's term state affine systems.
\begin{figure}[t]
\centering
\includegraphics[width=0.8\hsize]{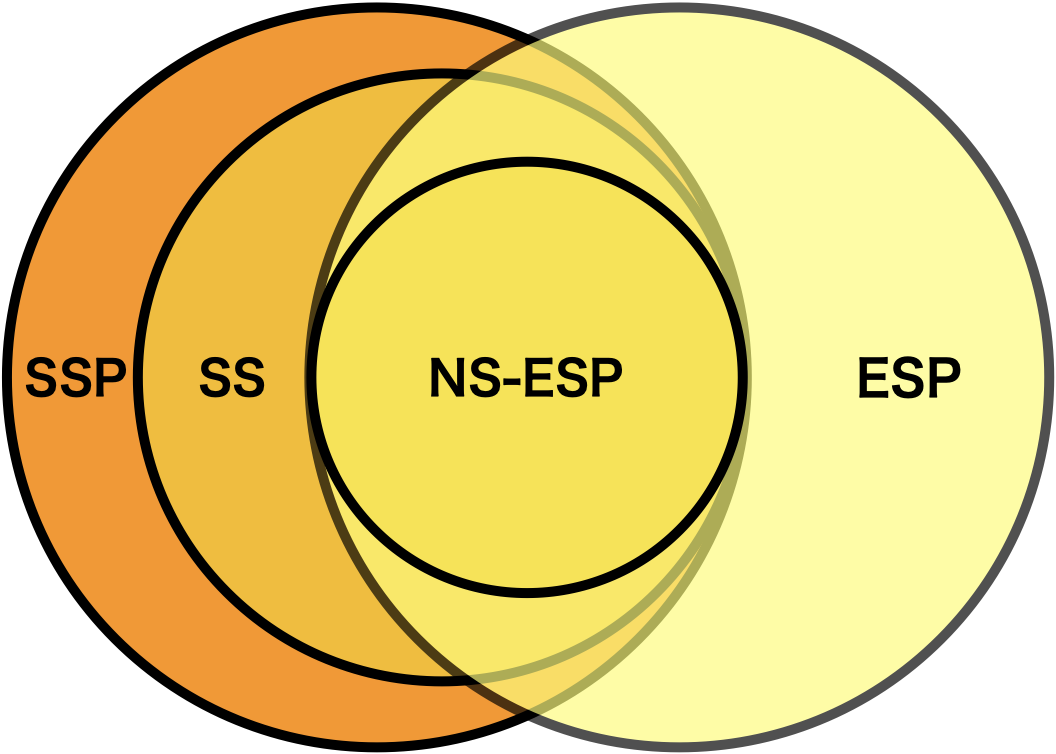}

\caption{Inclusion relationship of traditional ESP and the ESP defined in this paper. NS stands for \textit{non-stationary}, 
SS stands for \textit{subset} and SSP stands for \textit{subspace} ESP, respectively.}
\label{fig:esp-inclusion}
\end{figure}

Quantum systems have been attracting attention as promising substrates for PRC. However, quantum systems, in general, are not always stationary, and in some cases, the traditional definition of ESP does not adequately ensure their capability for temporal information processing. In this paper, we define and analyze new conditions that secure such a possibly non-stationary system to function effectively as a practical reservoir.

In this paper, we define two extensions of traditional ESP. One extension is \textit{non-stationary} ESP, which requires finite variance output signals relative to initial-state difference decay. Another extension is \textit{subset} and \textit{subspace} non-stationary ESP, which focuses on a situation in which a part of the system has a non-stationary ESP, leaving the entire system possibly initial-state dependent. Our analysis includes a numerical study of non-stationary ESP and subset non-stationary ESP in a typical QRC setup with a specific type of Hamiltonian system.

We have made the following contributions:
\begin{itemize}
	\item Defined non-stationary, subspace, and subset versions of ESP, which could be practical for QRC and other non-conventional systems.
	\item Showed a relationship between non-stationary ESP and the information processing capability of QRC using numerical experiments.
\end{itemize}

\section{Main results}

\subsection{Non-stationary ESP}
First, we present the definition of traditional ESP, which has been used extensively in the RC context.

\begin{dfn}{Echo state property \cite{Jaeger2001ESP, yildiz2012re}}

Let a compact system state space be $\mathcal{S}$, a compact input space be $\mathcal{X}$, and a set of time indexes be $\mathcal{T}$. For an input-driven dynamical system with dynamical map $s_t = f(\{\mathbf{u}_\tau\}_{\tau < t};s_0)$ such that $f:\mathcal{X}^{\mathcal{T}} \times \mathcal{S} \to \mathcal{S}$, where $s_0$ is the initial state and $\{\mathbf{u}_\tau\}$ is a sequence of inputs indexed by time $\tau$, the ESP holds if and  only if
\begin{equation}
\label{eqn:esp}
\begin{aligned}
	\forall \{\mathbf{u}_\tau\},\ &\forall (s_0, s_0'),\\
	 &\|f(\{\mathbf{u}_\tau\}_{\tau \leq t};s_0) -  f(\{\mathbf{u}_\tau\}_{\tau\leq t};s_0')\| \underset{t \to \infty} \to 0.
	\end{aligned}
\end{equation}
\end{dfn}

We argue that this ESP definition by state difference decay is general in that all known definitions of ESP are equivalent to this form. For further discussion, please refer to Appendix.~\ref{sec:esp_equiv}.

ESP is supposed to work on stationary systems. However, a quantum system, for instance, is not always stationary even if the system state does not explicitly depend on time. A trivial example is the case in which a Pauli noise exists. Let us depict an example where uniform depolarization of rate $\epsilon$ exists. When the depolarization is the only noise that exists in the system, the norm of the system state $\rho$ is measured using the trace norm; $D(\rho, I)$ 
vanishes when $t \to \infty$, namely, $D(\rho, I) \propto (1 -\epsilon)^t$. 

In this case, ESP does not mean fading memory because the state difference $\|f(\{\mathbf{u}_\tau\}_{\tau \leq t}; s_0) -  f(\{\mathbf{u}_\tau\}_{\tau \leq t}; s_0') \|$ relative to $\|f(\{\mathbf{u}_\tau\}_{\tau \leq t}; s_0')\|$ does not change. In our analysis, we propoesed the following modified definition of ESP to handle such a non-stationary system.

\begin{dfn}{Non-stationary ESP (NS-ESP)}
\begin{figure}[t]
\centering
\subfloat[]{
\label{subfig:non-stationary-esp}
	\includegraphics[width=0.8\hsize]{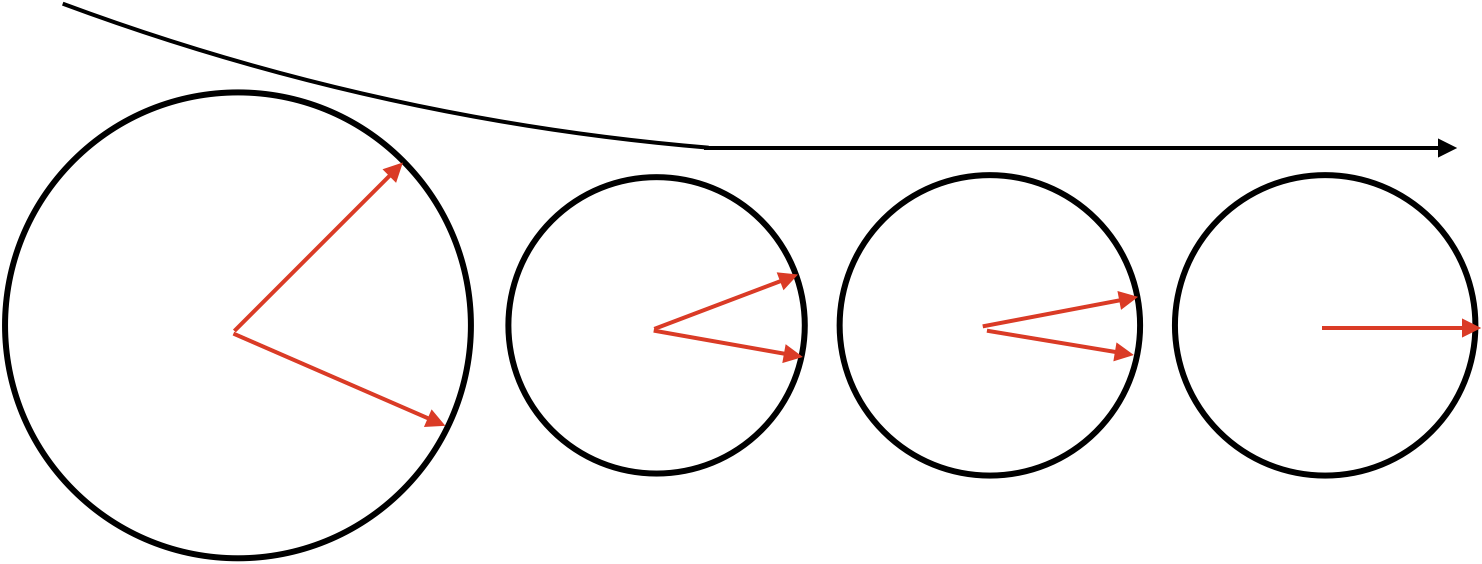}
}
\par
\subfloat[]{
\label{subfig:stationary-esp}
	\includegraphics[width=0.8\hsize]{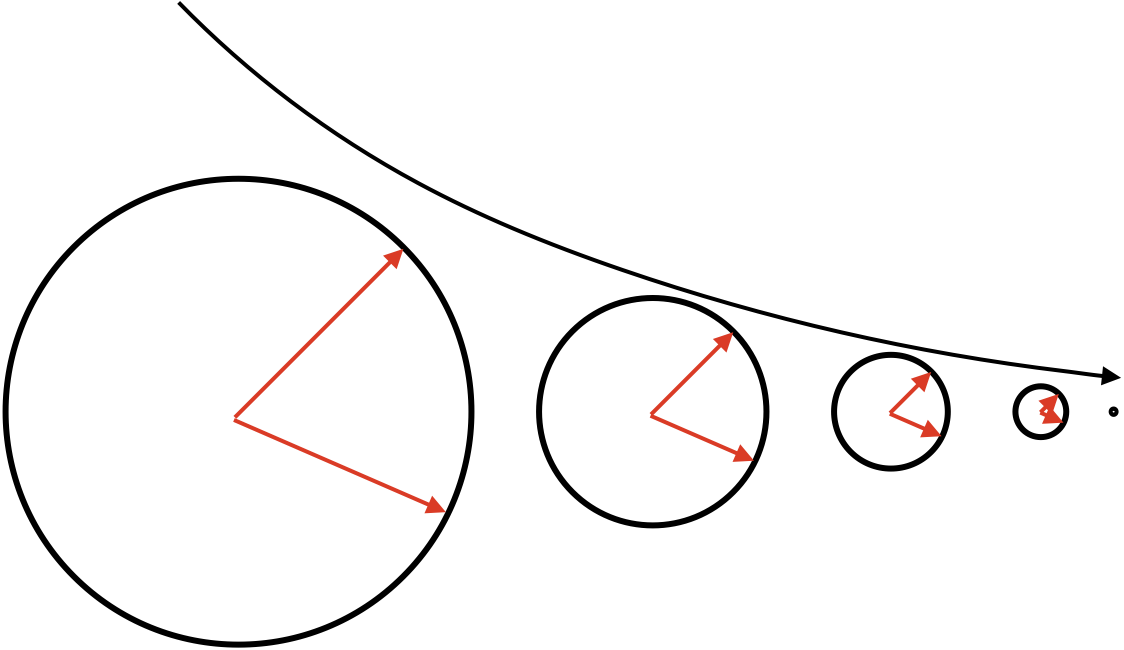}
}
\caption{Schematics of two types of ESPs. State space is illustrated as a circle, and different states are illustrated as red (gray) arrows. (a) Non-stationary ESP. State difference decays, but the state space remains finite. (b) With traditional ESP but not with the non-stationary ESP. It has an input-invariant fixed point. The state difference decays as fast as the state decay. Schematics show representative cases where two states compared in ESP calculation are on the same shrinking state space for simplicity. }
\label{fig:stationary-non-stationary-esp}
\end{figure}

Given system dynamics $f: \mathcal{X}^{\mathcal{T}} \times \mathcal{S} \to \mathcal{S}$, 
$f$ has \textit{non-stationary} ESP if the following condition holds:
\begin{equation}
\label{eqn:esp_ext2}
\begin{aligned}
	&\forall \{\mathbf{u}_\tau\} ,\ \forall (s_0, s_0'),\ \exists w \in \mathbb{N} < +\infty \text{ s.t. }\\
	&\quad \underset{t\to \infty}{\liminf} \left[\left\|\mathrm{Var}_w^t(\{\mathbf{u}_\tau\})\right\|\right] > 0 \Rightarrow\\
	&\frac{\|f(\{\mathbf{u}_\tau\}_{\tau \leq t}; s_0) -  f(\{\mathbf{u}_\tau\}_{\tau \leq t}; s_0') \|}{\sqrt{\min\left[\left\|\overline{\mathrm{Var}}_w^t(f; s_0)\right\|, \left\|\overline{\mathrm{Var}}_w^t(f; s'_0)\right\|\right]}} \underset{t \to \infty} \to 0,\\
	\end{aligned}
\end{equation}
where

\begin{equation}
\begin{aligned}
	\mathrm{E}_w^t(\{\mathbf{u}_\tau\}) &\equiv \frac{1}{w}\sum_{k=0}^{w-1}\mathbf{u}_{t - k},\\
	\left(\mathrm{Var}_w^t(\{\mathbf{u}_\tau\})\right)_i &\equiv{\mathrm{E}}_w^t\left(\{(\mathbf{u}_{\tau} - \mathrm{E}_w^t(\{\mathbf{u}_{\tau}\})_i^2\}\right),\\
	\overline{\mathrm{E}}_w^t(f; s_0) &\equiv \frac{1}{w}\sum_{k=0}^{w-1}f(\{\mathbf{u}_\tau\}_{\tau \leq t-k}; s_0),\\
	\left(\overline{\mathrm{Var}}_w^t(f; s_0)\right)_i &\equiv \overline{\mathrm{E}}_w^t\left(\left(f - \overline{\mathrm{E}}_w^t(f, s_0)\right)_i^2; s_0\right),\\
\end{aligned}
\end{equation}
in which $(\mathbf{v})_i$ denotes the $i$-th element of a vector $\mathbf{v}$.

\end{dfn}

The normalizing part on the denominator causes a non-stationary system, such as one with a strong depolarizing channel, to not satisfy the condition. Additionally, it follows that if non-stationary ESP holds, then ESP holds.

In stationary systems, satisfying non-stationary ESP excludes trivial scenarios in which constant output signals are observed. Additionally, in  non-stationary systems, we expect output signals to be utilized in temporal information processing tasks under suitable post-processing, given that non-stationary ESP is satisfied. The expected scenarios for non-stationary cases include the following:
\begin{enumerate}
	\item Suppose the state space itself shrinks, yet state difference shrinks faster than space itself. In this case, appropriately scaling up the states based on the input step cycle will generate a traditional ESP-compatible state sequence. 
	\item Suppose the state space size and the state difference diverge, yet the state difference diverges slower than the state space itself. In this case, appropriately scaling down the states based on the input step cycle will generate a traditional ESP-compatible state sequence.
	\item Suppose the state's mean monotonically varies, and non-stationary ESP holds. In this case, appropriately shifting the states based on the input step cycle will generate a traditional ESP-compatible state sequence.
\end{enumerate}
\add{An example for cases 1 and 2 above would be an input-driven dynamical system with explicit scaling at each time step. For instance, $f$ can be an echo state network \cite{Jaeger2001ESP}, and the following two-step state update rule can be written:
\begin{equation}
\left\{
\begin{aligned}
	x_{t+1} &= f(x_t, u_t),\\
	y_{t+1} &= c^{t+1}x_{t+1},
\end{aligned}
\right.
\end{equation}
can be converted into the state update rule of $y_t$ as below.
\begin{equation}
	y_{t+1} = c^{t+1} f(\frac{y_t}{c^t}, u_t).
\end{equation}
This will be an example for case 1 if $c \ll 1$ and case 2 if $c \gg 1$, provided that, when $c = 1$, it satisfies the non-stationary ESP without shrinking or expanding state space over time. Here, $x_t \in \mathbb{R}^N$ and $y_t \in \mathbb{R}^N$ represents reservoir states, and $u_t \in \mathbb{R}^M$ represents inputs.}

\add{An example for case 3 would be an input-driven dynamical system with explicit bias at each time step. For instance, the state update rule will be 
\begin{equation}
\left\{
\begin{aligned}
	x_{t+1} &= f(x_t, u_t),\\
	y_{t+1} &= x_{t+1} + b(t+1),
\end{aligned}
\right.
\end{equation}
again, provided that when $b = 0$, it satisfies the non-stationary ESP without shrinking or expanding state space over time.
The equation above can be converted into the single-step state update below.
\begin{equation}
	y_{t+1} = f(y_{t} - bt, u_t) + b(t+1).
\end{equation}
It should be noted that the traditional ESP is also a good indicator of fading memory in this case, although it is not a stationary system.
}

Overall, we expect the definition of non-stationary ESP to cover a more prominent family of dynamical systems that may have information processing capability.
\subsection{Subspace and subset non-stationary ESP}
\begin{figure}[t]
\centering
\subfloat[]{
\label{subfig:subspace-esp}
	\includegraphics[width=0.4\hsize]{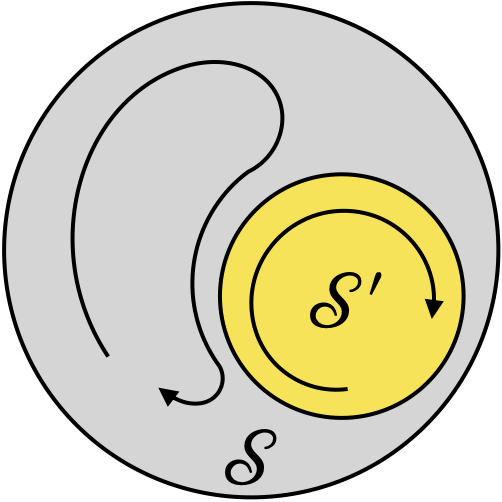}
}
\subfloat[]{
\label{subfig:subset-esp}
	\includegraphics[width=0.4\hsize]{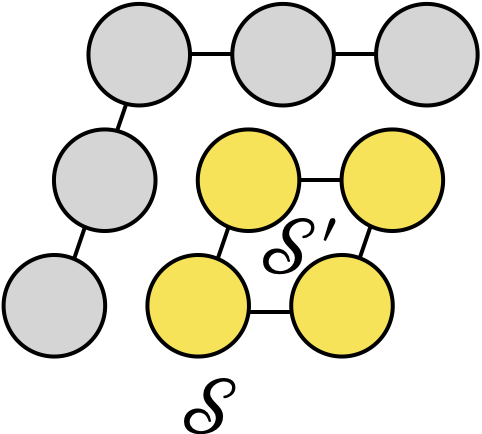}
}
\caption{(a) Subspace ESP. The full state space $\mathcal{S}$ has an invariant subspace $\mathcal{S'}\subseteq \mathcal{S}$ under the input-driven system dynamics $f$. Furthermore, $\mathcal{S'}$ satisfies non-stationary ESP under $f|_{\mathcal{S'}}$. (b) Subset ESP. A limited set of vector elements of each state $s\in \mathcal{S}$ has ESP.}
\label{fig:subspace-subset-esp}
\end{figure}
\done{\diff{motivation for the definition, compatibility to measurement, usefulness when defined like this, ESN and other reservoir compatibility }\\}
In the RC setup, we can post-process output signals from the reservoir. Simple post-processing methods include linear transformation and \textit{subset selection}. Here, \textit{subset selection} denotes the selection of $m \leq n$ elements from the system output $x \in \mathbb{R}^{n}$.

For instance, if an invariant subspace of the input-driven dynamics exists, and that subsystem holds non-stationary ESP, then it can obtain non-stationary ESP-compatible output signals by post-projecting the outputs to that invariant subspace. In other words, if the system dynamics has a disjoint structure among its elements and a non-stationary ESP-compatible subset of the outputs exists, we can post-restrict them to that subset to obtain non-stationary ESP-compatible output signals.

Following the observation above, we formally define a weak version of non-stationary ESP to preclude a situation where only a part of the system is initial-state dependent.

The \textit{subspace} non-stationary ESP holds if one or more linear system subspaces, including the trivial subspace, have non-stationary ESP.
\begin{dfn}{Subspace non-stationary ESP}
\label{def:subspace_esp}

Given system dynamics $f: \mathcal{X}^{\mathcal{T}} \times \mathcal{S} \to \mathcal{S}$,
 $f$ has \textit{subspace non-stationary ESP} if there exists a transformation $\mathbb{P}: \mathcal{S} \to \mathcal{S}'$ such that $\mathcal{S'} \subseteq \mathcal{S}$ and $\mathbb{P}\circ f$ satisfies non-stationary ESP.
\end{dfn}

Specifically, to treat cases in which a \textit{subset} $\mathcal{S}'$ of $\mathcal{S}$ satisfies non-stationary ESP, we define the following version.
\begin{dfn}{Subset non-stationary ESP}
\label{def:subset_esp}

Given system dynamics $f: \mathcal{X}^{\mathcal{T}} \times \mathcal{S} \to \mathcal{S}$, 
$f$ has \textit{subset non-stationary ESP} if there exists a subset selection procedure  $\mathcal{P}: \mathcal{S} \to \mathcal{S'}$ such that $\mathcal{S'} \leq \mathcal{S}$ and $\mathcal{P}\circ f$ satisfies non-stationary ESP. 

	The expression $B \leq A$ denotes that $B$ is a non-void element-wise subset of $A$. If $A \equiv \mathbb{R}^n$, then $B = \mathbb{R}^m$ $(m\leq n)$, for instance. 
\end{dfn}

It follows that if subset non-stationary ESP holds, then subspace non-stationary ESP holds because we can define $\mathbb{P}$ as a linear transformation represented by a diagonal matrix such that it has $1$ in the dimension included in $\mathcal{S}'$ and $0$ otherwise. In addition, if non-stationary ESP holds, then every element of the system state has non-stationary ESP. Therefore, the subset non-stationary ESP retains. These relationships can be written as 
\begin{equation}
\label{eqn:subset-subspace}
\begin{aligned}
	&\text{NS-ESP} \subsetneq \text{Subset NS-ESP} \subsetneq \text{Subspace NS-ESP},
	\end{aligned}
\end{equation}
where NS-ESP stands for non-stationary ESP.

More generally, we have the inclusion relationship of the ESP variants, as seen in Fig.~\ref{fig:esp-inclusion}.

The definition of the subset non-stationary ESP is natural for practical QRC because we can select any observable for our system output. That is usually a subset of all Pauli strings or a linear combination of them, which can be reconstructed by measuring some of the observables. 

If the subset non-stationary ESP holds for some non-trivial subset of a system, a linear regression automatically makes emphasis on that subset guided by a loss function, provided that initial-state sensitive variables do not contribute sufficiently to temporal information processing.

Def.~\ref{def:subspace_esp} and Def.~\ref{def:subset_esp} are meant to be used for systems in which some parts have ESP while the remaining portion does not. An example of such a system in a quantum case is when the system dynamics are a tensor product of unitary and dissipative evolution. Ensuring a subset of non-stationary ESP guarantees that such a system can be used as a reservoir with a simple transformation of output signals.

\subsection{Non-stationary ESP of QRC}
\label{sec:ns_esp_qrc}
\begin{figure}[t]
\centering
\subfloat[]{
\label{subfig:sk_esp_sphere}
	\includegraphics[width=0.5\hsize]{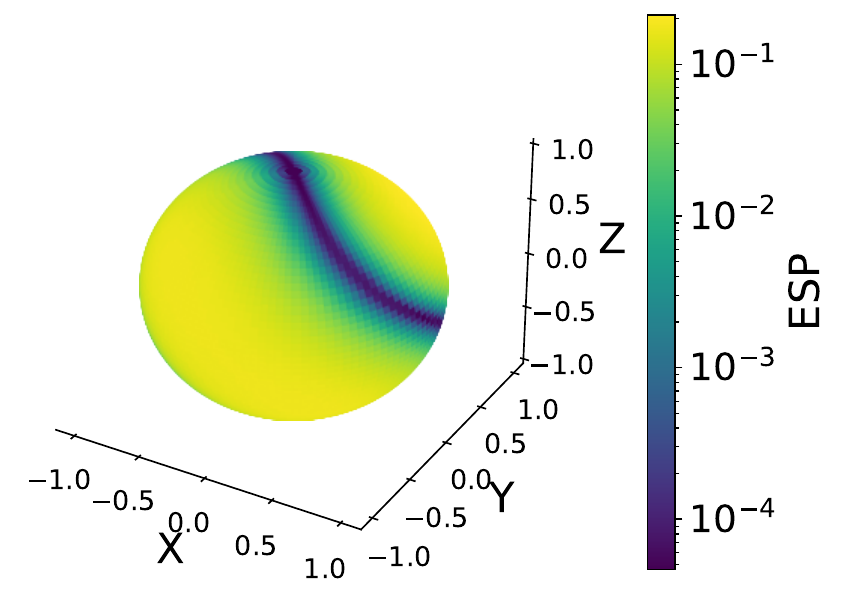}
}
\subfloat[]{
\label{subfig:sk_ns_esp_sphere}
	\includegraphics[width=0.5\hsize]{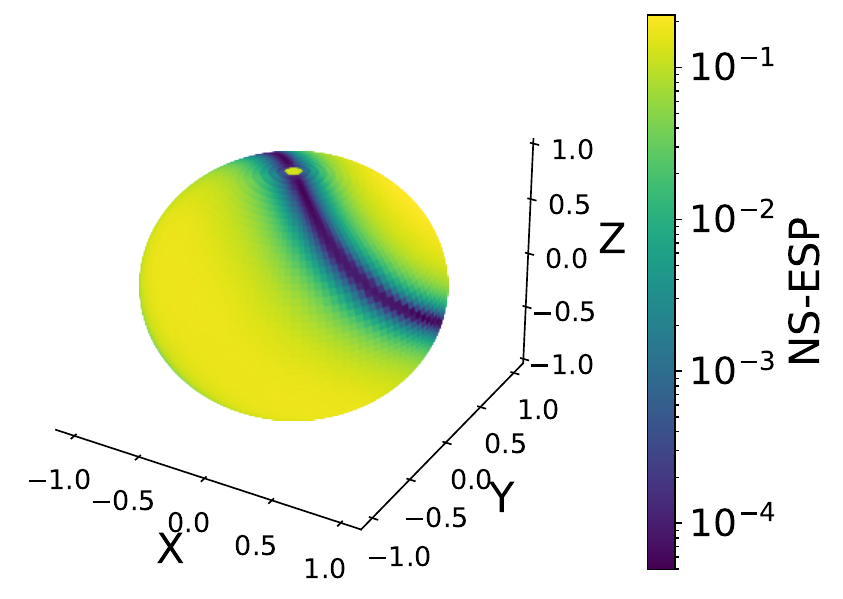}
}
\par
\subfloat[]{
\label{subfig:sk_esp_uv}
	\includegraphics[width=0.7\hsize]{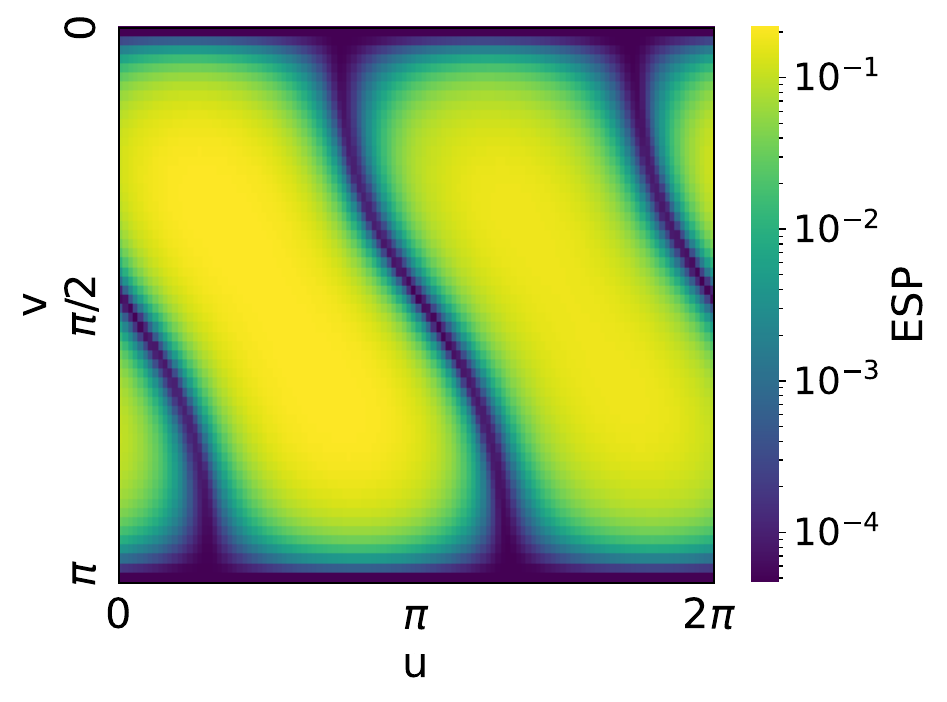}
}
\par
\subfloat[]{
\label{subfig:sk_ns_esp_uv}
	\includegraphics[width=0.7\hsize]{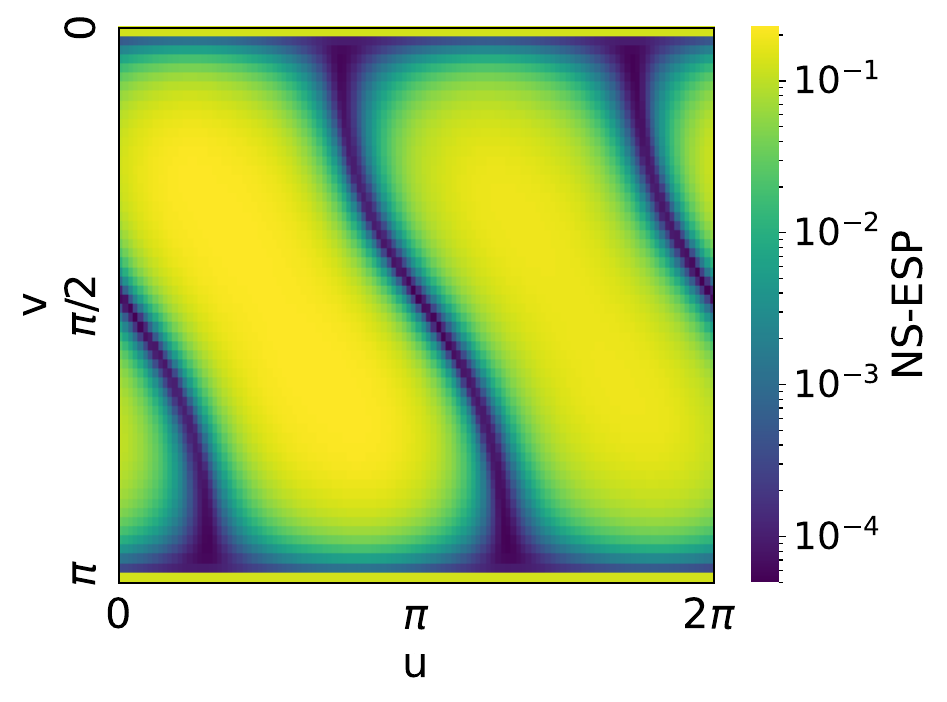}
}
\caption{
Grid search result of input rotation axis $\theta$ for a QRC with SK Hamiltonian with external field and qubit-reset for input encoding. Each spherical coordinate corresponds to the input rotation axis $\theta$ described in the main text. (a) ESP characterized by the ESP indicator drawn on the Bloch sphere. (b) Non-stationary ESP characterized by the non-stationary ESP indicator, drawn on the Bloch sphere. Our experiments are done with $|\{\mathbf{u}_\tau\}|=200$ and $w=10$, where $w$ is the variable defined in Eq.~\eqref{eqn:esp_ext2}. (c) The flattened surface of the Bloch sphere of traditional ESP in (a). (d) The flattened surface of the Bloch sphere of non-stationary ESP in (d). The top and bottom lines in (c-d) correspond to the north and south poles, respectively. }

\label{fig:esp}
\end{figure}
\label{subsec:ns-esp}
To numerically examine the correspondence between non-stationary ESP and the actual information processing capability of QRC, we prepared a QRC of the following Sherrington–Kirkpatrick (SK) Hamiltonian \cite{Sherrington_1975} with an external field:
\add{\begin{equation}
\label{eqn:sk_hamiltonian_resp}
	H=\sum_{i>j=1}^N J_{i j} \sigma_i^x \sigma_j^x+\frac{1}{2} \sum_{i=1}^N(h + D_i) \sigma_i^z,
\end{equation}}
where $N$ was the number of qubits in the system, $\sigma^x_i$ and $\sigma^z_i$ were Pauli X and Z operators of the $i$-th qubit, respectively, and $J_{ij}$ and $h_i$ were sampled from the following probability distribution:

\begin{equation}
\label{eqn:sk_sampling_resp}
\begin{aligned}
 J_{ij} &\sim \mathrm{Uniform}([-J_s/2, J_s/2])\\
D_i &\sim \mathrm{Uniform}([-WJ_s/2, WJ_s/2]),
\end{aligned}
\end{equation}
where $J_s=1$, $W = 0.312$, and $h =0.013$, in precision up to the third digit after the decimal point. \add{Here $h$ represents constant global external field, and each $D_i$ represents random local external field.}

The input sequence $\mathcal{U} \equiv \{\mathbf{u}_t\}_{t \in \mathcal{T}} \in \mathbb{R}^{|\mathcal{T}|}$ were fed into the reservoir using the following input encoding method named \textit{reset-input encoding}:
\begin{equation}
\begin{aligned}
	\rho' &= \mathcal{E}(\rho, \mathbf{u}; \theta) \\
	&= \mathrm{tr}_{\mathrm{A}}(\rho) \otimes \sigma_{\mathrm{A}}(\mathbf{u}; \theta),
	\end{aligned}
\end{equation}
where $\mathrm{A}$ was the subsystem that we used for qubit state replacement by subsystem state $\sigma_\mathrm{A}(\mathbf{u})$ of form
\begin{equation}
	\sigma_A(\mathbf{u} ; \theta) \equiv U(\mathbf{u}; \theta)\left(|0\rangle\langle 0|^{\otimes |A|}\right)U^\dagger(\mathbf{u}; \theta),
\end{equation}
where
\begin{equation}
\begin{aligned}
	U(\mathbf{u}; \theta) &\equiv U_3^\dagger(\theta)R_Z(\add{\mathrm{arccos}}(\mathbf{u}))U_3(\theta)\\
	U_3(\theta) &\equiv \begin{pmatrix}
		\cos\left(\frac{\Theta(\theta)}{2}\right) & -e^{i\Lambda(\theta)} \sin\left(\frac{\Theta(\theta)}{2}\right) \\
		e^{i\Phi(\theta)}\sin\left(\frac{\Theta(\theta)}{2}\right) & e^{i\left(\Phi(\theta) + \Lambda(\theta)\right)}\cos\left(\frac{\Theta(\theta)}{2}\right)
	\end{pmatrix}
\end{aligned}
\end{equation}
where $\Theta(\theta),\ \Phi(\theta)$, and $\Lambda(\theta)$ were Euler angles from our rotation axis $\theta \in \mathbb{R}^3$ to the $Z$ axis on the Bloch sphere.
Therefore, the overall state update was
\begin{equation}
	\rho_{t+1} = e^{-iH}\mathcal{E}(\rho_t, \mathbf{u}_t; \theta)e^{iH},
\end{equation}
and reservoir output signals were 
\begin{equation}
	\left\{\mathrm{tr}(P\rho_t) \mid P \in \{I, X, Y, Z\}^{\otimes N}\right\}
\end{equation}
for each $t$. \add{Here $I$ is a $2\times 2$ identity matrix, and $X$, $Y$ and $Z$ are Pauli matrices defined as follows: 
\begin{equation}
	X = \begin{pmatrix}
		0 & 1\\
		1 & 0
	\end{pmatrix},\quad Y = \begin{pmatrix}
		0 & i\\
		-i & 0
	\end{pmatrix},\text{ and } Z = \begin{pmatrix}
		1 & 0\\
		0 & -1
	\end{pmatrix},
\end{equation}
and an example for $P$ would be $ I \otimes X \otimes Y \in \mathbb{R}^{8\times 8}$ for a 3-qubit system for instance.}

Here, we parametrized the input encoding unitary $U(\cdot\ ; \theta)$ by a parameter of $\theta$ to explore different configurations to ensure that the QRC had different non-stationary ESP, while system Hamiltonian $H$ was fixed. The experiment was done on a 2-qubit setup, with $|A|$ = 1, while $\theta$ stood for the axis of single-qubit rotation in the Bloch sphere.
\begin{figure}[t]
\centering
\subfloat[]{
\label{subfig:sk_narma2_uv}
	\includegraphics[width=0.7\hsize]{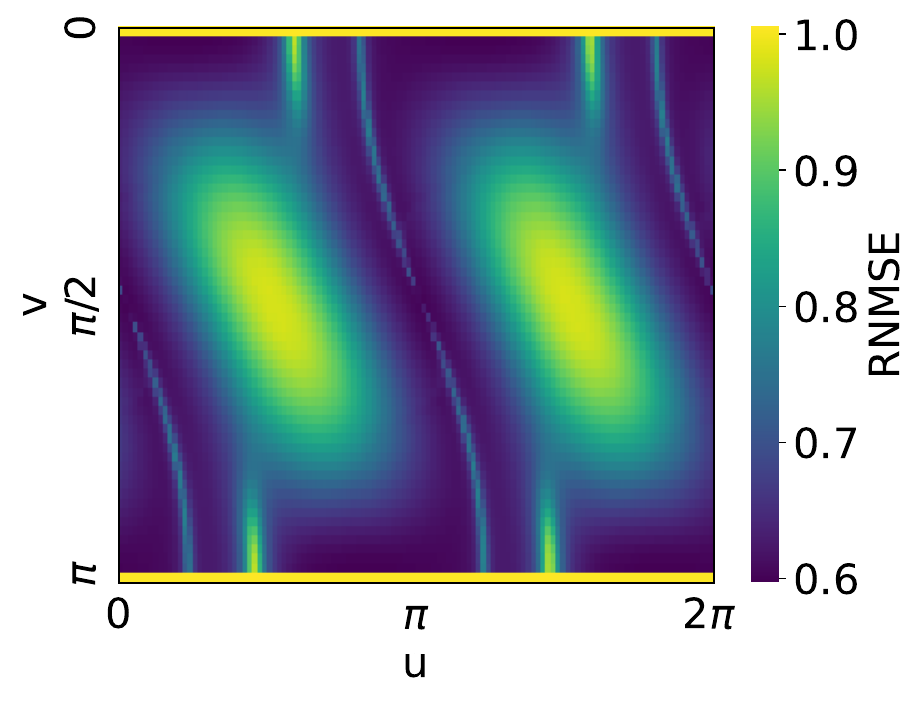}
}
\par
\subfloat[]{
\label{subfig:sk_narma10_uv}
	\includegraphics[width=0.7\hsize]{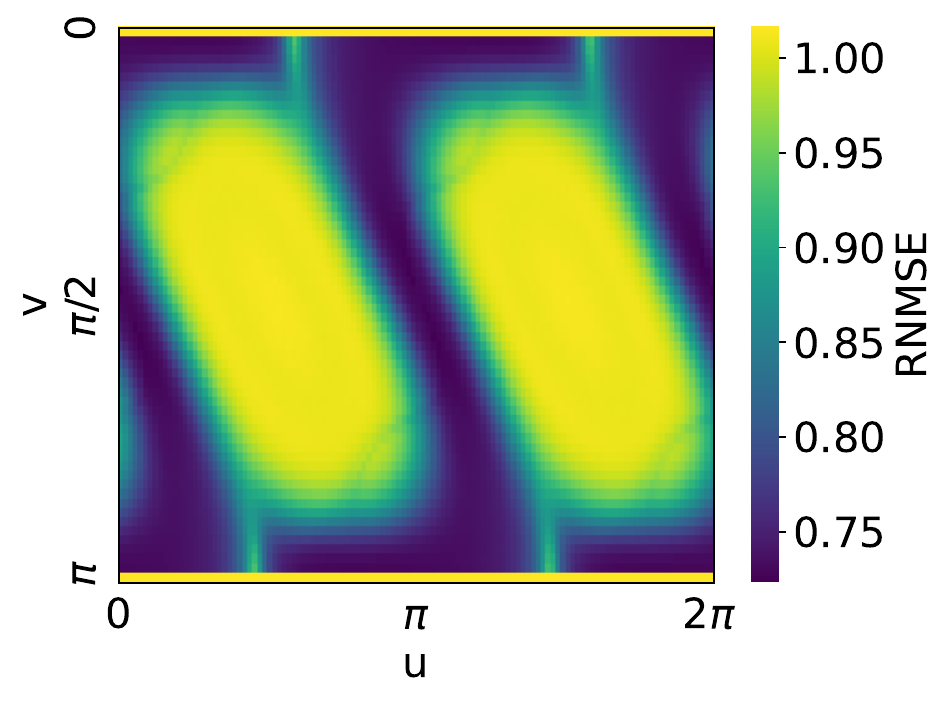}
}

\caption{Root normalized mean squared error (RNMSE) of (a) NARMA2, (b) NARMA10 tasks. \add{The settings for the Hamiltonian, parametrized input-encoding, and the output signal are the same as the experiment of ESP indicator results in Fig.~4. Please refer to the main text for how we generated input and target signals. }}
\label{fig:qrc-narma}
\end{figure}
\subsubsection{ESP and non-stationary ESP}
\label{subsubsec:ns_esp}

\add{First, we computed the ESP indicator, defined as follows:
\begin{equation}
\label{eqn:esp_indicator}
	\mathcal{I}_{ESP}(t, s_0, s_0') \equiv \frac{\|f(\{\mathbf{u}_\tau\}_{\tau \leq t};s_0) -  f(\{\mathbf{u}_\tau\}_{\tau\leq t};s_0')\|}{\|s_0 -s_0'\|},
\end{equation}
and the non-stationary ESP indicator defined as follows:
\begin{equation}
\label{eqn:ns_esp_indicator}
\begin{aligned}
	&\mathcal{I}_{NS}(t, s_0, s_0', w) \equiv \\
	&\quad \mathcal{I}_{ESP}(t, s_0, s_0') \times \frac{\sqrt{\min\left[\left\|\overline{\mathrm{Var}}_w^w(f; s_0)\right\|, \left\|\overline{\mathrm{Var}}_w^w(f; s'_0)\right\|\right]}}{\sqrt{\min\left[\left\|\overline{\mathrm{Var}}_w^t(f; s_0)\right\|, \left\|\overline{\mathrm{Var}}_w^t(f; s'_0)\right\|\right]}}
	\end{aligned}
\end{equation}
for every configuration of the parameter $\theta$.}  We randomly sampled four input sequences of size $200$ from $\mathrm{Uniform}([-1, 1])$ and three initial reservoir states from Haar random distribution of 2-qubit pure states.

Output signals of QRC were collected for every initial reservoir state and input sequence combination. For each input sequence, the ESP indicator and non-stationary ESP indicator were examined for every different initial state combination. The result of Fig.~\ref{fig:esp} was the average of all such twelve ($12 = 4 \times {{3}\choose {2}}$)calculations.

The flattened surface plots in Fig.~\ref{subfig:sk_esp_uv} and Fig.~\ref{subfig:sk_ns_esp_uv} showed an expanded Bloch sphere surface, where the upper line and bottom line, respectively, correspond to the north and south poles. We can see the difference between ESP and non-stationary ESP in those poles, where ESP is satisfied, while non-stationary ESP is not. \add{These poles correspond to the input configurations in which an input encoding axis direction and the direction of the system state Bloch vector coincide after a reset operation. In such settings, input encoding does not modulate the system's fixed point, and the system falls into constant output. That is why the traditional ESP was satisfied at these poles while the non-stationary ESP was not. In simpler words, because qubit reset generates a state that is fixed under input encoding, the system becomes input-independent at those poles.}

Please note that ESP and non-stationary ESP indicators \erase{are upper bound by one because indicators}greater than one imply that there is no possibility of convergence, even if we increase the input sequence length in the experiments. The boundaries between yellow (light gray) and navy purple (dark gray) regions colored by white are not intentional because we did not cap values by any lower bounds. \add{In addition, the strong and weak ESP regions' pattern, as shown in Fig.~4, is Hamiltonian-dependent. Namely, this result is specific to our selected random Hamiltonian. However, the following features are invariant under the choice of Hamiltonians:
\begin{enumerate}
	\item The top and bottom yellow (light gray) lines for the non-stationary ESP, as shown in Fig.~4(d), appear when we use the reset-input encoding because of the reason discussed above.
	\item There always exist gradients of strong and weak non-stationary ESP regions. This indicates that the input-encoding method is an important factor in determining the fading memory property of QRC. Please see Appendix B for other examples of the ESP and non-stationary ESP results with different Hamiltonians.
\end{enumerate}}

\add{Here, the ESP indicators in Fig.~\ref{subfig:sk_esp_uv} are not the same as the largest conditional Lyapunov exponent \cite{Pecora1990}, which determines the diverging or converging behavior of input-driven dynamics around attractors, which can be calculated by numerical methods such as \cite{Clinton2003}. Because the ESP indicators are not calculated after sufficient washout time, it is not ensured that the input-driven dynamics of the QRC reach an attractor when calculating ESP indicators. Instead, they are calculated for random initial states, which are not guaranteed to be placed near attractors.}

\add{However, we argue that the upper bound of the largest Lyapunov exponent in QRC is 0 because any quantum channel is linear and the quantum state space is bounded. It should be noted that this fact also means that there is no classical chaos in quantum systems. It can be expected that the upper bound of zero will be observed in QRC with completely unitary subsystems, including a trivial subsystem of the total system. In our QRC settings described in Sec.~\ref{subsec:ns-esp}, completely unitary systems do not appear in any parameter configurations. Therefore, every region in Fig.~\ref{fig:esp} is expected to have a negative largest Lyapunov exponents. }

\add{Because the Lyapunov exponent is always negative in QRC, and we divide the ESP indicator by the variance decay as shown in the definition Eq.~\eqref{eqn:ns_esp_indicator}, there is no simple relationship between the non-stationary ESP indicator and the largest Lyapunov exponent. The non-stationary ESP indicator catches the relative strength of relative divergence between different trajectories and the divergence of each trajectory. Here, we did not calculate the Lyapunov exponents of the QRC systems because of the computational difficulty that arose from the fact that the phase space of the 2-qubit dissipative quantum system is a 16-dimensional complex completely-positive and trace-preserving manifold.}

\subsubsection{NARMA tasks}
\label{subsubsec:narma2}

Next, to check the correspondence between QRC's temporal information processing capability and non-stationary ESP, we conducted experiments on non-linear autoregressive moving-average (NARMA) tasks \cite{Atiya_2000}. Given an input sequence $\{\mathbf{u}_t\}$, a NARMA sequence $\{\mathbf{y}_t\}$ of order $k$, termed NARMA$k$, is defined as a nonlinear combination of $\{\mathbf{y}_t\}$ and $\{\mathbf{u}_t\}$ in the past, as follows:
\begin{figure*}[t]
\subfloat[]{
\label{subfig:x-input-raw}
	\includegraphics[width=0.37\hsize]{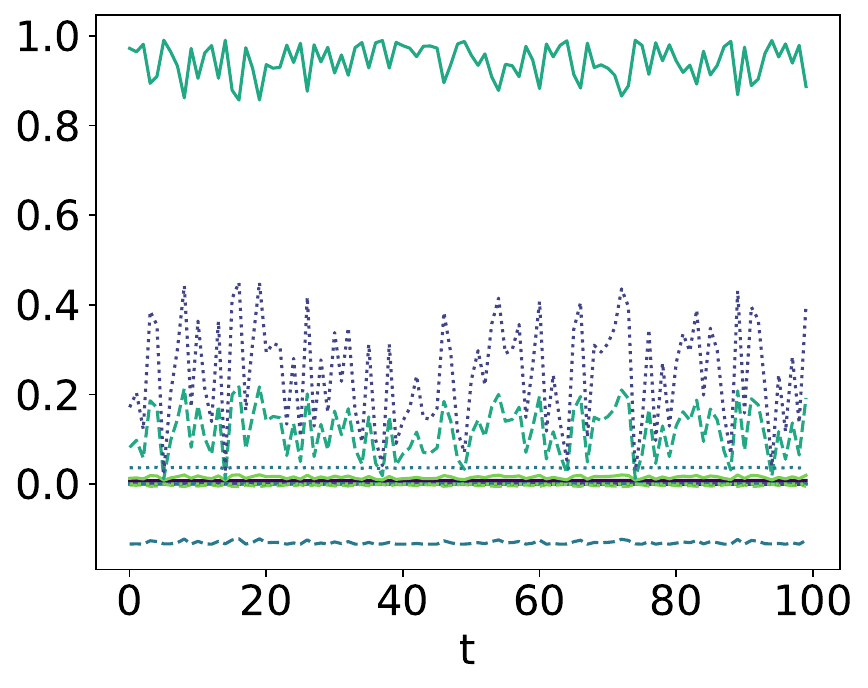}
}
\subfloat[]{
\label{subfig:y-input-raw}
	\includegraphics[width=0.6\hsize]{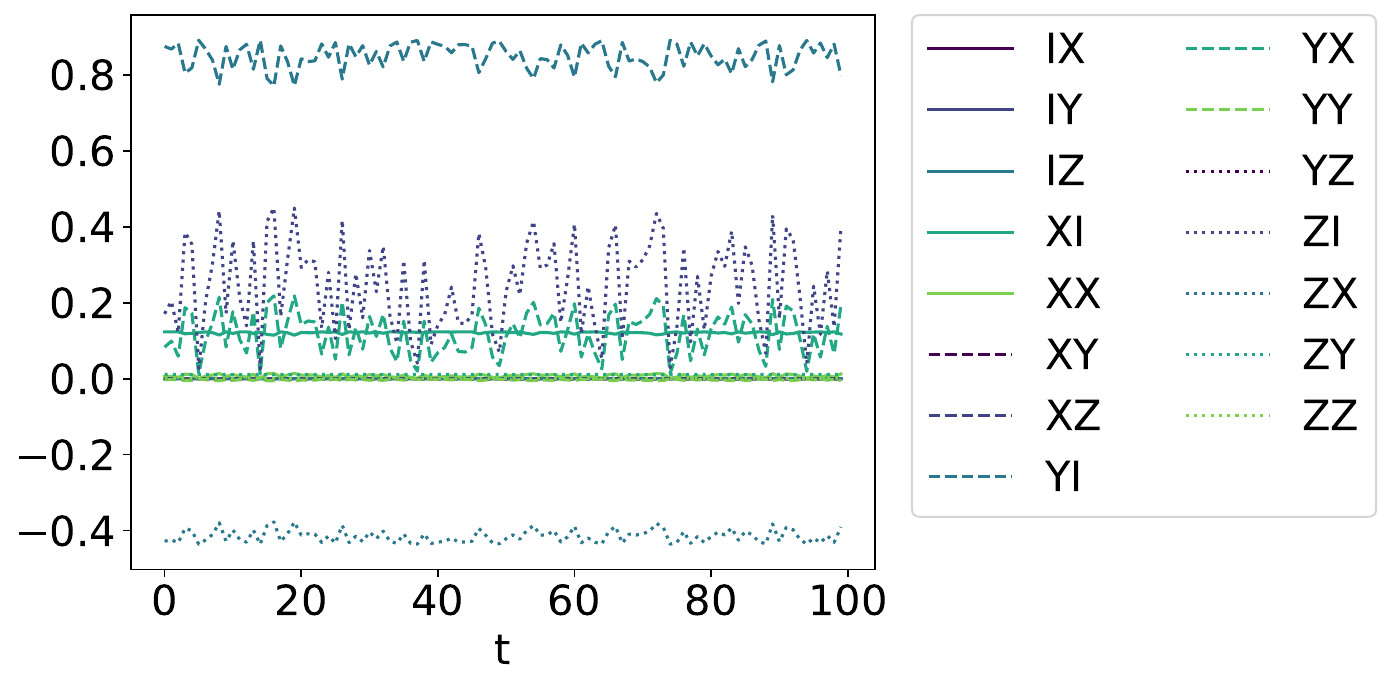}
}
\par
\subfloat[]{
\label{subfig:x-input-pred}
	\includegraphics[width=0.4\hsize]{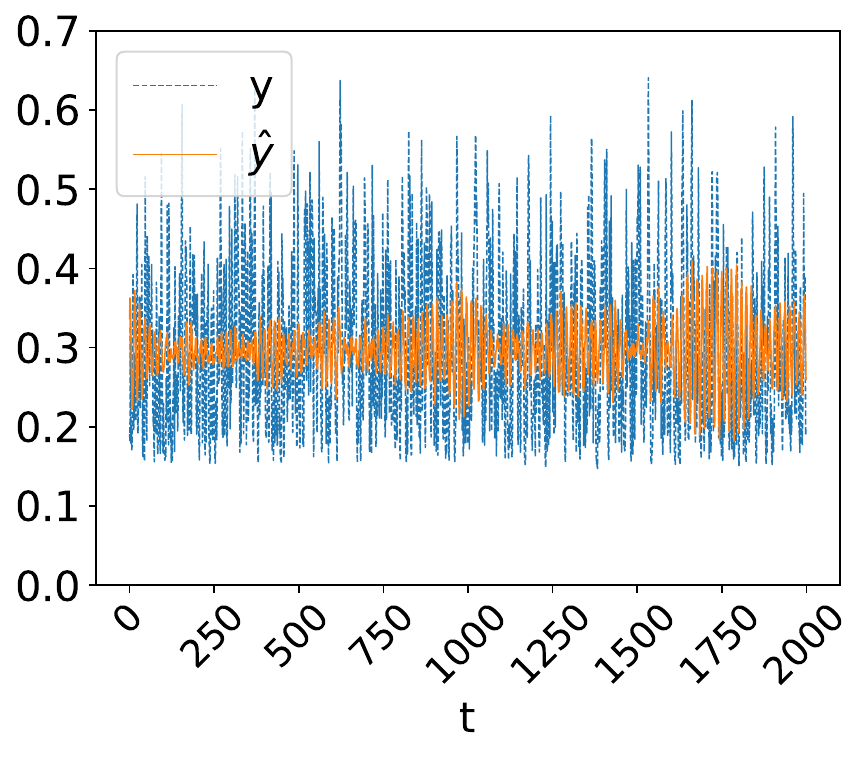}
}
\hspace{0.1\hsize}
\subfloat[]{
\label{subfig:y-input-pred}
	\includegraphics[width=0.4\hsize]{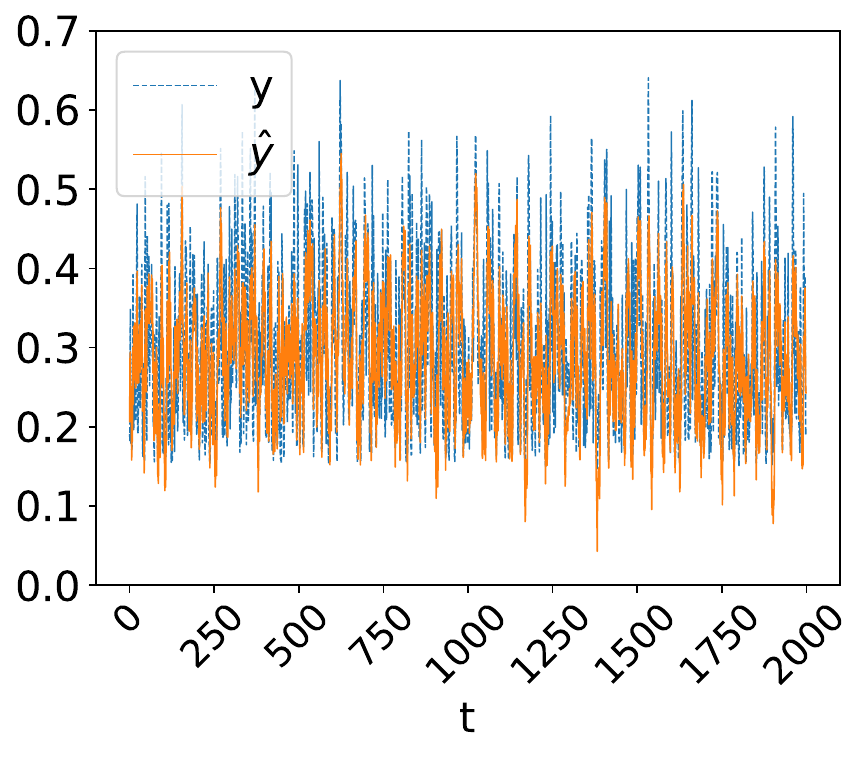}
}
\par
\subfloat[]{
\label{subfig:x-input-pred-zoomed}
	\includegraphics[width=0.4\hsize]{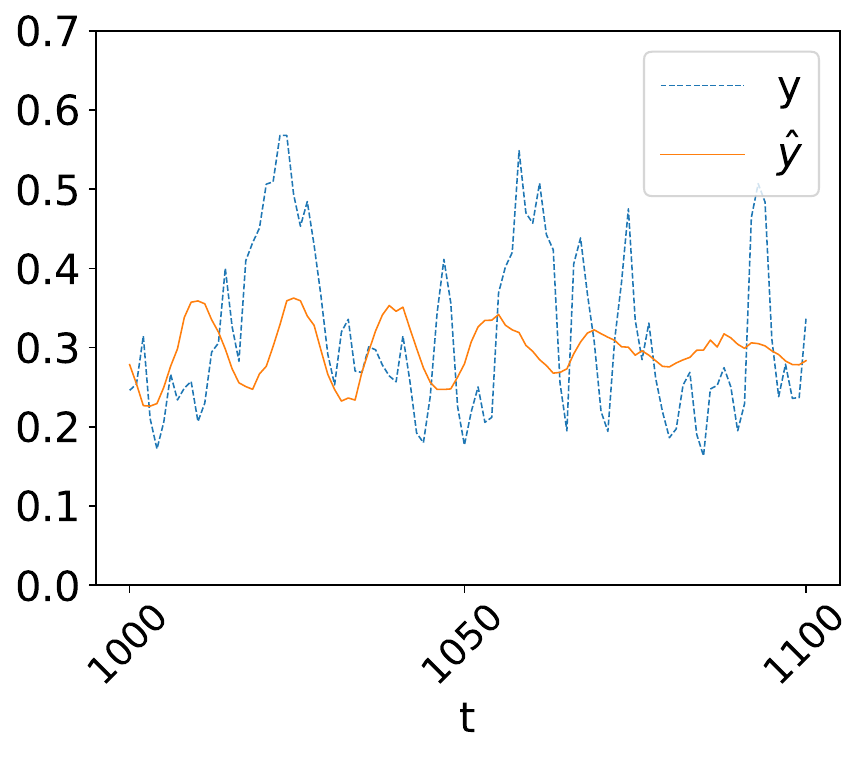}
}
\hspace{0.1\hsize}
\subfloat[]{
\label{subfig:y-input-pred-zoomed}
	\includegraphics[width=0.4\hsize]{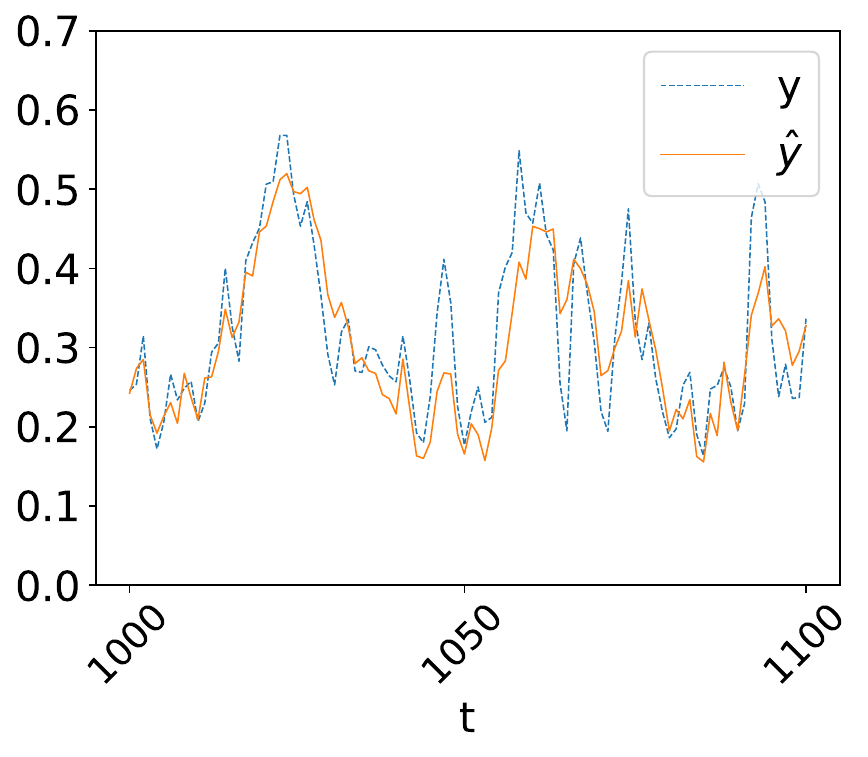}
}
\caption{
Raw (a, b) and linear regression (LR) outputs (c-f) of 2 different input rotation axis ($\theta$) configurations: (a, c, e) X-axis and (b, d, f) Y-axis. (a, b) Raw measurement results. (c, d) Target sequence (dotted blue) and linear regression results (solid orange). \add{(e, f) shows the zoomed signals of (c, d) at time steps $t\in[1000, 1100]$ for clarification of the fact that strong non-stationary ESP has positive effects in the NARMA2 task.}
}
\end{figure*}
\begin{equation}
	\mathbf{y}_{t} = 0.3 \mathbf{y}_{t-1} + 0.05 \mathbf{y}_{t-1} \sum_{i=t-k}^{t-1}\mathbf{y}_i + 1.5 \mathbf{u}_{t-1}\mathbf{u}_{t-k} + 0.1,
\end{equation}
where each $u_t$ was identically drawn from $\mathrm{Uniform}([0, 0.5])$.

Our sequences had a length of $2\times 10^5$. The first 50\% of $\{\mathbf{u}_t\}$ were discarded as washout, and 80\% of the remainder was used to fit a linear regression to optimize MSE against the first 80\% of the remainder. The test results were computed using the remaining 20\% of the sequences after washout. We had 5 NARMA sequences for each order $k$, and the results in Fig.\ref{fig:qrc-narma} were averaged over the sequences.

We computed the root normalized mean square error (RNMSE) 
\begin{equation}
	\mathcal{L}_{RNMSE}(\left\{\mathbf{y}_t\right\}, \left\{\mathbf{\hat{y}}_t\right\}) = \sqrt{\frac{\mathbb{E}_t\left[\left(\hat{\mathbf{y}}_t - \mathbf{y}_t\right)^2\right]}{\mathrm{Var}_t(\mathbf{y}_t)}}
\end{equation}
for target sequence $\left\{\mathbf{y}_t\right\}$ and linear regression result $\left\{\hat{\mathbf{y}}_t\right\}$.

We observed that the results of both NARMA2 and NARMA10 tasks in Fig.~\ref{subfig:sk_narma2_uv} and Fig.~\ref{subfig:sk_narma10_uv}, respectively, nicely corresponded to non-stationary ESP results in Fig.~\ref{subfig:sk_ns_esp_uv}. It should be noted that the yellow (light gray) line at the top and bottom of Fig.~\ref{subfig:sk_ns_esp_uv}, which corresponds to the convergence of the dynamics to a single fixed point, and results in large RNMSE in Fig.~\ref{subfig:sk_narma2_uv} and Fig.~\ref{subfig:sk_narma10_uv}, cannot be reproduced by the traditional ESP indicator, as shown in Fig.~\ref{subfig:sk_esp_uv}.

To clarify the actual dynamics and prediction behavior of different input encoding axes in the NARMA2 experiment of Sec.~\ref{subsubsec:narma2}, we show actual output signals of two separate input axis configurations depicted in Fig.~\ref{subfig:x-input-raw} and Fig.~\ref{subfig:y-input-raw}, in which all Pauli string measurement results from $t=3000$ to $t=3100$ are plotted for both setups. In Fig.~\ref{subfig:x-input-pred} and Fig.~\ref{subfig:y-input-pred}, the target sequence $y$ and the predicted sequence $\hat{y}$ from an optimized linear regression readout for the first $2000$ steps of the test sequence are shown.

In this comparison, the X-axis input demonstrates a configuration where non-stationary ESP does not hold, while the Y-axis input is for non-stationary ESP-compatible demonstration. We cannot identify which has non-stationary ESP by simply seeing the raw output signal in Fig.~\ref{subfig:x-input-raw} and Fig.~\ref{subfig:y-input-raw}. However, linear readout results in Fig.~\ref{subfig:x-input-pred} and Fig.~\ref{subfig:y-input-pred} clearly show the performance difference between the two configurations, \add{in the sense that the output signal transformed by the linear readout has better fit to the target signal as shown in Fig.~\ref{subfig:x-input-pred-zoomed} and Fig.~\ref{subfig:y-input-pred-zoomed}.}

\begin{figure*}[t]
\centering
	\subfloat[]{
\label{subfig:sk_mc1_uv}
	\includegraphics[width=0.33\hsize]{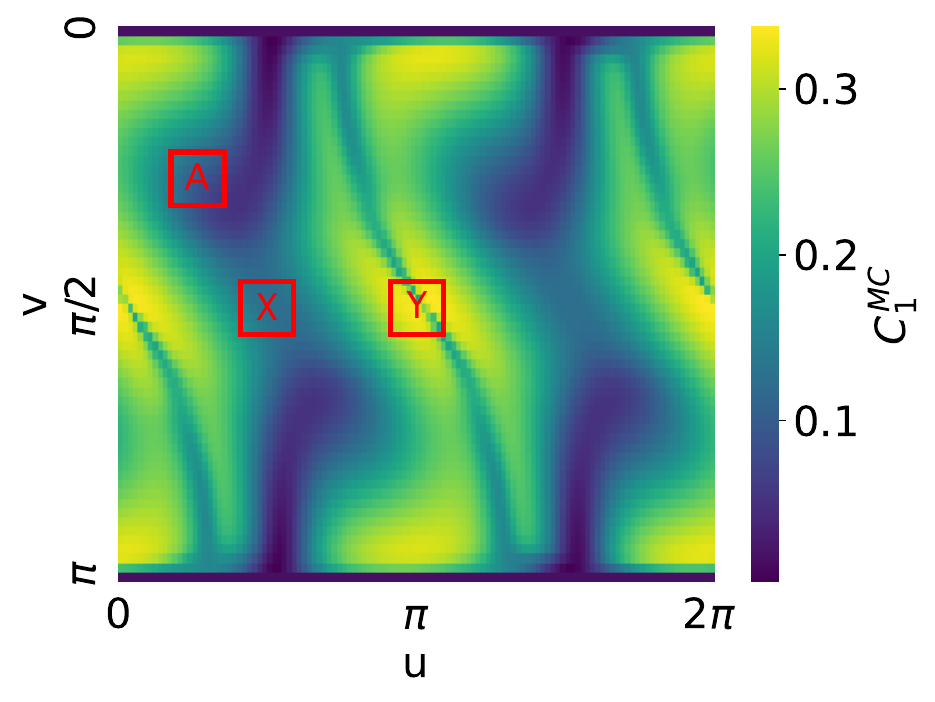}
}
\subfloat[]{
\label{subfig:sk_ipc_uv}
	\includegraphics[width=0.33\hsize]{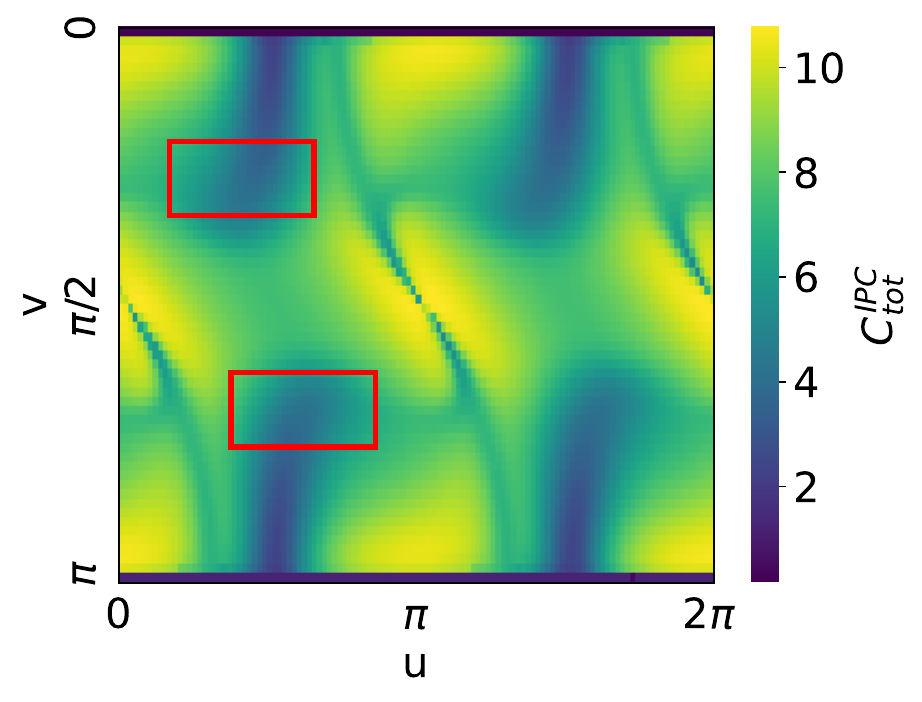}
}
\subfloat[]{
\label{subfig:sk_memory_decay_long}
	\includegraphics[width=0.33\hsize]{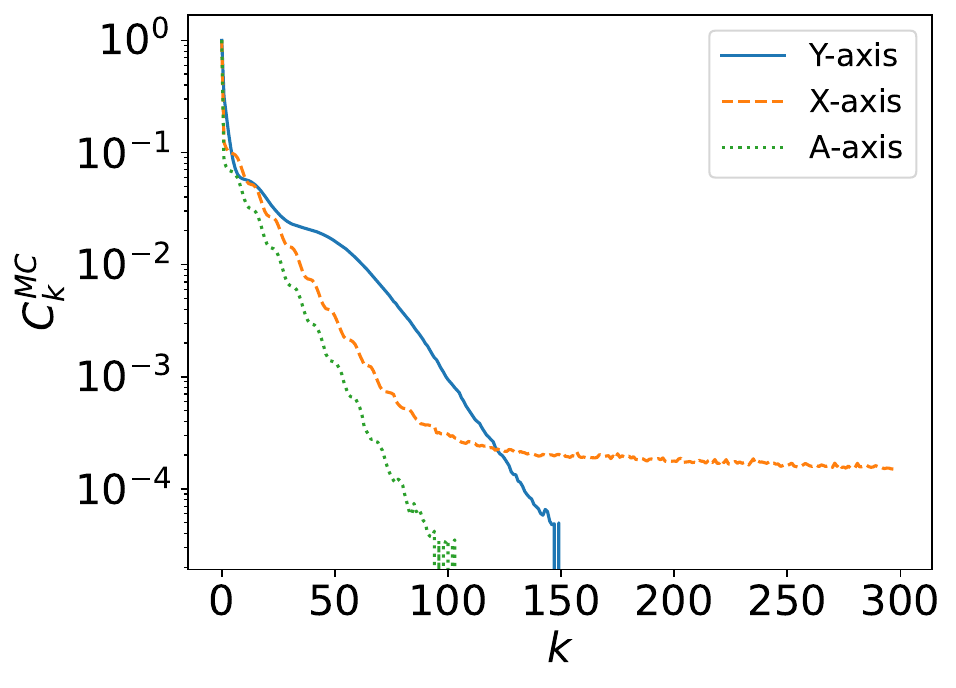}
}
\par
\subfloat[]{
\label{subfig:sk_max_delay}
	\includegraphics[width=0.33\hsize]{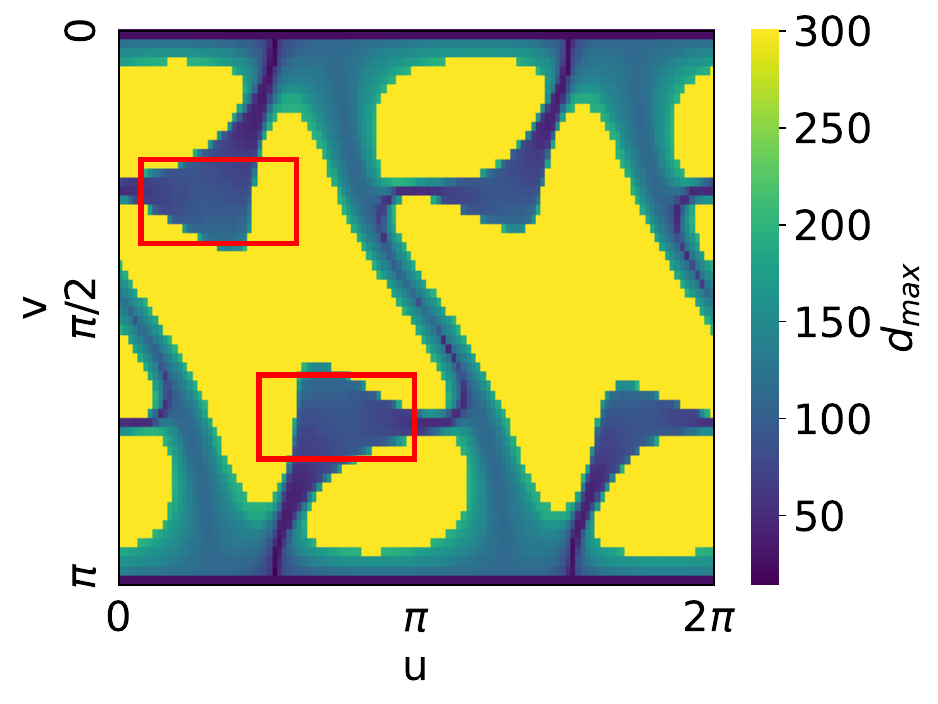}
}
\subfloat[]{
\label{subfig:sk_rank}
	\includegraphics[width=0.33\hsize]{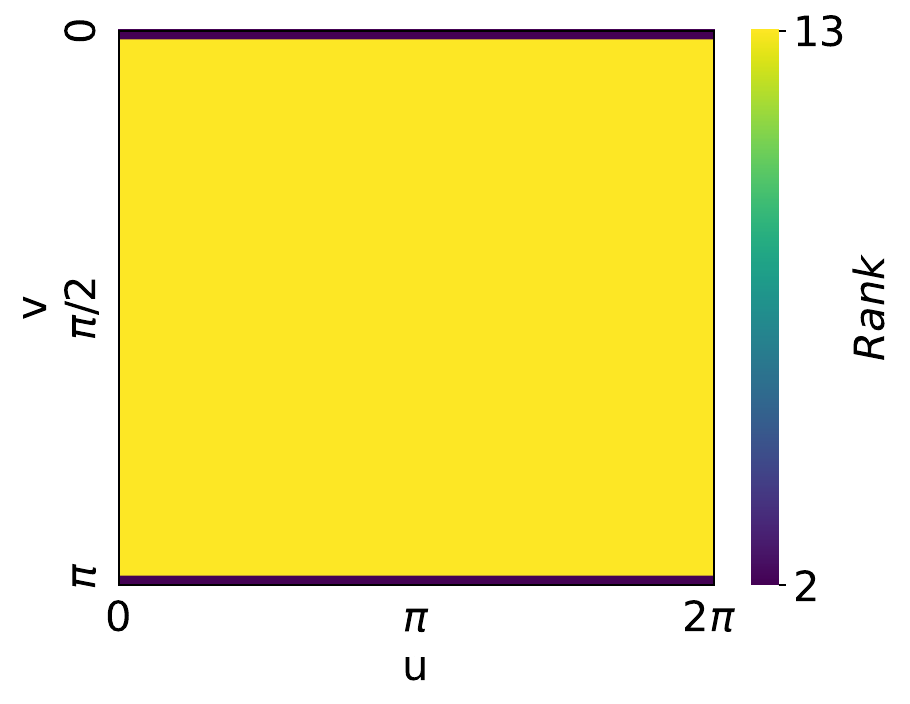}
}
\subfloat[]{
\label{subfig:sk_memory_decay_short}
	\includegraphics[width=0.33\hsize]{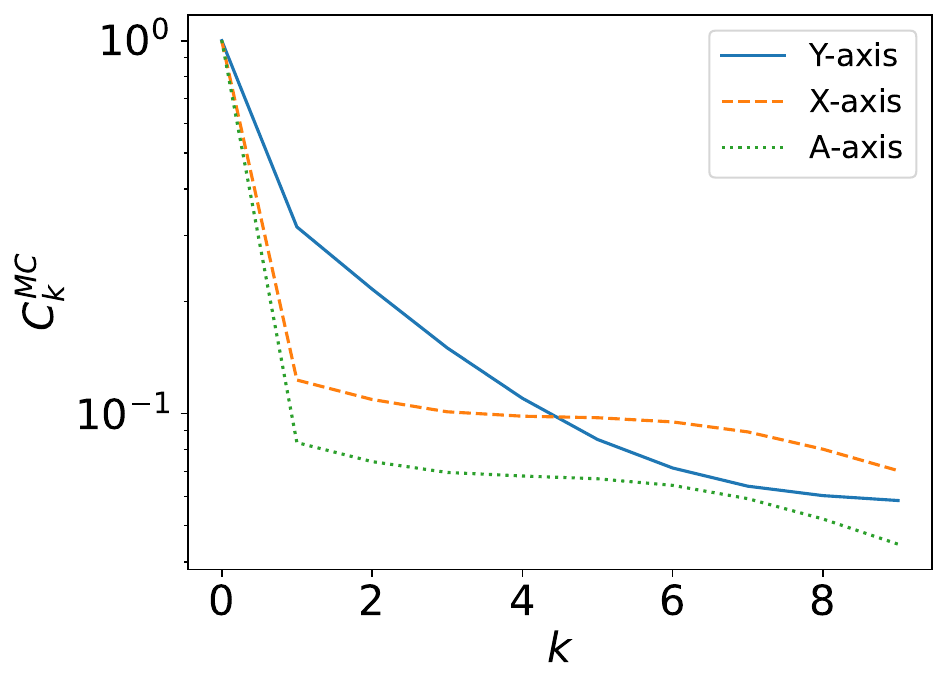}
}
\caption{(a) Memory function of delay 1. X, Y, and A denote different parameter configurations that are further explored in (c) and (f). (b) Total IPC.  (c,f) Memory function decay with respect to $k$ for different axes configurations within a timescale of (c) 300 steps and (f) 10 steps. The X-axis has a longer memory with a weaker short-term memory. The Y-axis has a strong short-term memory. The A-axis has a shorter and a weaker memory. (d) Maximum linear delay, computed by $\mathrm{argmax}_k( C^{MC}_{k} > 10^{-4})$. (e) Rank of the reservoir's output signals of length $10^6$ after washout of $3\cdot 10^5$ steps. Rank is $2$ at the north and south poles and $13$ for other configurations. Boxes in (b) and (d) denote typical parameter configurations exhibiting non-consistent behavior concerning NS-ESP in Fig.~\ref{subfig:sk_ns_esp_uv}.}

\end{figure*}

\subsubsection{MC and IPC}

Finally, we computed memory capacity (MC) \cite{Jaeger_2001} and information processing capacity (IPC) \cite{dambre2012} that measure how much linear and non-linear input dependency is within the output signal. Given a random input sequence $\{u_t \in \mathbb{R}\}$ and corresponding reservoir outputs $\{x_t \in \mathbb{R}^d\}$, where $d$ is the number of the observables involved, we computed MC and IPC for all parameter configurations as the ESP experiment. The total memory capacity $C^{MC}_{tot}$ can be calculated as
\begin{equation}
\begin{aligned}
	C^{MC}_{tot}(\{u_t\}, \{x_t\}) &= \sum_k C^{MC}_{k}(\{u_t\}, \{x_t\})\quad\text{where}\\
	C^{MC}_{k}(\{u_t\}, \{x_t\}) &= \frac{\mathbb{E}_t[u_{t-k}x_t^T] \mathbb{E}_t[x_tx_t^T] \mathbb{E}_t[u_{t-k}x_t]}{\mathbb{E}_t[\left(u_t - \mathbb{E}_t(u_t)\right)^2]}.
	\end{aligned}
\end{equation}

Here $C^{MC}_{k}$ is called a memory function of delay $k$. IPC can be calculated using products of orthogonal polynomial transformations of $\{u_k\}$ and its delayed sequences as target sequences. Namely, total IPC and degree $d$ IPC, $C^{IPC}_{d}$, can be calculated as
\begin{equation}
\begin{aligned}
	C^{IPC}_{tot}&\left(\{u_t\}, \{x_t\};\mathcal{Y}\right) = \sum_{d}C^{IPC}_{d}\left(\{u_t\}, \{x_t\}; \mathcal{Y}\right)\\
	C^{IPC}_{d}&\left(\{u_t\}, \{x_t\}; \mathcal{Y}\right) = \\
	&\sum_{\{v_t\} \in \mathcal{Y}\left(\{u_t\}; d\right)}
	\frac{\mathbb{E}_t[v_{t}x_t^T] \mathbb{E}_t[x_tx_t^T] \mathbb{E}_t[v_{t}x_t]}{\mathbb{E}_t[\left(v_t - \mathbb{E}_t(v_t)\right)^2]},
\end{aligned}
\end{equation}

where $\mathcal{Y}$ is a system of orthogonal polynomial and $\mathcal{Y}\left(\{u_t\}; d\right)$ is a set of transformed inputs, whose elements are a sequence having form in each time step $t$ as
\begin{equation}
	v_t = \prod_i Y_{d_i}(u_{t-k_i})\quad \text{s.t.}\quad \sum_{i}d_i = d.
\end{equation}
Here, $Y_{d}$ is a degree-$d$ polynomial in $\mathcal{Y}$, and $0 \leq k_i$ is the delay for $i$th component in the product. Ideally, $C^{IPC}_{tot}$ and $C^{IPC}_{d}$ will be calculated using polynomials of every degree and inputs of every delay. However, since this is impractical, we usually limit degree and delay range. In this experiment, IPC calculations are limited to maximum-delay/degree pairs denoted by $k$-$d$ for maximum degree $k$ and maximum delay $d$ as follows: $1$-$300$, $2$-$100$, $3$-$30$, $4$-$10$ and $5$-$10$. We further select a subset of delay-degree configurations by computing IPC with all configurations for the aforementioned maximum-delay/degree pairs for some of the parameter configurations, then selecting delay-degree configurations that significantly contribute to $C^{IPC}_{tot}$.

We observed that the low values of memory function of delay $k=1$ ($C^{MC}_{1}$) depicted as the navy purple (dark gray) region in Fig.~\ref{subfig:sk_mc1_uv} partially corresponds to the yellow (light gray) high-error region in Fig.~\ref{subfig:sk_narma2_uv}. Furthermore, the maximum linear delay, computed by $\mathrm{argmax}_k( C^{MC}_{k} > 10^{-4})$, shown in Fig.~\ref{subfig:sk_max_delay}, has global consistency with the non-stationary ESP  depicted in Fig.~\ref{subfig:sk_ns_esp_uv}. However, regions in Fig.~\ref{subfig:sk_max_delay} indicated by red (gray) rectangles are inconsistent with the corresponding regions in Fig.~\ref{subfig:sk_ns_esp_uv}. It can be speculated that those regions have longer memory in non-linear terms, as we can see from the non-saturating IPC values in the region, also indicated by red (gray) rectangles in Fig.~\ref{subfig:sk_ipc_uv}.

Figure~\ref{subfig:sk_rank} depicts the rank of a state sequence, computed using the number of significant singular values of a matrix formed by arranging every state in the time direction. It is expected that the total IPC ($C^{IPC}_{tot}$) in Fig.~\ref{subfig:sk_ipc_uv} should saturate at the value of these rank values. For instance, as the rank is $13$ for every configuration except the northern and southern poles in Fig.~\ref{subfig:sk_rank}, we anticipate that the low IPC region in Fig.~\ref{subfig:sk_ipc_uv} would vanish if a sufficient number of higher-degree/longer-delay nonlinear memory functions were collected. However, this is computationally infeasible in our current environment due to the high calculation cost of IPC for higher-degree capacities.

The annotated parameter region, X, Y, and the A-axis in Fig.~\ref{subfig:sk_mc1_uv}, has different memory decay properties. As we can see in Fig.~\ref{subfig:sk_memory_decay_long} and Fig.~\ref{subfig:sk_memory_decay_short}, the X-axis input corresponds to small short-term memory and large long-term memory, the Y-axis input corresponds to large short-term memory and no long-term memory, and the A-axis input corresponds to small short-term and long-term memories.

\subsection{Subset non-stationary ESP of QRC}
\label{subsec:subset}
\begin{figure}[h]
\centering
\includegraphics[width=0.8\hsize]{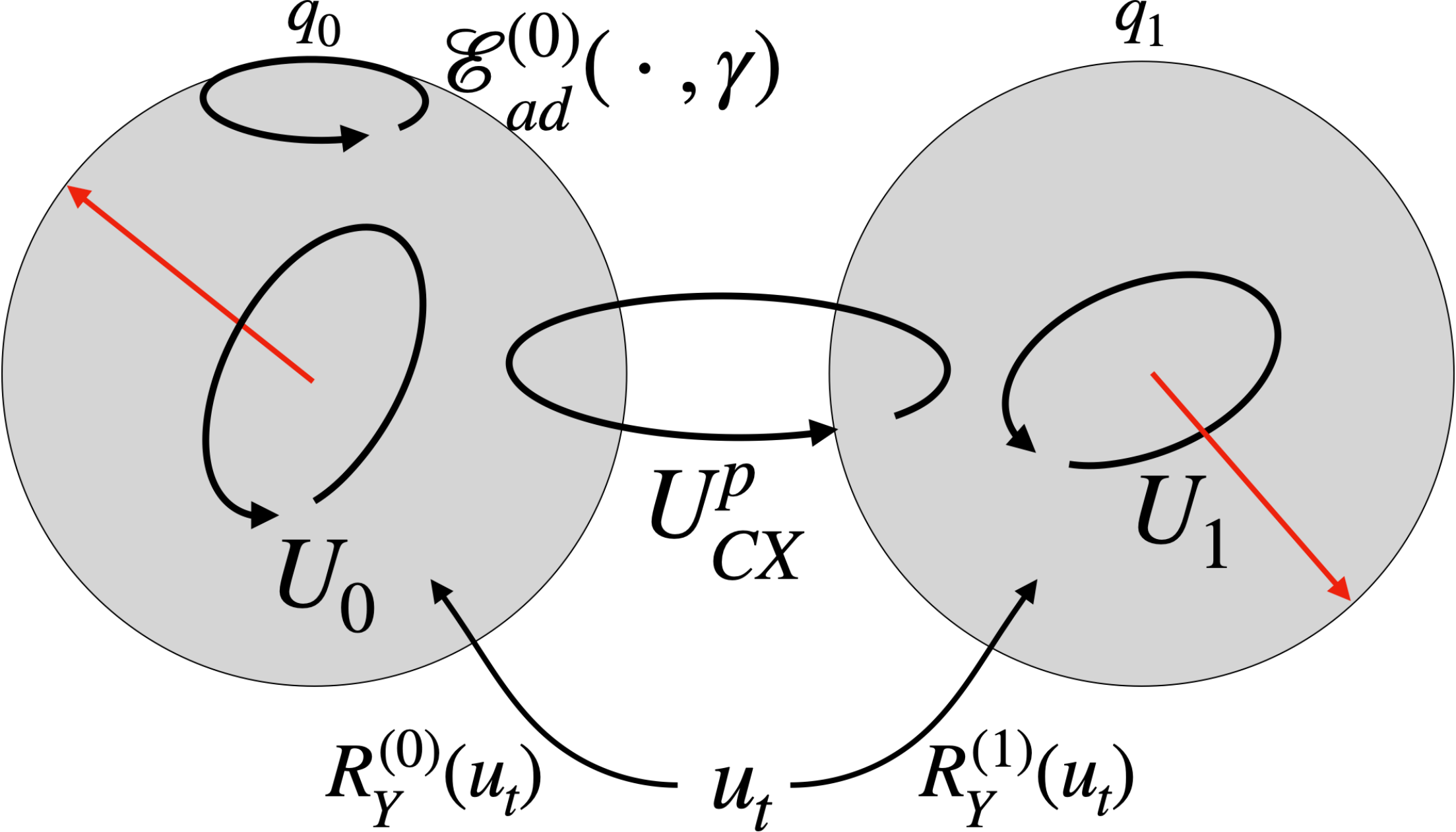}
\caption{Schematics of subset non-stationary ESP experiment setup. We have two qubits: $q_0$ and $q_1$. Only $q_0$ has amplitude damping channel $\mathcal{E}_{ad}^{(0)}$. Although $U_0$ and $U_1$ are local to each qubits, $q_0$ and $q_1$, respectively, $U_{CX}^p$ induces entangling dynamics.}	
\label{fig:subset-esp-exp}
\end{figure}
\begin{figure}[t]
\centering

\subfloat[]{
\label{subfig:subset_ns_esp_all}
	\includegraphics[width=0.6\hsize]{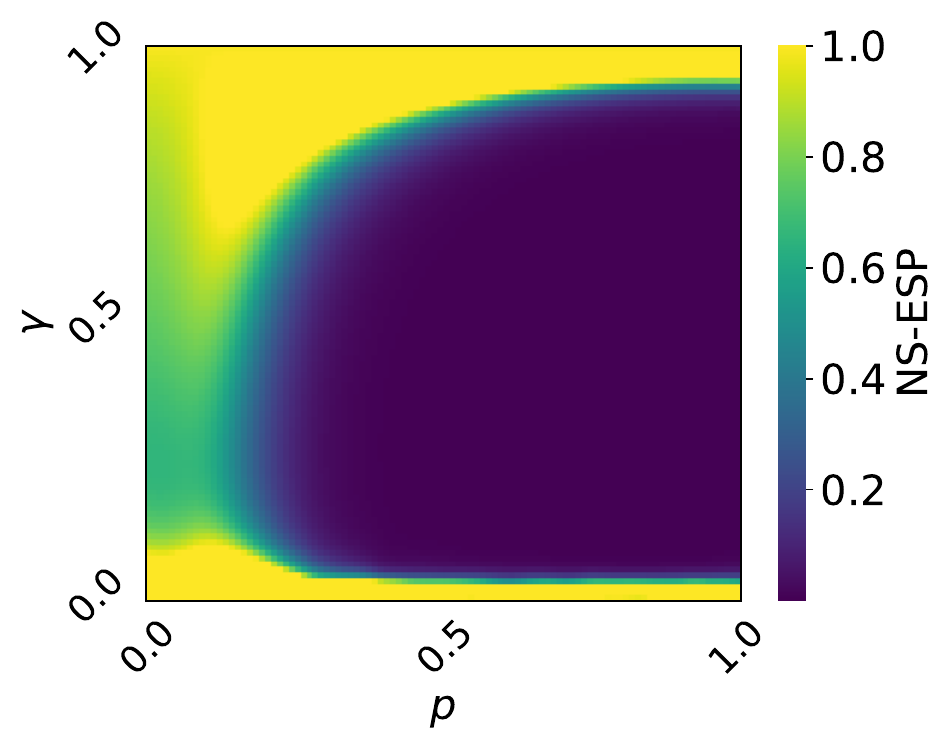}
}
\par
\subfloat[]{
\label{subfig:subset_ns_esp_q0}
	\includegraphics[width=0.6\hsize]{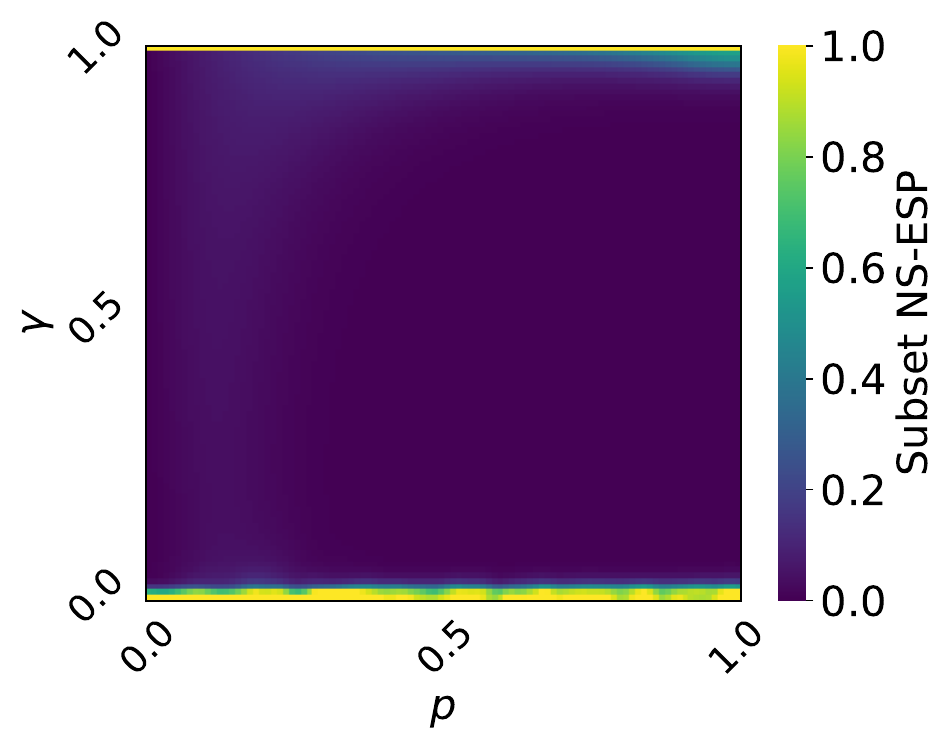}
}
\par
\subfloat[]{
\label{subfig:subset_ns_esp_q1}
	\includegraphics[width=0.6\hsize]{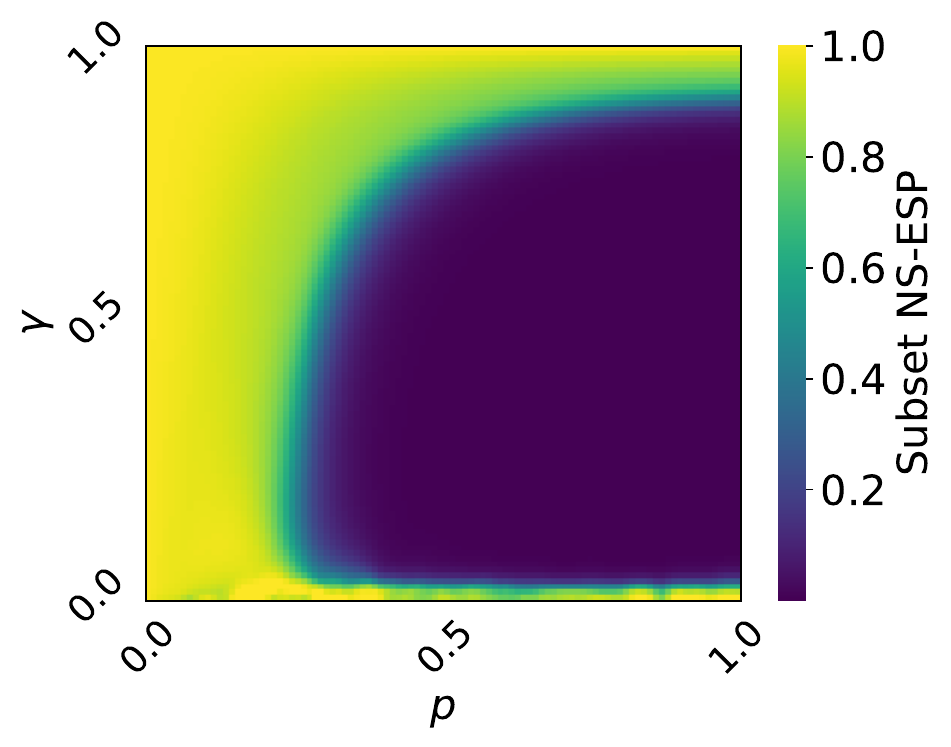}
}

\caption{
(a) Non-stationary ESP of the total system. Subset non-stationary ESP of (b) the damping subsystem, and (c) the non-damping subsystem.
}
\label{fig:qrc-subset-ns-esp}
\end{figure}
We also demonstrated numerically that the subset non-stationary ESP can characterize the information processing power of systems with partial fading memory. We devised a two-qubit system; each of its qubits had random unitary $U_0$ and $U_1$, respectively, as its dynamics. One of the qubits had amplitude damping channel $\mathcal{E}_{ad}^{(0)}(\cdot, \gamma)$ of damping rate $\gamma$ applied locally to the subsystem $0$ after local unitary evolution. Here, an amplitude damping channel $\mathcal{E}_{ad}^{(0)}(\cdot, \gamma)$ was defined as 
\begin{equation}
\begin{aligned}
	\mathcal{E}_{ad}^{(0)}(\rho, \gamma) &= \left(K_0(\gamma)^{(0)} \otimes I^{(1)} \right)\rho \left(K_0^\dagger(\gamma)^{(0)} \otimes I^{(1)}\right)\\
	&+ \left(K_1(\gamma)^{(0)} \otimes I^{(1)} \right)\rho \left(K_1^\dagger(\gamma)^{(0)} \otimes I^{(1)}\right)
	\end{aligned}
\end{equation}
where 
\begin{equation}
	\begin{aligned}
		K_0(\gamma) &= \begin{pmatrix}
			1 & 0\\
			0 & \sqrt{1 - \gamma}
		\end{pmatrix}\quad \text{and}\\
		K_1(\gamma) &= \begin{pmatrix}
			0 & \sqrt{\gamma}\\
			0 & 0
		\end{pmatrix}.
	\end{aligned}
\end{equation}
 Furthermore, we modulated the entanglement between two qubits by applying \add{two-qubits unitary that is inspired by a controlled-NOT (CNOT) operation in the QC framework. A controlled-NOT operation is a multiplication of a matrix $U_{CX}$ defined as follows:}
\begin{equation}
U_{CX} = \begin{pmatrix}
	1 & 0 & 0 & 0\\
	0 & 1 & 0 & 0\\
	0 & 0 & 0 & 1\\
	0 & 0 & 1 & 0\\
\end{pmatrix}.
\end{equation}
\add{For instance, by applying $U_{CX}$, a tensor product density matrix $\rho_{+0} = \begin{pmatrix}
	\frac{1}{2} & \frac{1}{2}\\
	\frac{1}{2} & \frac{1}{2}
\end{pmatrix} \otimes \begin{pmatrix}
	1 & 0\\
	0 & 0
\end{pmatrix}$ becomes
\begin{equation}
\begin{aligned}
	&(U_{CX})\rho_{+0}(U_{CX}^\dagger)\\ &= \begin{pmatrix}
	1 & 0 & 0 & 0\\
	0 & 1 & 0 & 0\\
	0 & 0 & 0 & 1\\
	0 & 0 & 1 & 0\\
\end{pmatrix}\begin{pmatrix}
		\frac{1}{2} & 0 & \frac{1}{2} & 0\\
		0 & 0 & 0 & 0\\
		\frac{1}{2} & 0 & \frac{1}{2} & 0\\
		0 & 0 & 0 & 0\\
	\end{pmatrix}\begin{pmatrix}
	1 & 0 & 0 & 0\\
	0 & 1 & 0 & 0\\
	0 & 0 & 0 & 1\\
	0 & 0 & 1 & 0\\
\end{pmatrix}\\
	&= \begin{pmatrix}
		\frac{1}{2} & 0 & 0 & \frac{1}{2}\\
		0 & 0 & 0 & 0\\
		0 & 0 & 0 & 0\\
		\frac{1}{2} & 0 & 0 & \frac{1}{2}\\
	\end{pmatrix},
\end{aligned}
\end{equation}
which cannot be written as a tensor product of single qubit density matrices. A two-qubit density matrix that cannot be written as a tensor product of single-qubit density matrices is called \textit{entangled}. Therefore, the CNOT operation is called an \textit{entangling} unitary. We modified our entangling unitary by taking a matrix power of $U_{CX}$. We applied $U_{CX}^p$ in each time step, where $p$ represents an exponent of the matrix power.}
Namely, if $p=0$, it equaled an identity, and if $p=1$, it equaled a CNOT operation that took $q_0$ as the control qubit. Overall, non-input-driven system dynamics became

\begin{equation}
\label{eqn:subset_exp_eq	}
\begin{aligned}
	\rho &\to \mathcal{E}_{sys}(\rho)\ \  \textrm{where} \\
 \mathcal{E}_{sys}(\rho) &\equiv U_{CX}^p\mathcal{E}_{ad}^{(0)}\left((U_0 \otimes U_1) \rho (U_0^\dagger \otimes U_1^\dagger), \gamma\right)\left(U_{CX}^\dagger\right)^p.
 \end{aligned}
\end{equation}
\begin{figure*}[t]
\centering
\subfloat[]{
\label{subfig:subset_narma2_all}
	\includegraphics[width=0.3\hsize]{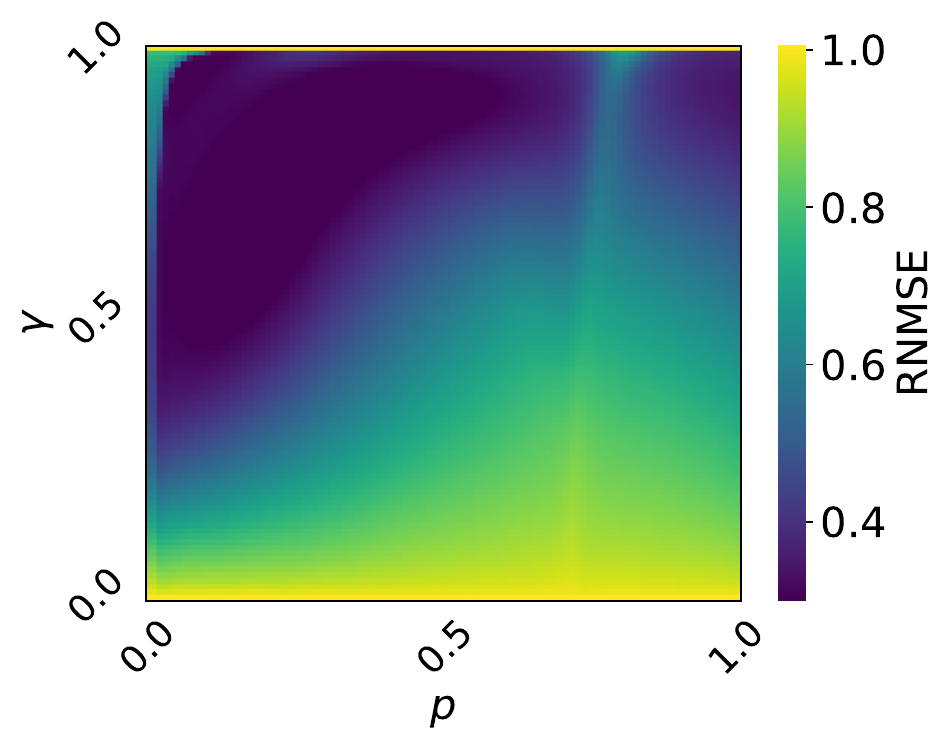}
}
\subfloat[]{
\label{subfig:subset_narma2_q1}
	\includegraphics[width=0.3\hsize]{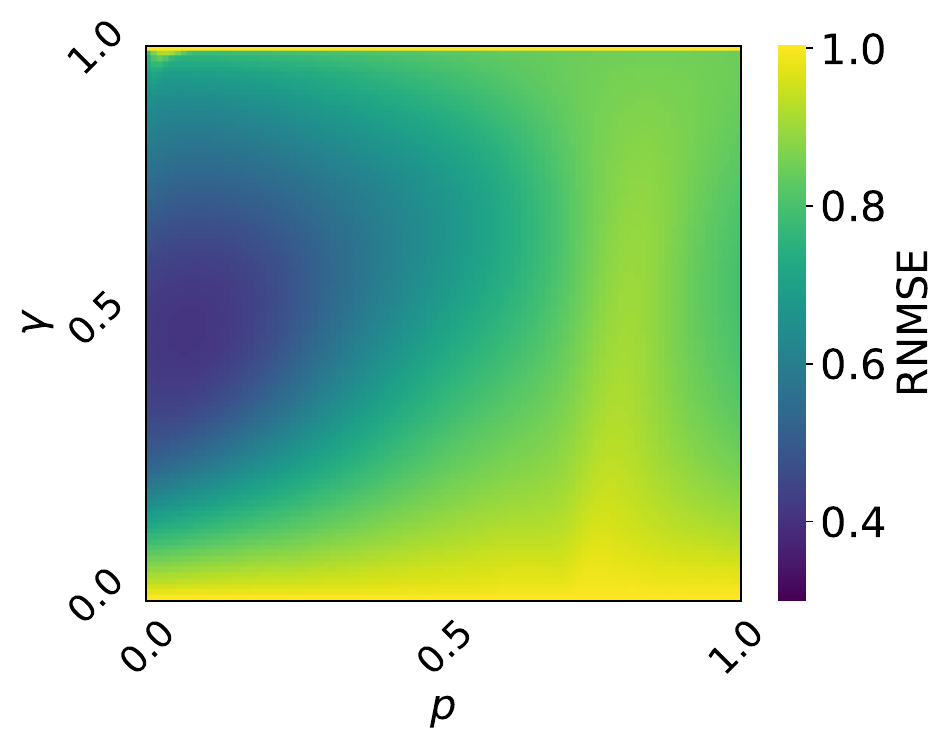}
}
\subfloat[]{
\label{subfig:subset_narma2_q0}
	\includegraphics[width=0.3\hsize]{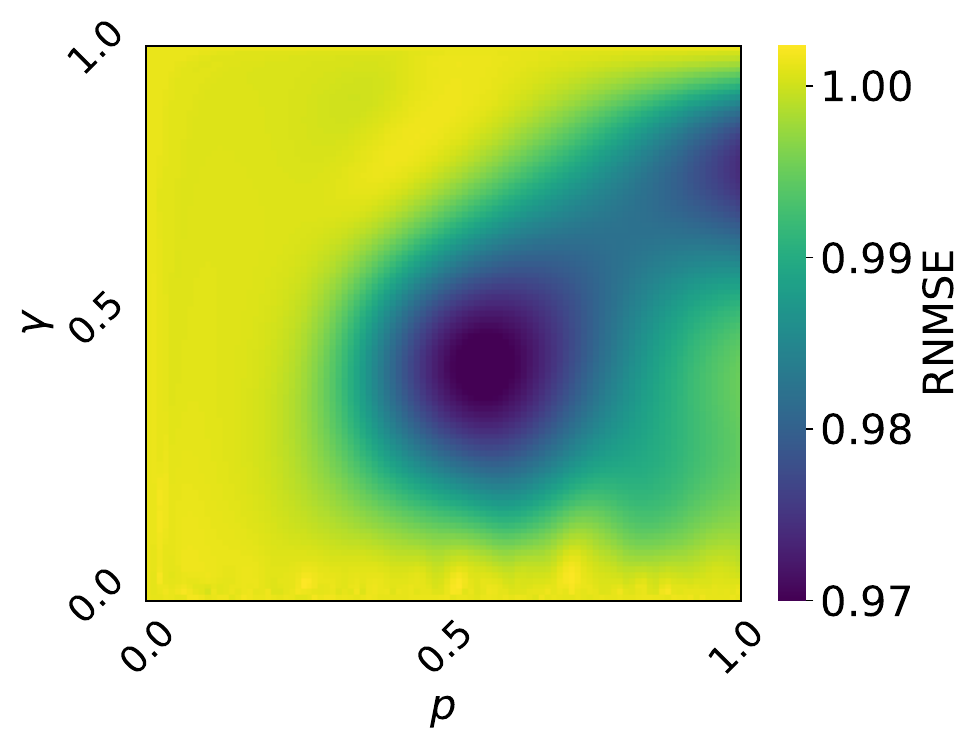}
}
\par
\subfloat[]{
\label{subfig:subset_mc_all}
	\includegraphics[width=0.3\hsize]{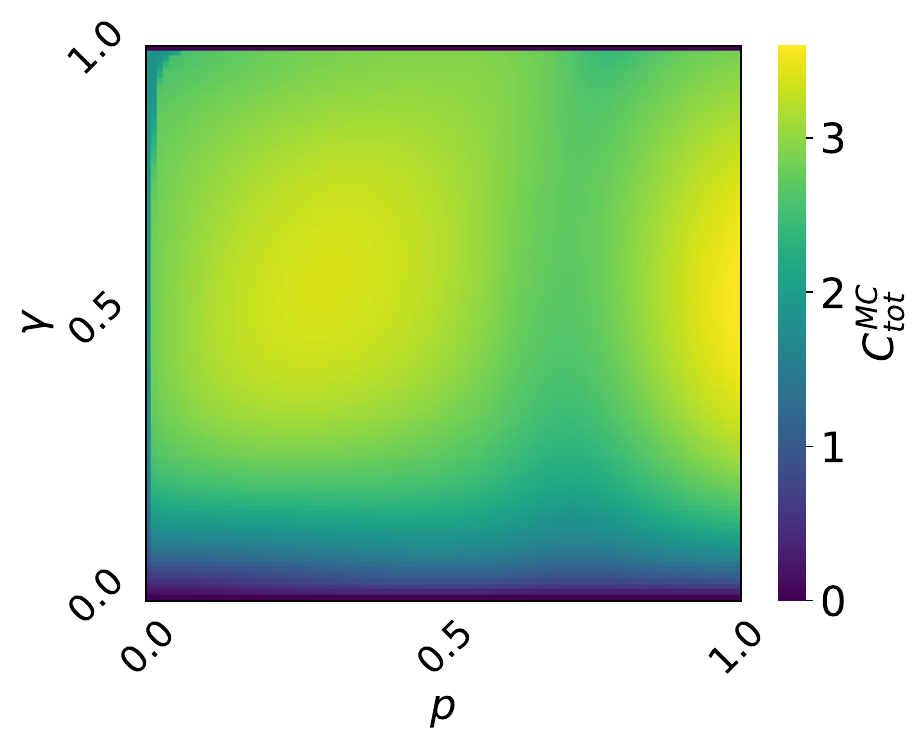}
}
\subfloat[]{
\label{subfig:subset_mc_q1}
	\includegraphics[width=0.3\hsize]{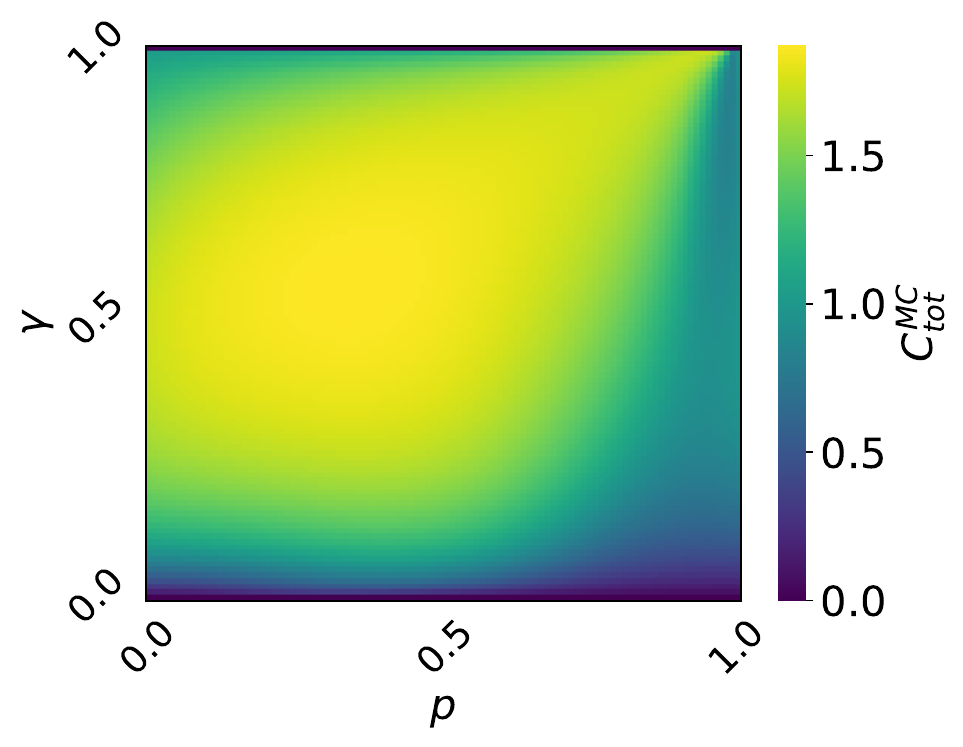}
}
\subfloat[]{
\label{subfig:subset_mc_q0}
	\includegraphics[width=0.3\hsize]{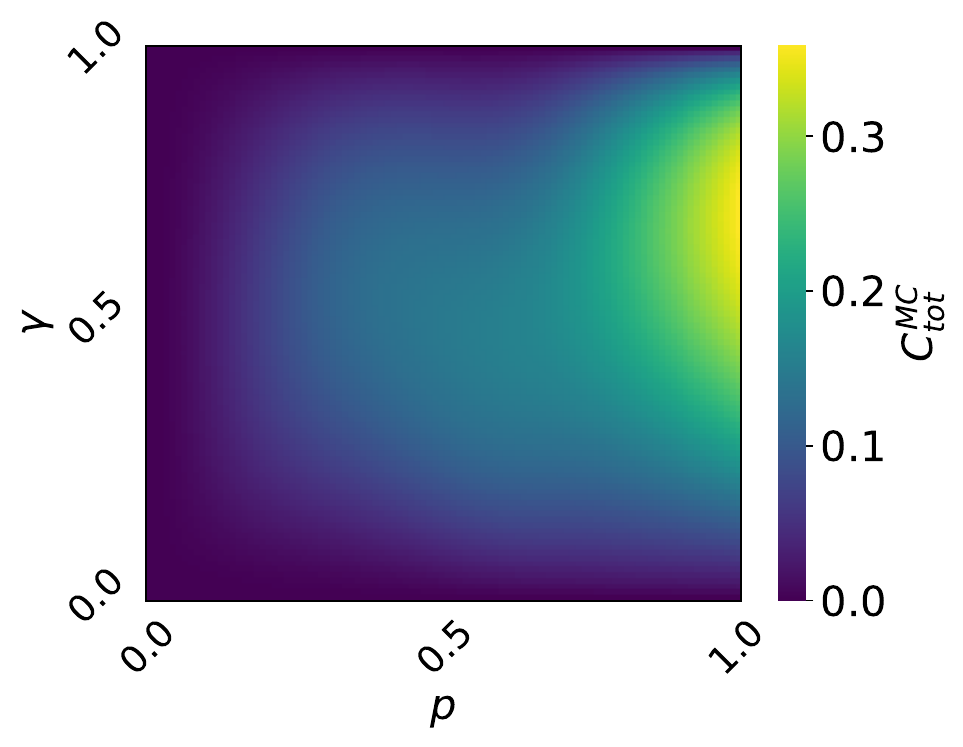}
}
\par
\subfloat[]{
\label{subfig:subset_ipc_2+_all}
	\includegraphics[width=0.3\hsize]{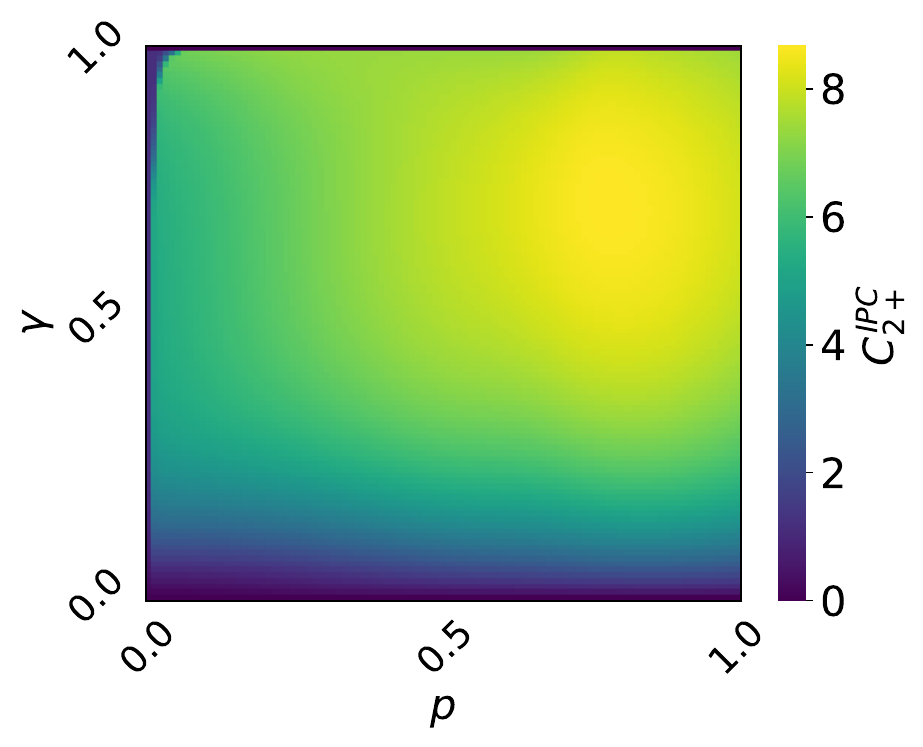}
}
\subfloat[]{
\label{subfig:subset_ipc_2+_q1}
	\includegraphics[width=0.3\hsize]{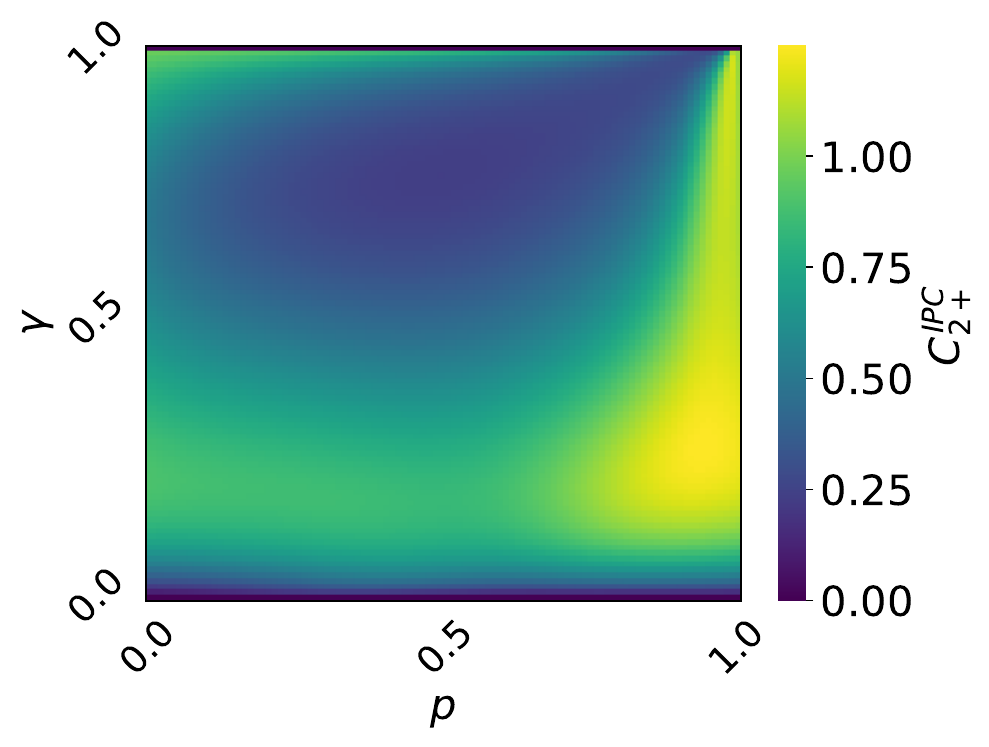}
}
\subfloat[]{
\label{subfig:subset_ipc_2+_q0}
	\includegraphics[width=0.3\hsize]{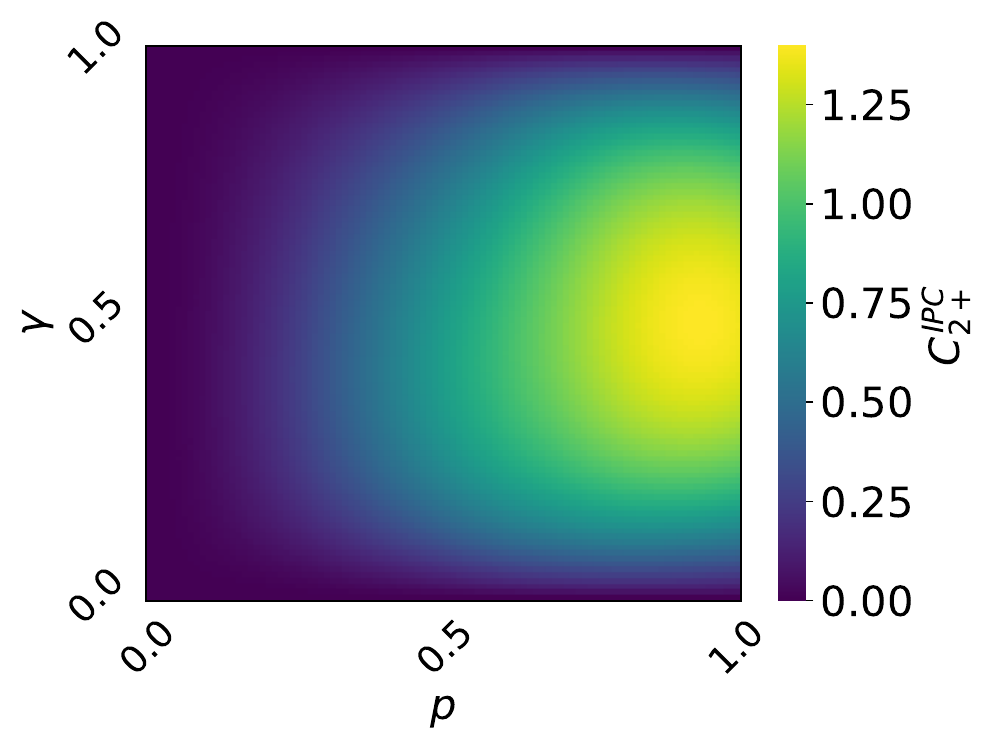}
}
\par
\subfloat[]{
\label{subfig:subset_rank_all}
	\includegraphics[width=0.3\hsize]{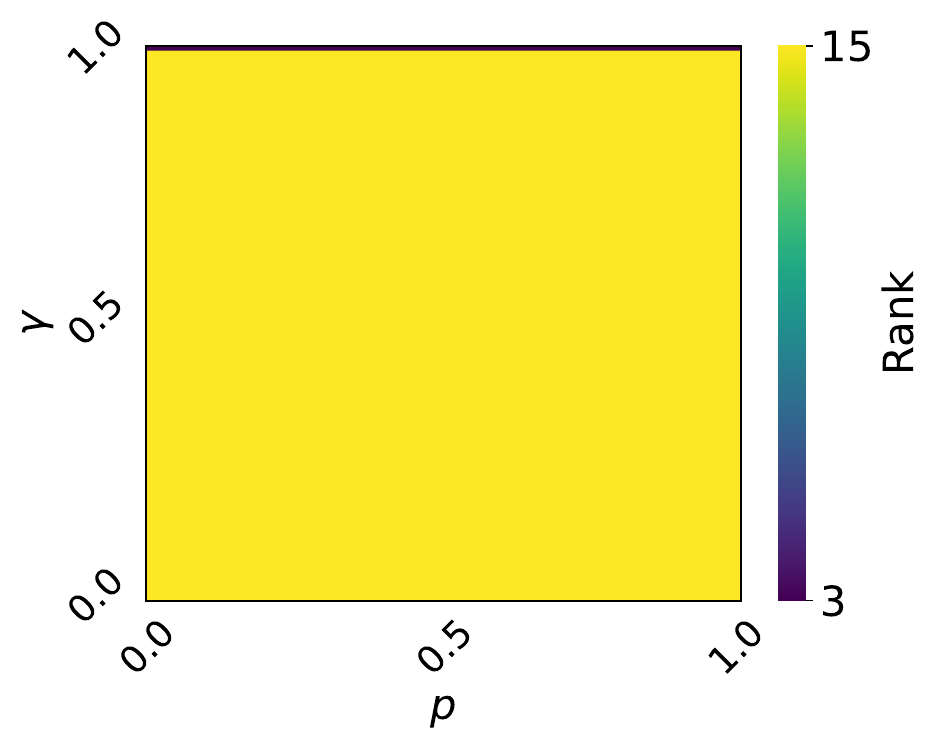}
}
\subfloat[]{
\label{subfig:subset_rank_q1}
	\includegraphics[width=0.3\hsize]{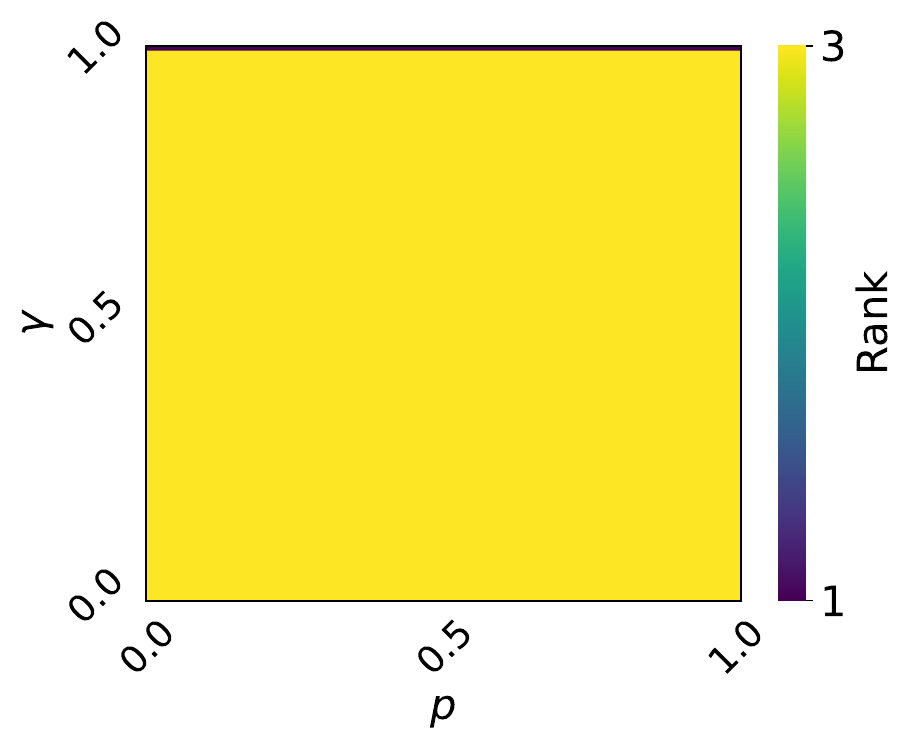}
}
\subfloat[]{
\label{subfig:subset_rank_q0}
	\includegraphics[width=0.3\hsize]{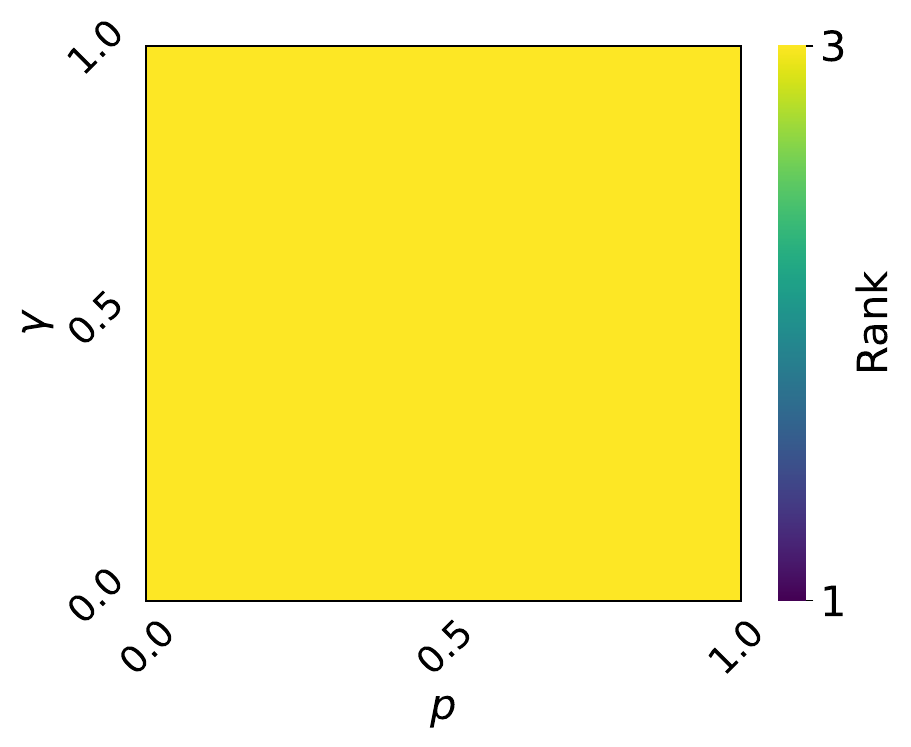}
}
\caption{
(a-c) RNMSE of NARMA2 task, (d-f) linear memory capacity ($C^{MC}_{tot}$), (g-i) non-linear memory capacity ($C^{IPC}_{2+}$) and rank of reservoir's output signals (j-l) of (a,d,g,j) total system, (b,e,h,k) the damping subsystem, and (c,f,i,l) the non-damping subsystem.
}
\label{fig:subset_narma_mc}
\end{figure*}

In this setup, the entanglement between two qubits can be devised by changing $p$; the system dynamics is a complete tensor product of single qubit dynamics when $p=0$ and $p$-dependent entanglements are introduced based on the relationship between $U_{CX}^p$, $U_0$, and $U_1$ when $p>0$.

The input encoding also has a tensor product structure. Namely, given an input $\mathbf{u}\in[-1, 1]$, our input encoding unitary $U_{in}$ was
\begin{equation}
    U_{in}(u) \equiv R_Y^{(0)}(\mathrm{arccos}(\mathbf{u})) \otimes R_Y^{(1)}(\mathrm{arccos}(\mathbf{u})).
\end{equation}

Therefore, the final input-driven dynamics became
\begin{equation}
\begin{aligned}
    \rho_{t+1} &= \mathcal{E}_{in}\left(\mathcal{E}_{sys}\left(\rho\right), \mathbf{u}_{t}\right)\ \ \ \textrm{where}\\
    &\mathcal{E}_{in}(\rho, \mathbf{u}_{t}) \equiv  U_{in}(\mathbf{u}_t)\rho  U_{in}^\dagger(\mathbf{u}_t).
\end{aligned}
\end{equation}

\subsubsection{Subset ESP and subset non-stationary ESP}

Similar to the setup in the non-stationary ESP experiments, we computed the
 non-stationary ESP indicator (Eq.~\eqref{eqn:ns_esp_indicator}) for every configuration of $\gamma$ and $p$. The other setups are the same in the non-stationary ESP experiment in Sec~\ref{subsubsec:ns_esp}. We randomly sampled four input sequences of size $200$ from a uniform distribution over the interval $[-1,1]$ and three initial reservoir states from Haar random distribution of 2-qubit pure states.

Figure~\ref{fig:qrc-subset-ns-esp} depicts the non-stationary ESP of the total system (Fig.~\ref{subfig:subset_ns_esp_all}), the subset non-stationary ESP of the damping subsystem (Fig.~\ref{subfig:subset_ns_esp_q0}), and the non-damping subsystem (Fig.~\ref{subfig:subset_ns_esp_q1}). As we can see, the damping subsystem has a strong subset non-stationary ESP in almost every parameter region except when there is no amplitude damping ($\gamma=0$). Therefore, we can expect that the damping subsystem has fading memory in almost every parameter configurations of $\gamma$ and $p$ even when the total system does not in a certain configuration.

\subsubsection{NARMA tasks, MC and IPC}

Next, in Fig.~\ref{fig:subset_narma_mc}, we examined the relationship between the subset non-stationary ESP and the information processing capabilities as indicated by performance on the NARMA2 task, MC, and IPC. The input, target, and metrics settings were identical to those used in the non-stationary ESP experiments in Sec.~\ref{sec:ns_esp_qrc}. We observed that the NARMA2 task performance and MC were more closely related to the subset non-stationary ESP of the damping subsystem than to that of the total system. Specifically, in regions of small $p$ and large $\gamma$, both NARMA2 performance and MC were good, even if the non-stationary ESP of the total system was not maintained, as shown in Fig.~\ref{subfig:subset_ns_esp_all}. In the NARMA2 task, the performance of the total system was also affected by the performance of the damping subsystem, as evidenced in the lower right part of Fig.~\ref{subfig:subset_narma2_all} and Fig.~\ref{subfig:subset_narma2_q0}.

We hypothesize that the low-performance range around $p\sim 0.7$ in Fig.~\ref{subfig:subset_narma2_all} results from a sweet spot in the trade-off between the fading memory in the damping subsystem and the information propagation between entangling and non-entangling bases, which is modulated by the completeness of the CNOT gate: $p$.

\label{subsubsec:tradeoff_entanglement_fading_memory}
\begin{figure}[t]
\centering
\subfloat[]{
\label{subfig:subset_mc0_q1}
	\includegraphics[width=0.5\hsize]{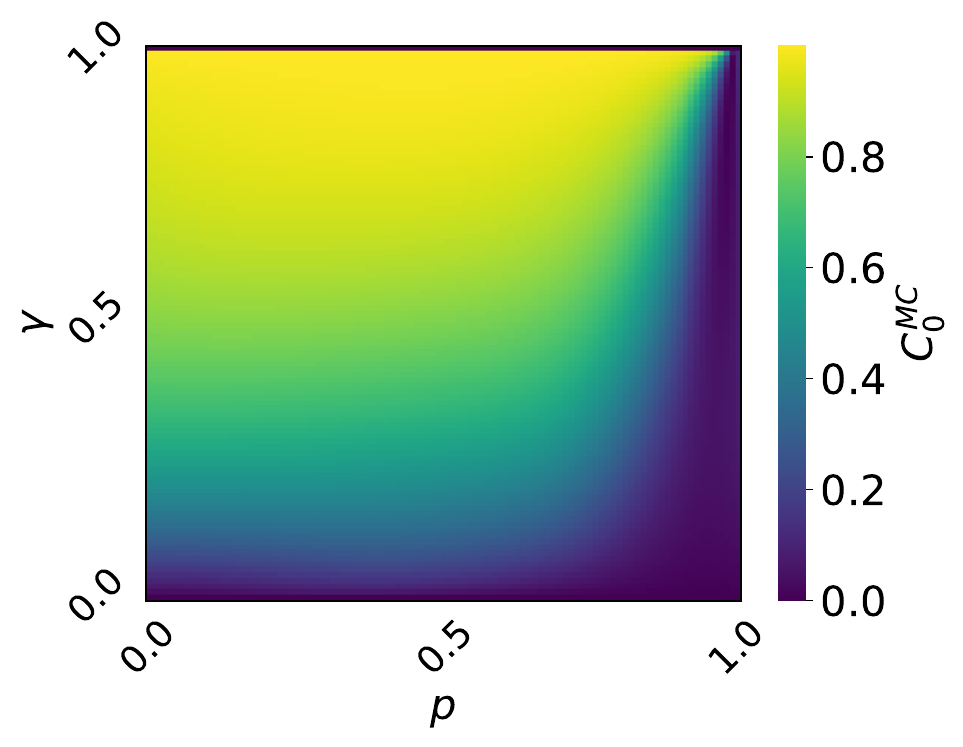}
}
\subfloat[]{
\label{subfig:subset_mc1_q1}
	\includegraphics[width=0.5\hsize]{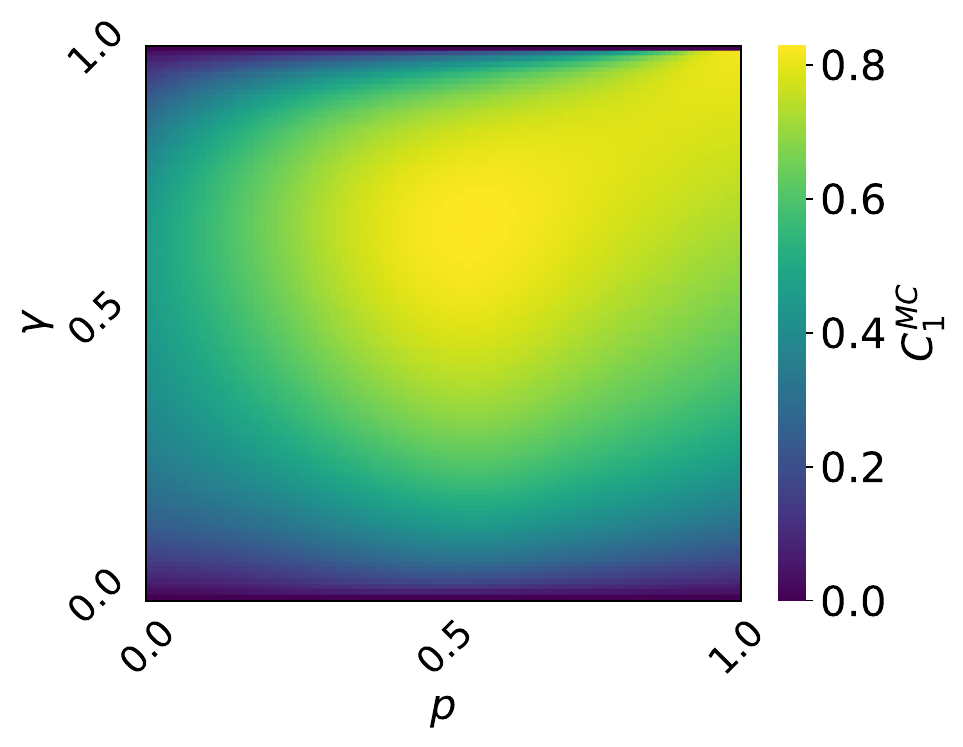}
}
\par
\subfloat[]{
\label{subfig:subset_mc2_q1}
	\includegraphics[width=0.5\hsize]{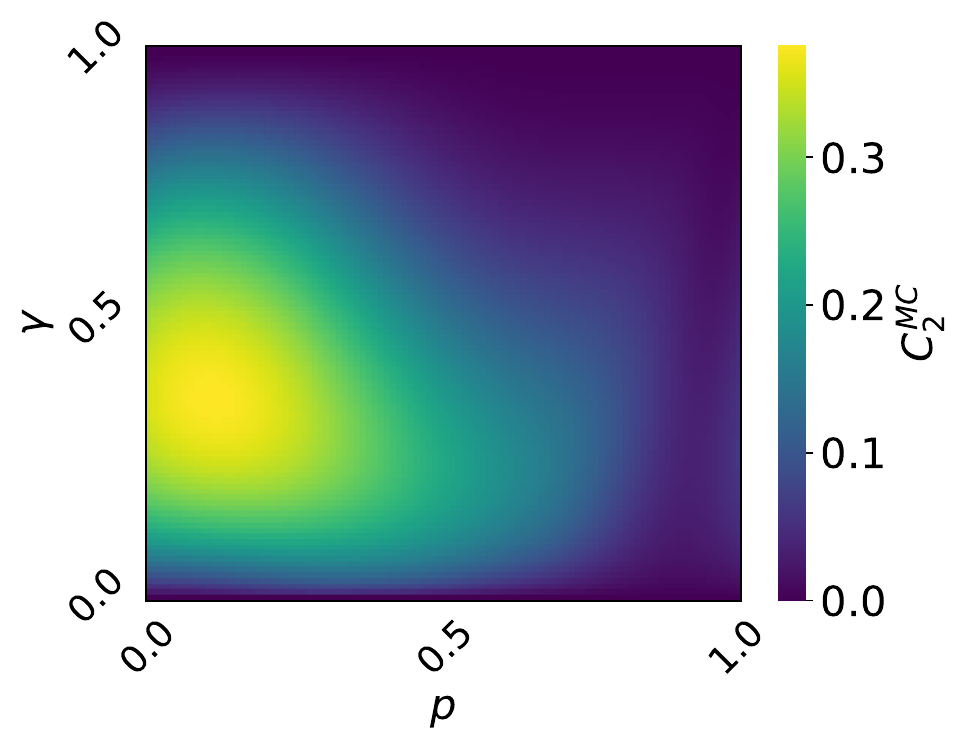}
}
\subfloat[]{
\label{subfig:subset_mc3_q1}
	\includegraphics[width=0.5\hsize]{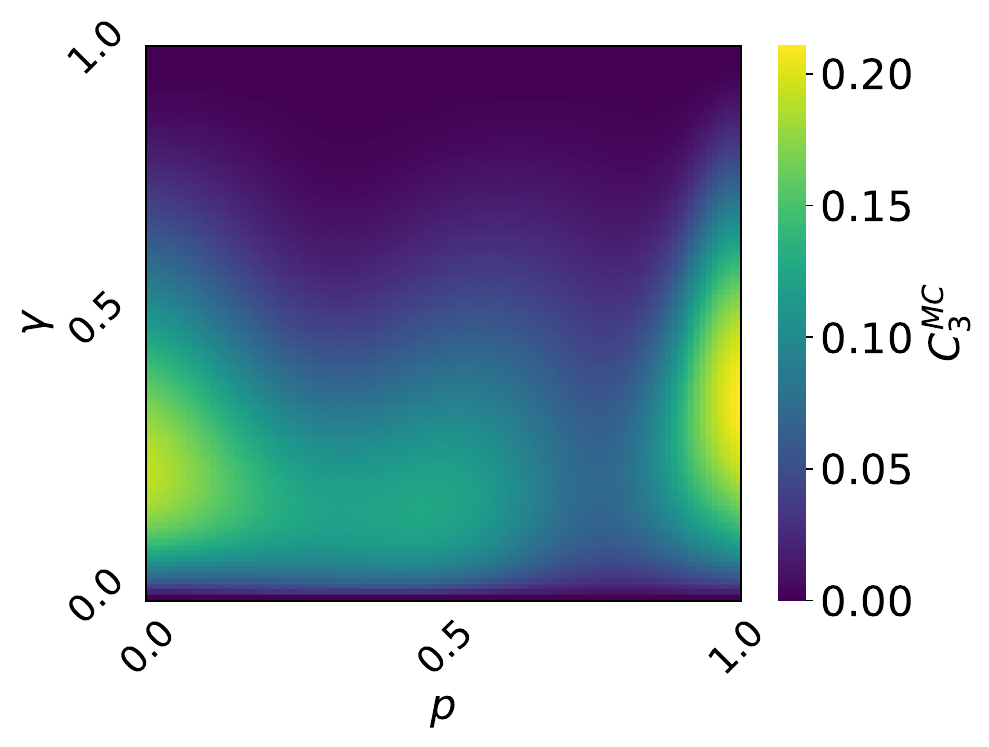}
}
\par
\subfloat[]{
\label{subfig:subset_mc_even_q1}
	\includegraphics[width=0.5\hsize]{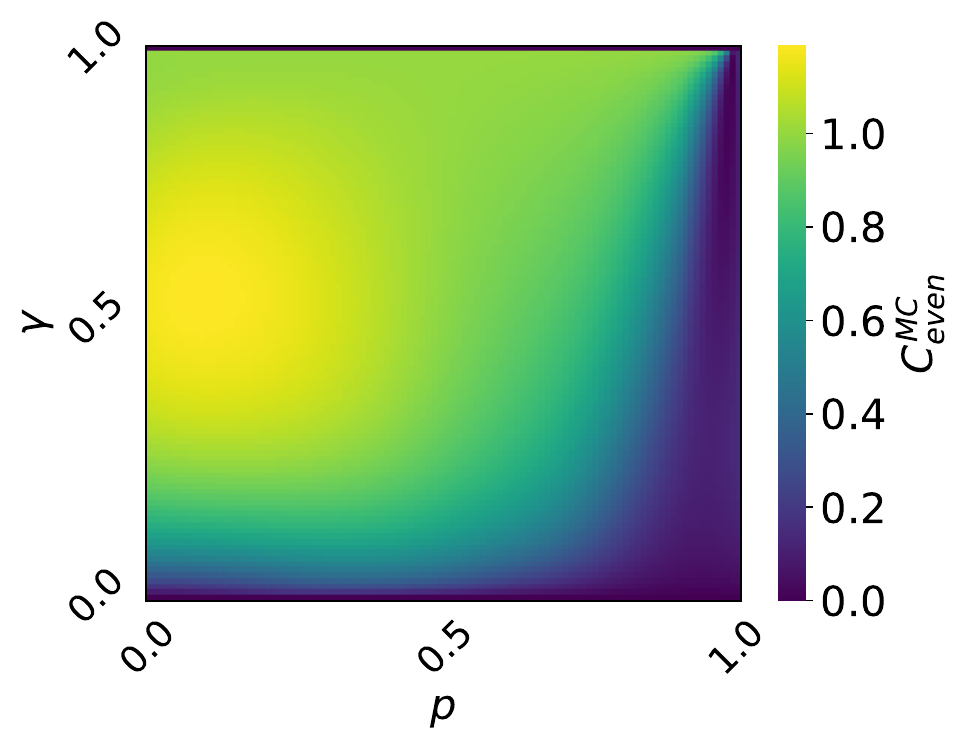}
}
\subfloat[]{
\label{subfig:subset_mc_odd_q1}
\includegraphics[width=0.5\hsize]{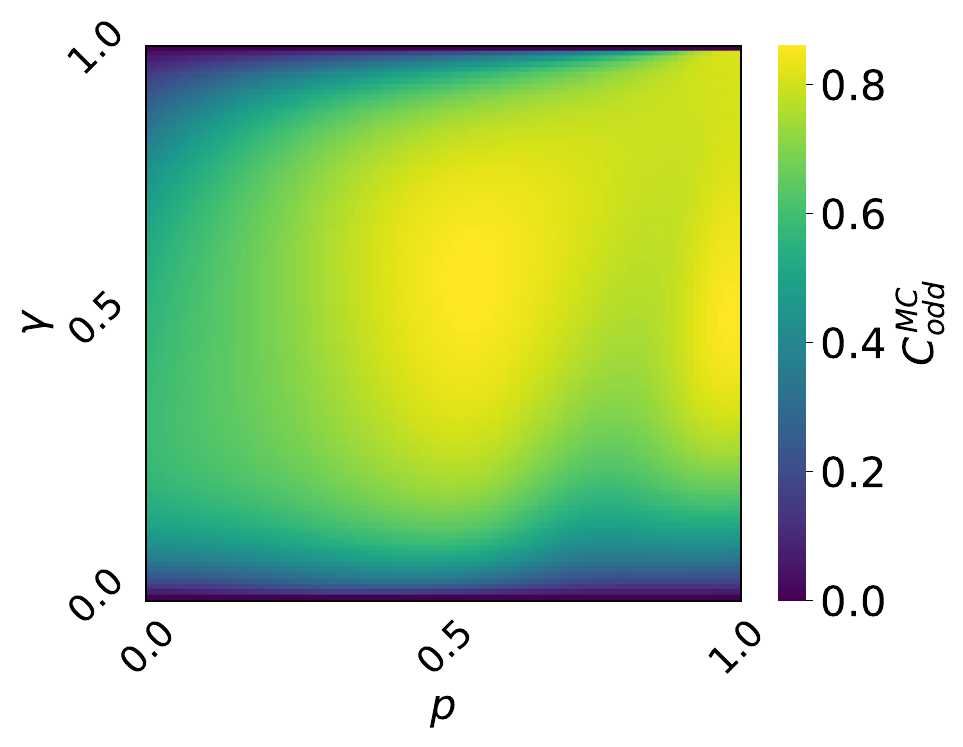}
}

\caption{
(a) $C^{MC}_{0}$, (b) $C^{MC}_{1}$, (c) $C^{MC}_{2}$, (d) $C^{MC}_{3}$, (e) $C^{MC}_{even}$, and (f) $C^{MC}_{odd}$ of the damping subsystem. 
}
\label{fig:q1_mc_0_3}
\end{figure}
The mechanism of the trade-off includes utilizing entangling basis: $\left\{\bigotimes P_i \mid P_i \in \{X, Y, Z\}\right\}$ as memory as follows. ESP does not hold if $\gamma$ is small. Also, the information encoded in this system is almost fully transferred to the entangling basis when $p$ is large. This is the mechanism of $C^{MC}_{0}$ in Fig.~\ref{subfig:subset_mc0_q1}. The next application of CNOT brings information encoded in the entangling basis back to this system. Therefore, $C^{MC}_{1}$ has a large value even if $p$ is large (Fig.~\ref{subfig:subset_mc1_q1}). However, yet another application of the CNOT brings the information back to the entangling basis again. Therefore, $C^{MC}_{2}$ does not have value in the parameter range of interest (Fig.~\ref{subfig:subset_mc2_q1}). Successive application of CNOT is the repetition of the process above  (Fig.~\ref{subfig:subset_mc3_q1}). These differences between even and odd delay memory capacities are clear in Fig.~\ref{subfig:subset_mc_even_q1} and Fig.~\ref{subfig:subset_mc_odd_q1} where $C^{MC}_{odd} \equiv \sum_{k:odd}C^{MC}_{k}$ and $C^{MC}_{even} \equiv \sum_{k:even}C^{MC}_{k}$ are plotted. 

\begin{figure}[t]
\centering

\par
\subfloat[]{
\label{subfig:subset_mc2+_ent}
\includegraphics[width=0.7\hsize]{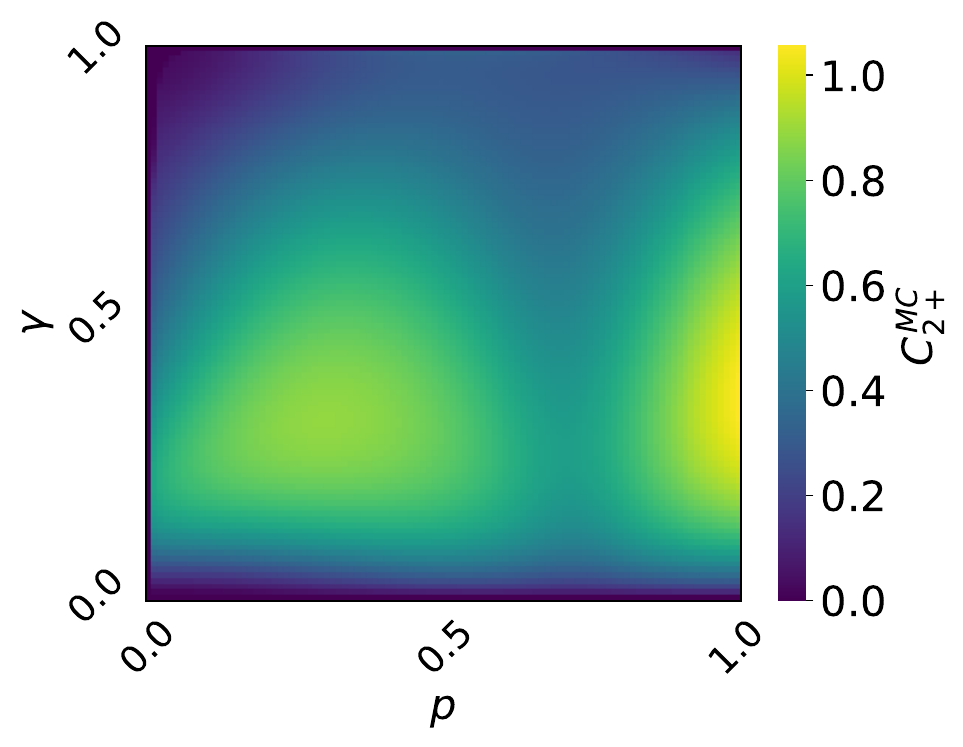}
}
\par
\subfloat[]{
\label{subfig:subset_mc2+_all}
\includegraphics[width=0.7\hsize]{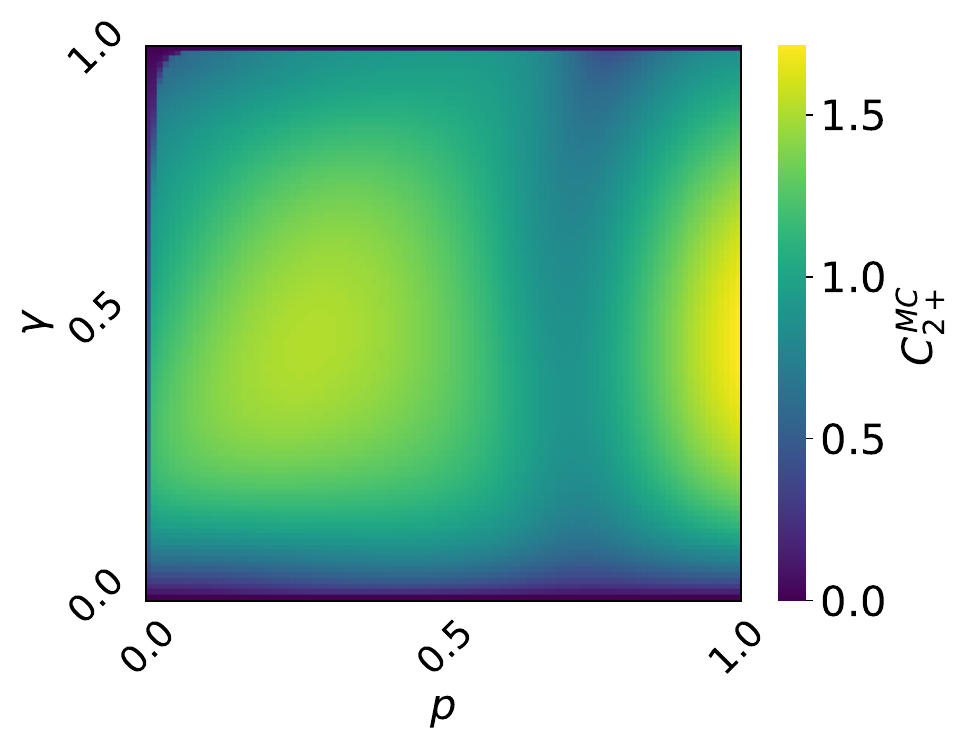}
}

\caption{
$C^{MC}_{2+}$ of (a) the entangling basis, and (b) the total system.
}
\label{fig:subset_mc2_plus}
\end{figure}

We imagine that a gap exists between the parameter ranges in which the entangling basis can behave as memory and cannot. Thus, as depicted in Fig.~\ref{fig:subset_mc2_plus}, $k \geq 2$ memory capacities, $C^{MC}_{2+} \equiv \sum_{k\geq 2}C^{MC}_{k}$, is composed of two regions where entangling basis can behave as memory.

\begin{figure*}[t]
	\centering
\subfloat[]{
\label{subfig:subset_mc0_xz}
	\includegraphics[width=0.25\hsize]{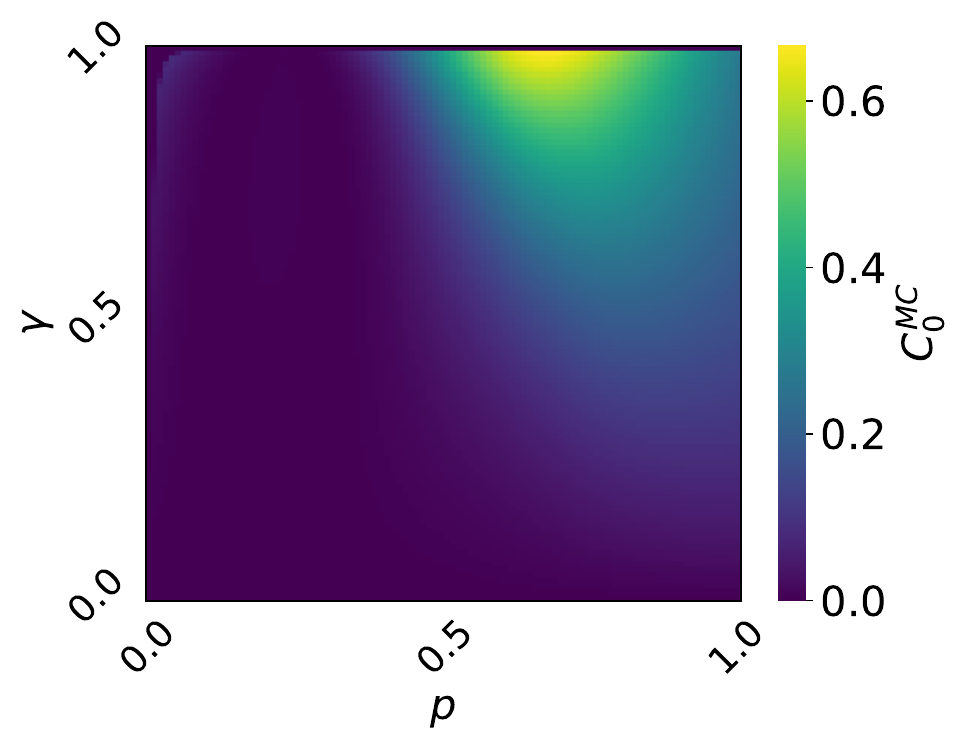}
}
\subfloat[]{
\label{subfig:subset_mc1_xz}
	\includegraphics[width=0.25\hsize]{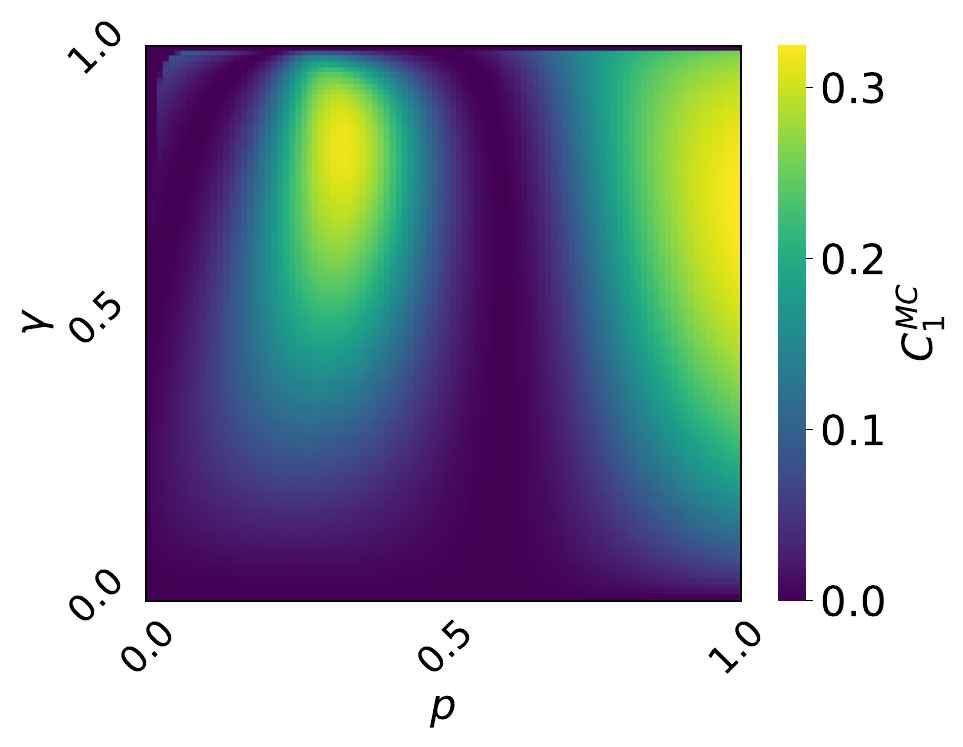}
}
\subfloat[]{
\label{subfig:subset_mc2_xz}
	\includegraphics[width=0.25\hsize]{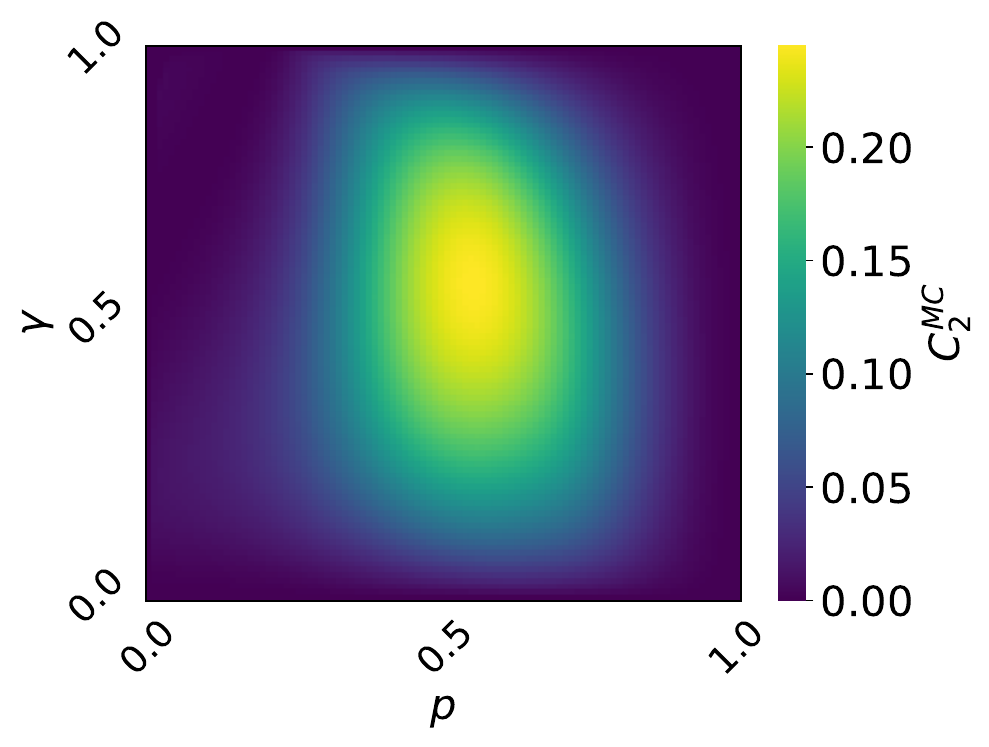}
}
\subfloat[]{
\label{subfig:subset_mc3_xz}
	\includegraphics[width=0.25\hsize]{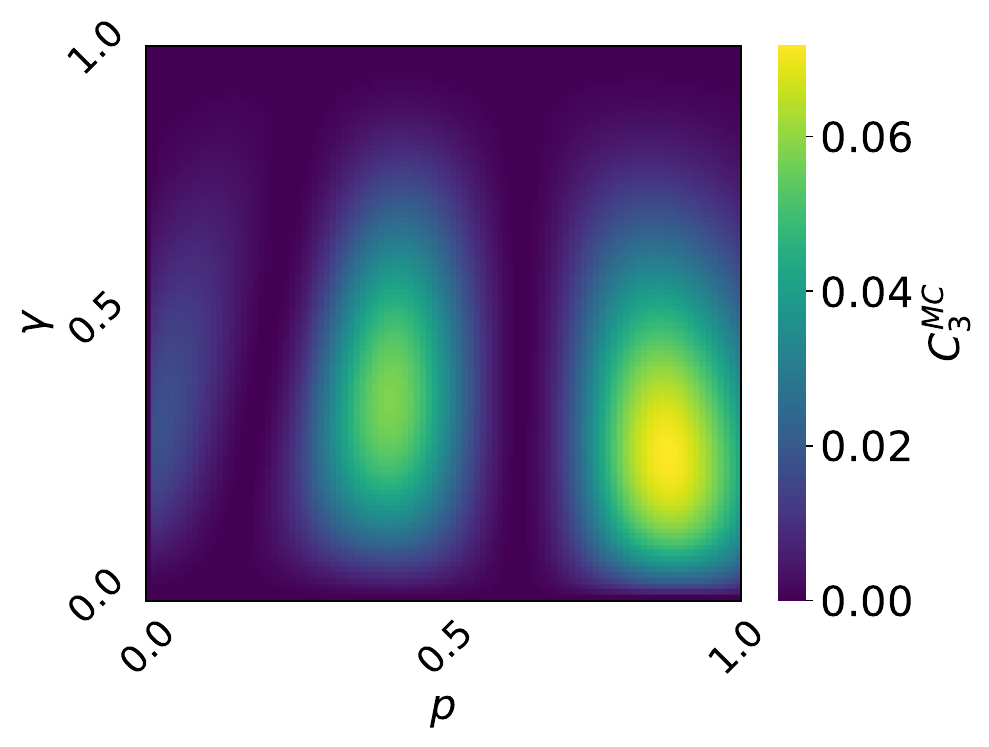}
}
\par
\subfloat[]{
\label{subfig:subset_mc4_xz}
	\includegraphics[width=0.25\hsize]{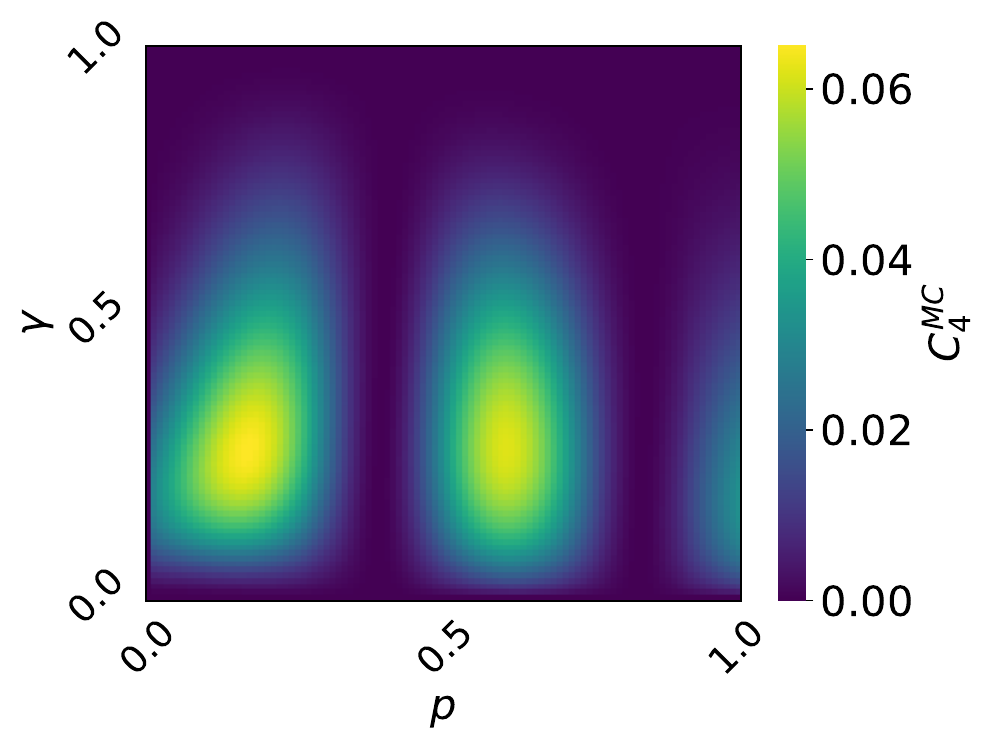}
}
\subfloat[]{
\label{subfig:subset_mc5_xz}
	\includegraphics[width=0.25\hsize]{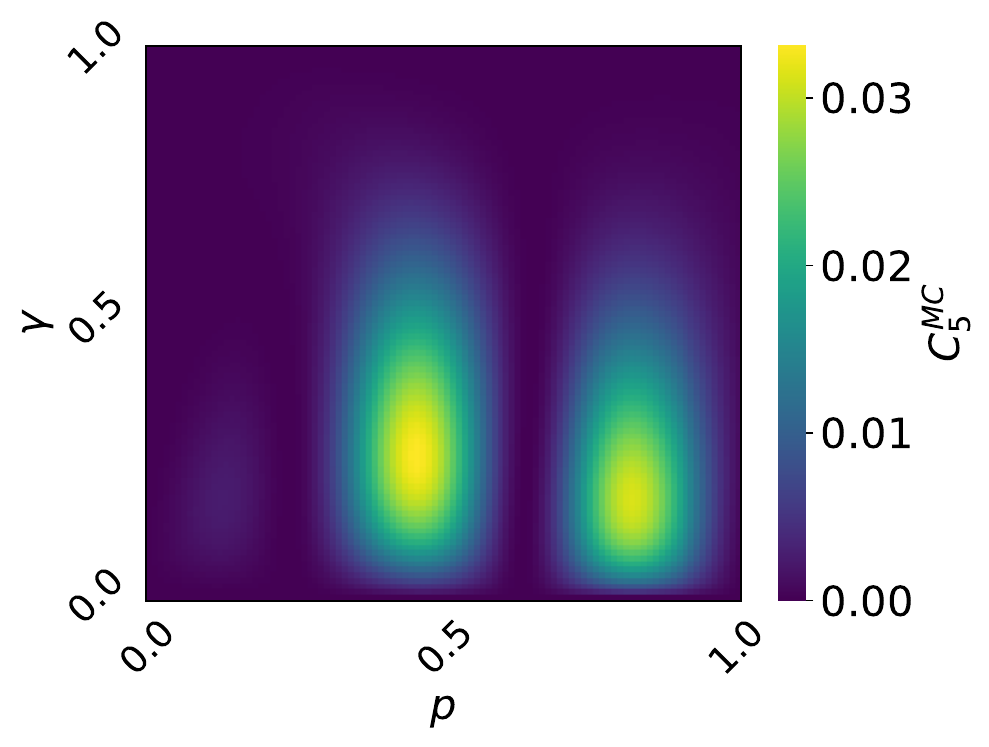}
}
\subfloat[]{
\label{subfig:subset_mc6_xz}
	\includegraphics[width=0.25\hsize]{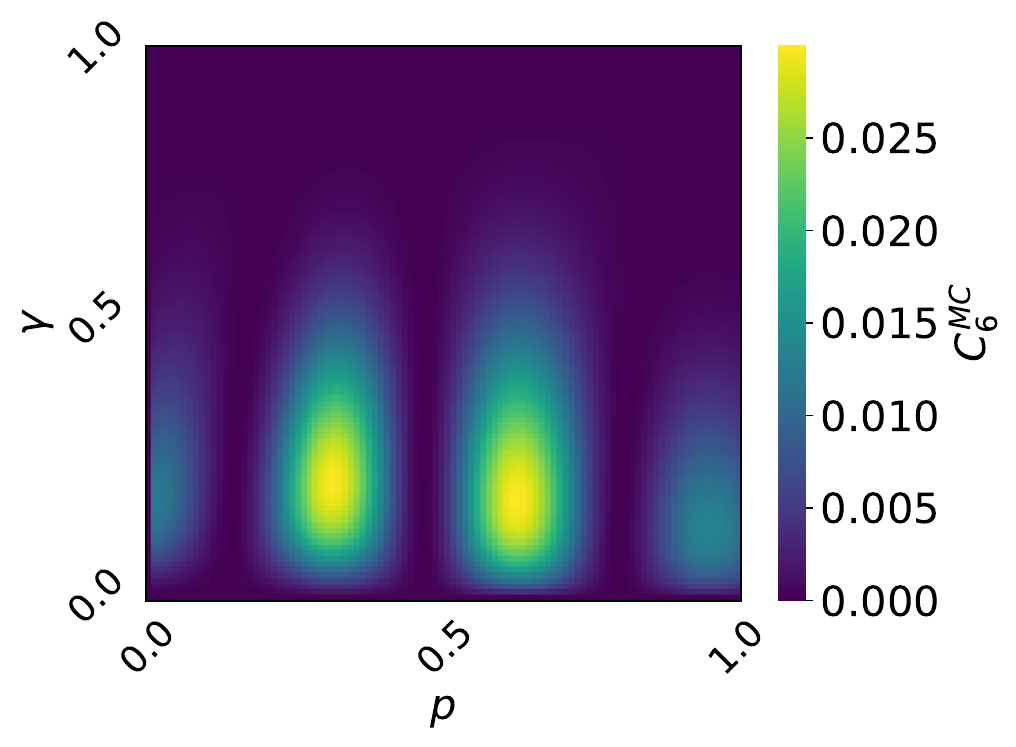}
}
\subfloat[]{
\label{subfig:subset_mc7_xz}
	\includegraphics[width=0.25\hsize]{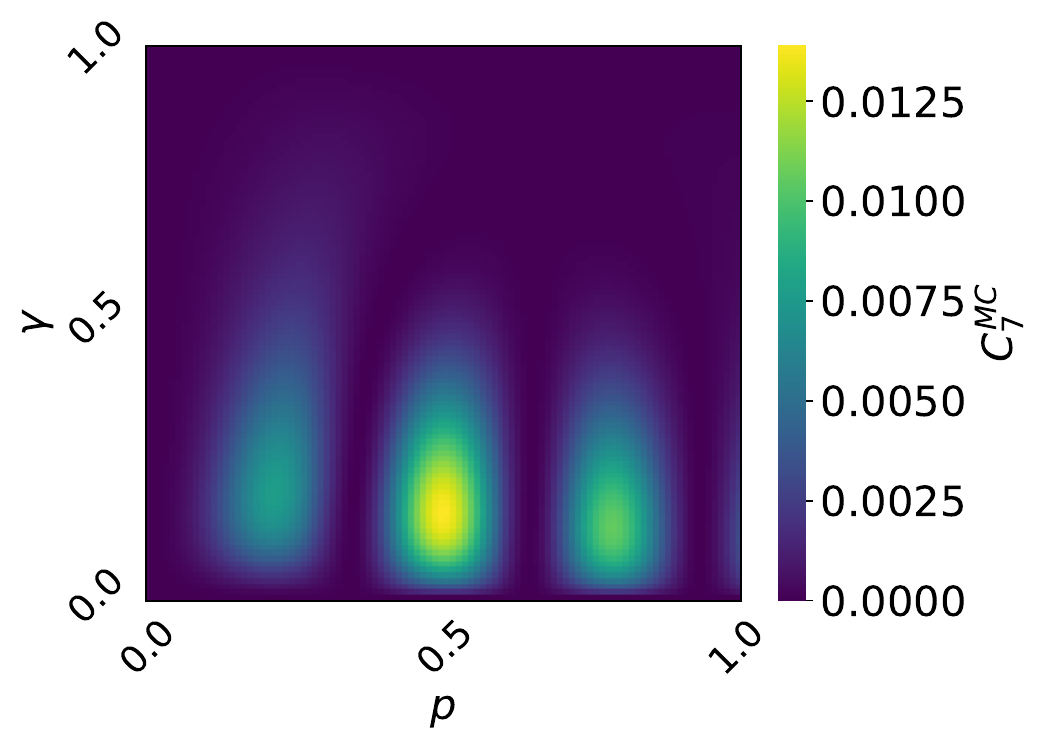}
}
\par
\subfloat[]{
\label{subfig:subset_mc8_xz}
	\includegraphics[width=0.25\hsize]{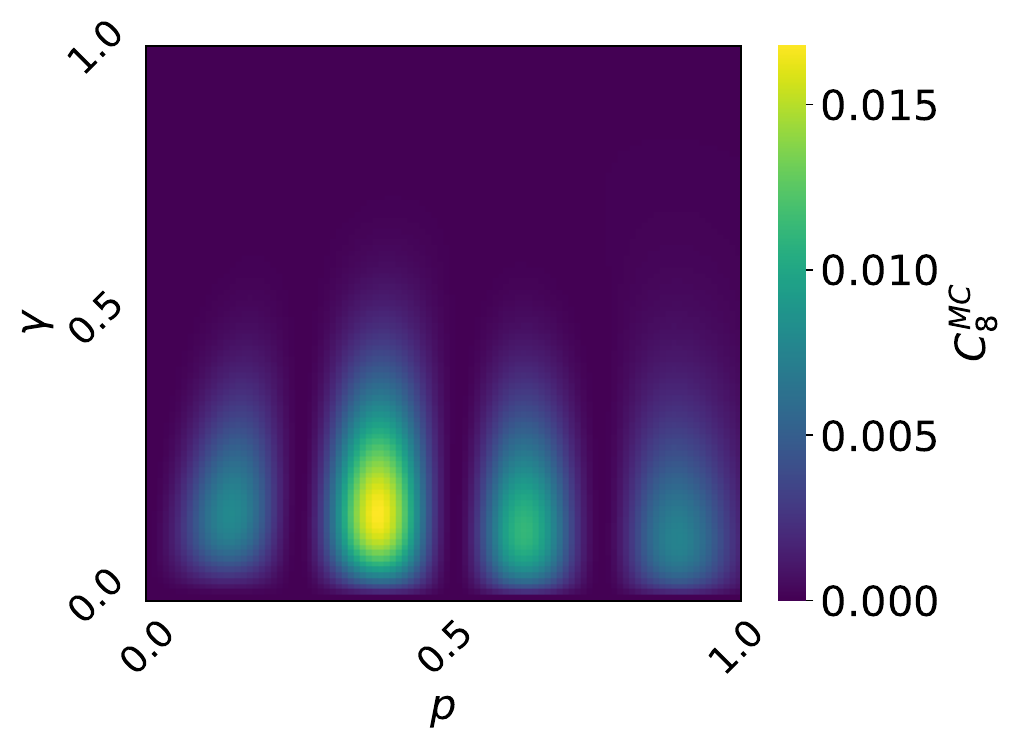}
}
\subfloat[]{
\label{subfig:subset_mc9_xz}
	\includegraphics[width=0.25\hsize]{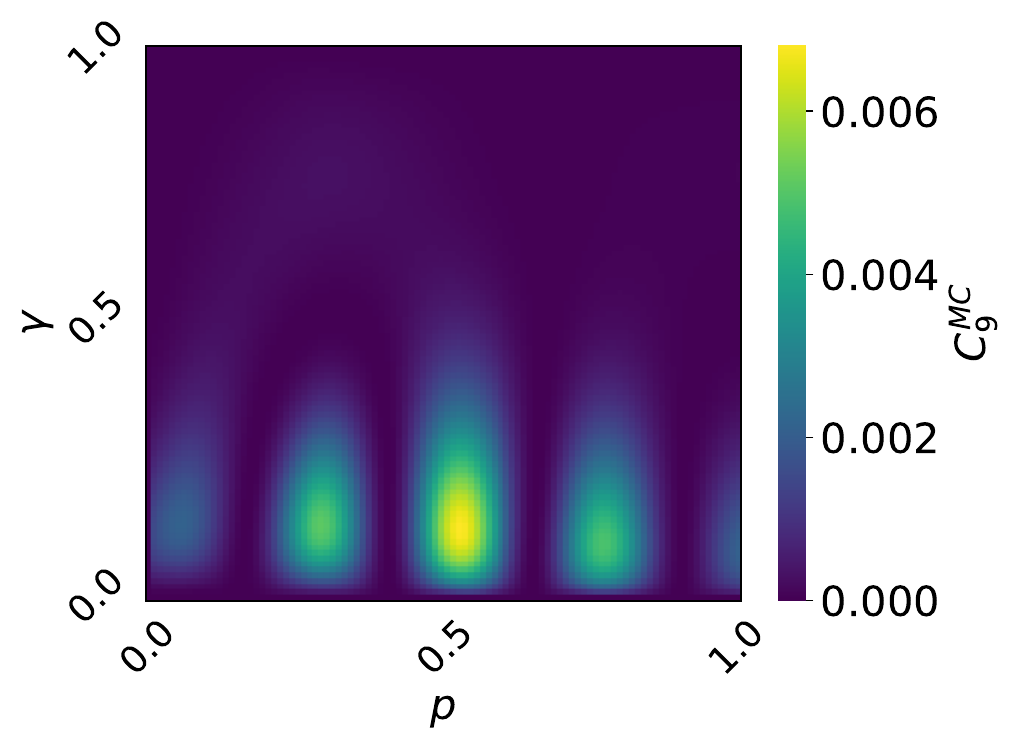}
}
\subfloat[]{
\label{subfig:subset_mc10_xz}
	\includegraphics[width=0.25\hsize]{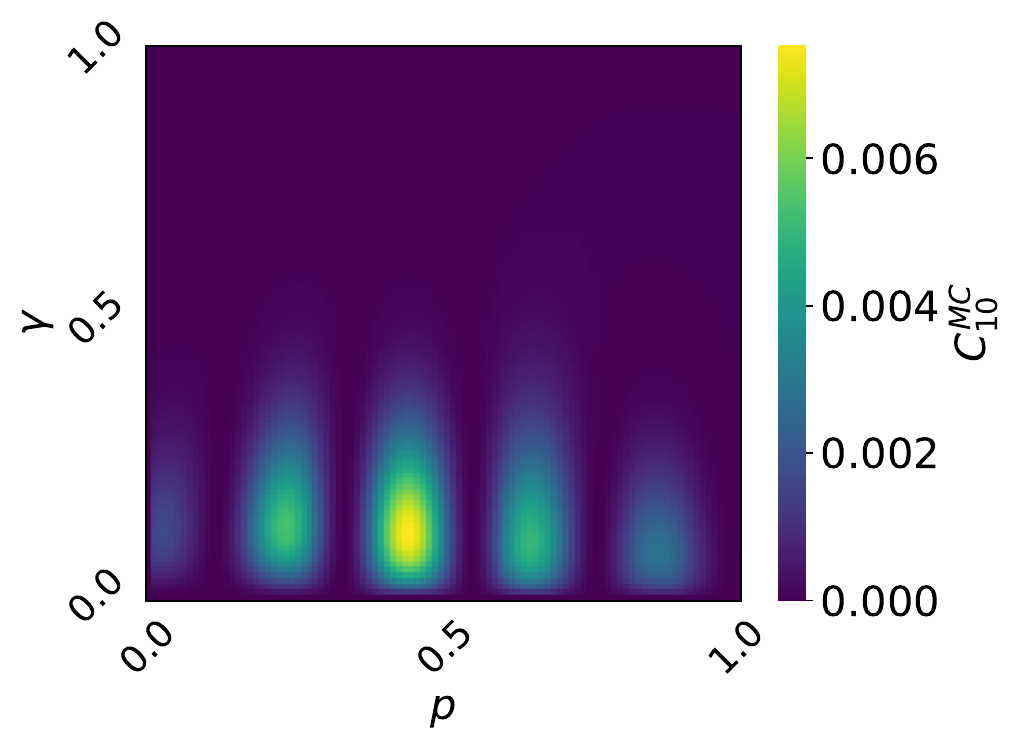}
}
\subfloat[]{
\label{subfig:subset_mc11_xz}
	\includegraphics[width=0.25\hsize]{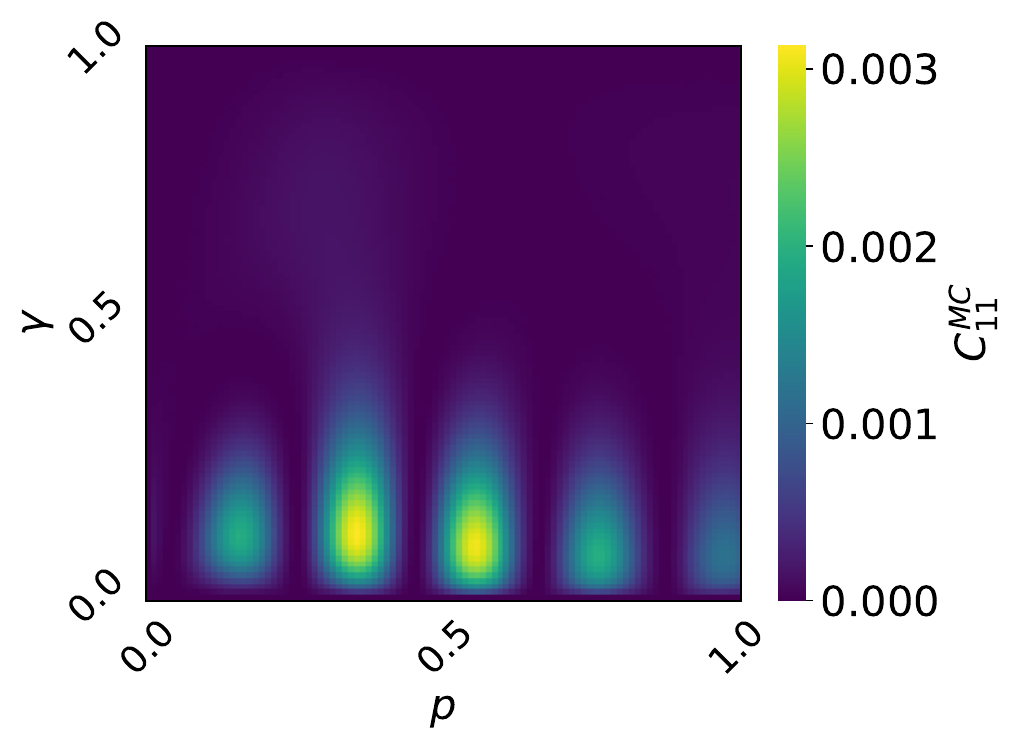}
}
\par
\subfloat[]{
\label{subfig:subset_mc12_xz}
	\includegraphics[width=0.25\hsize]{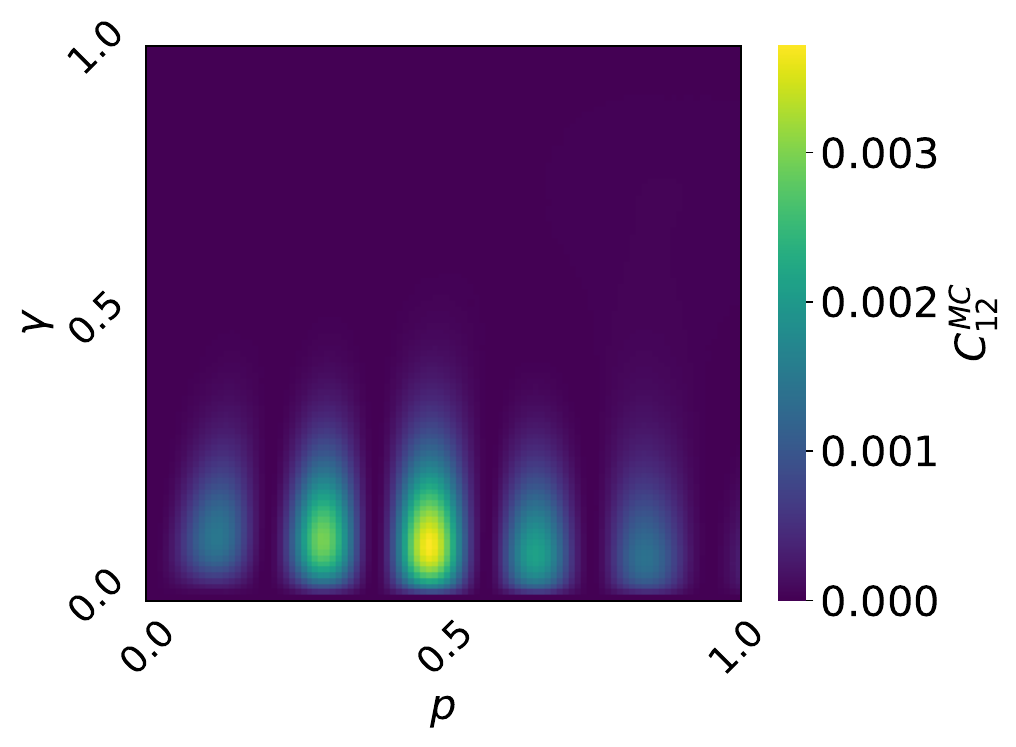}
}
\subfloat[]{
\label{subfig:subset_mc13_xz}
	\includegraphics[width=0.25\hsize]{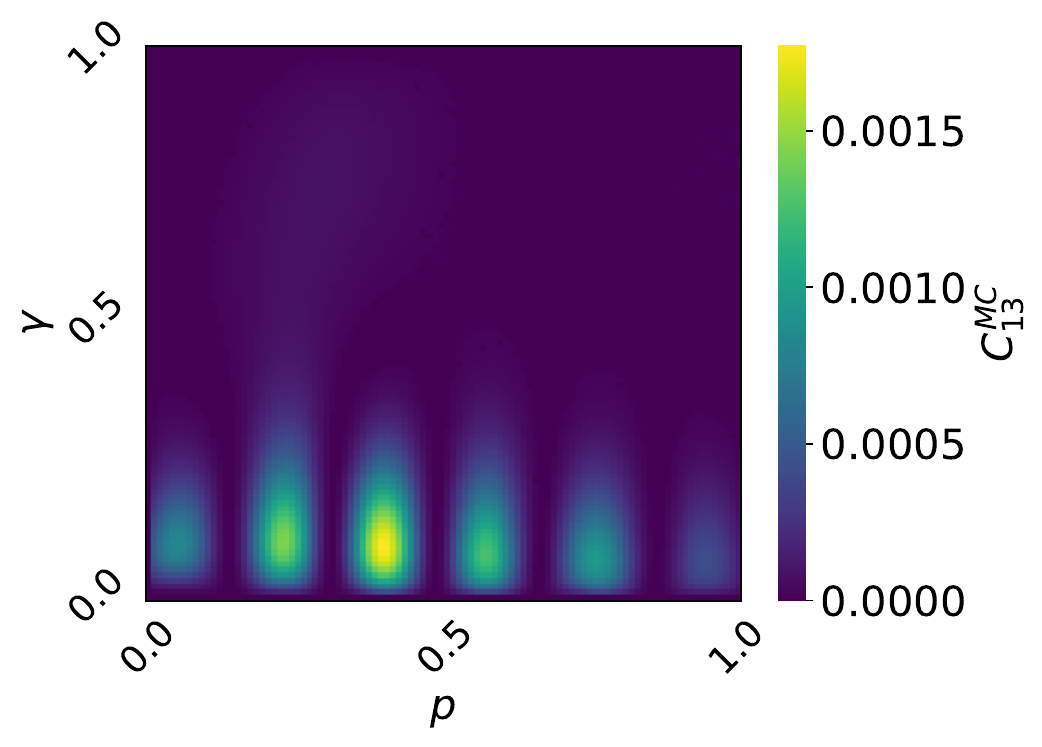}
}
\subfloat[]{
\label{subfig:subset_mc14_xz}
	\includegraphics[width=0.25\hsize]{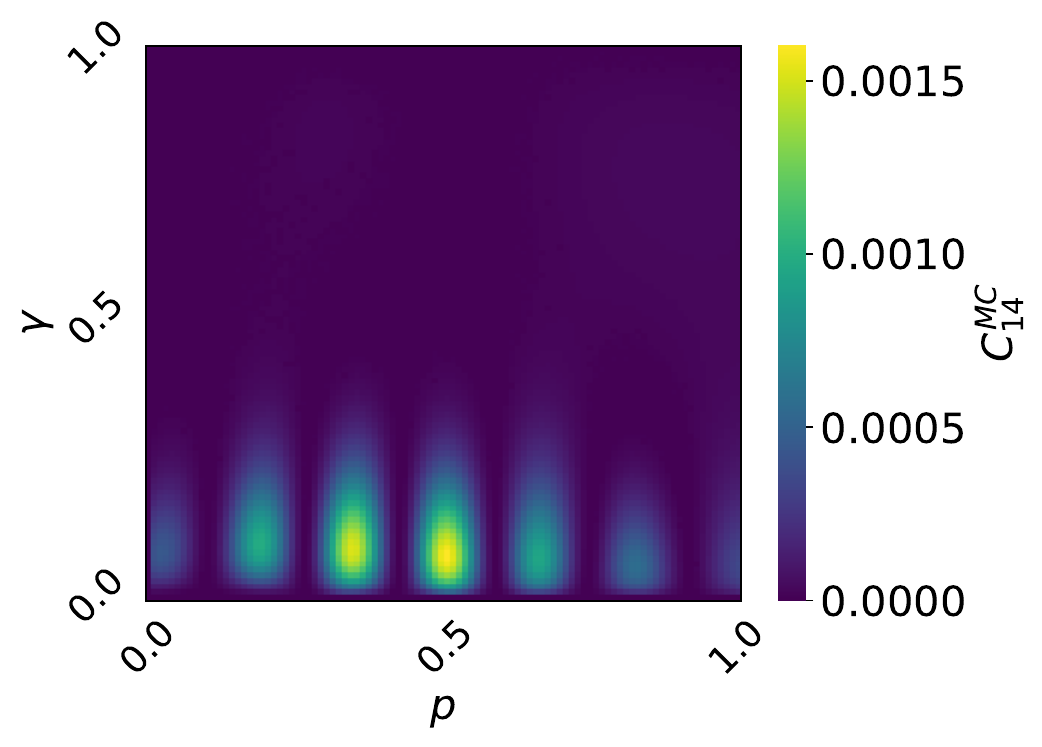}
}
\subfloat[]{
\label{subfig:subset_mc15_xz}
	\includegraphics[width=0.25\hsize]{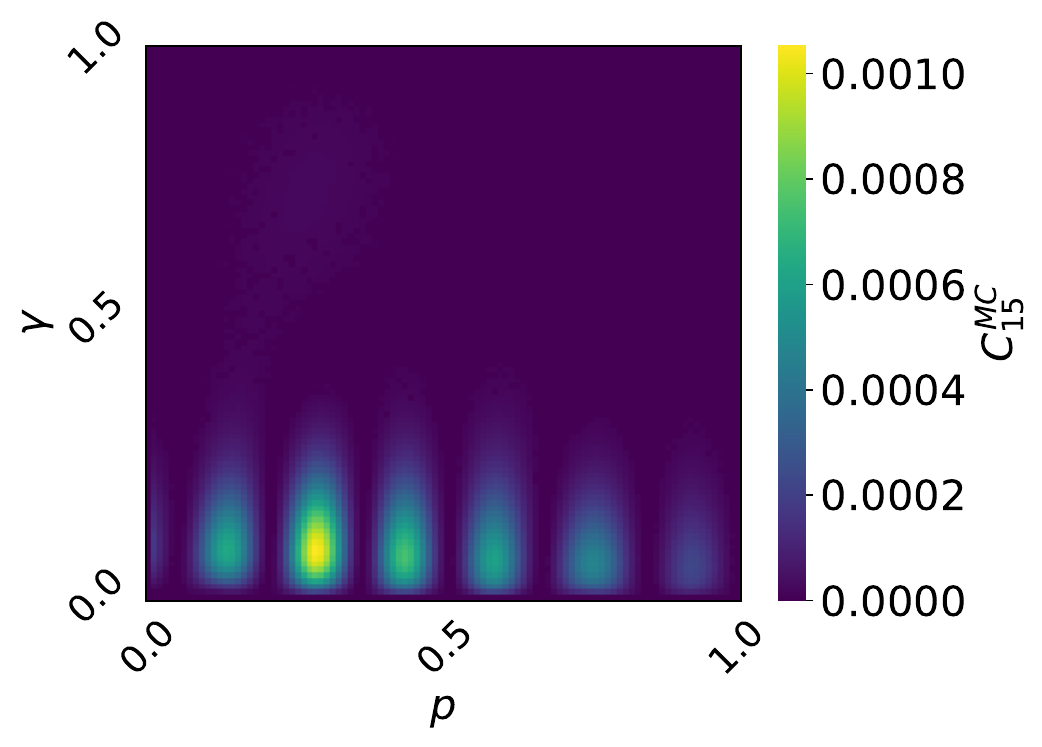}
}
\par
\subfloat[]{
\label{subfig:subset_mc16_xz}
	\includegraphics[width=0.25\hsize]{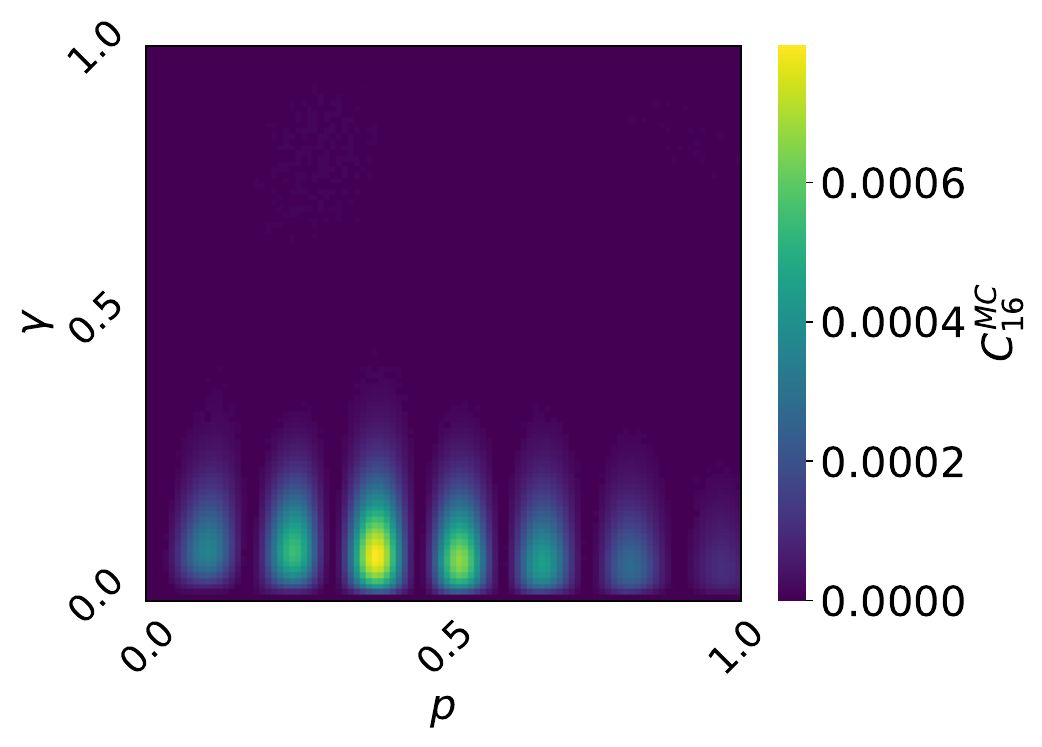}
}
\subfloat[]{
\label{subfig:subset_mc17_xz}
	\includegraphics[width=0.25\hsize]{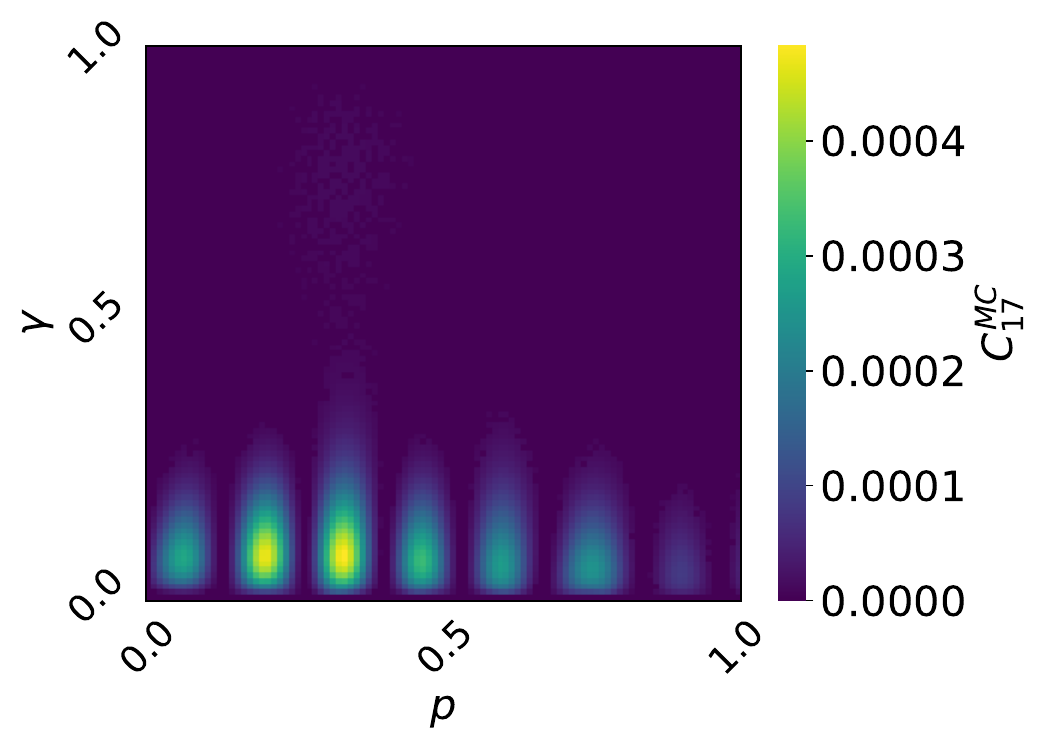}
}
\subfloat[]{
\label{subfig:subset_mc18_xz}
	\includegraphics[width=0.25\hsize]{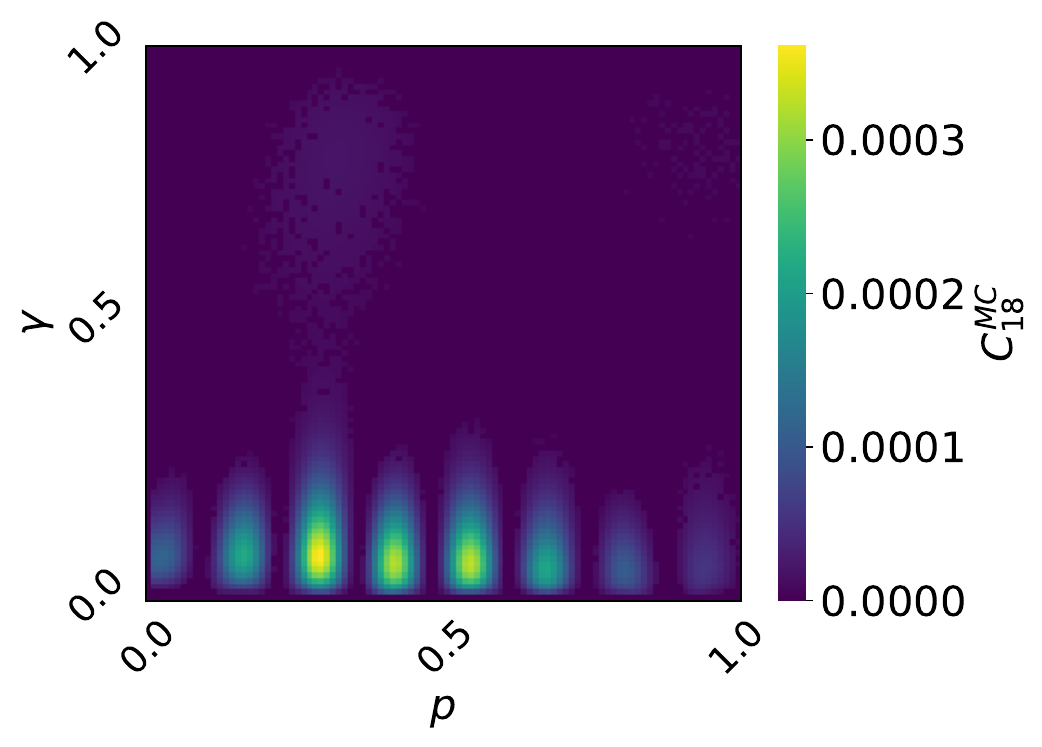}
}
\subfloat[]{
\label{subfig:subset_mc19_xz}
	\includegraphics[width=0.25\hsize]{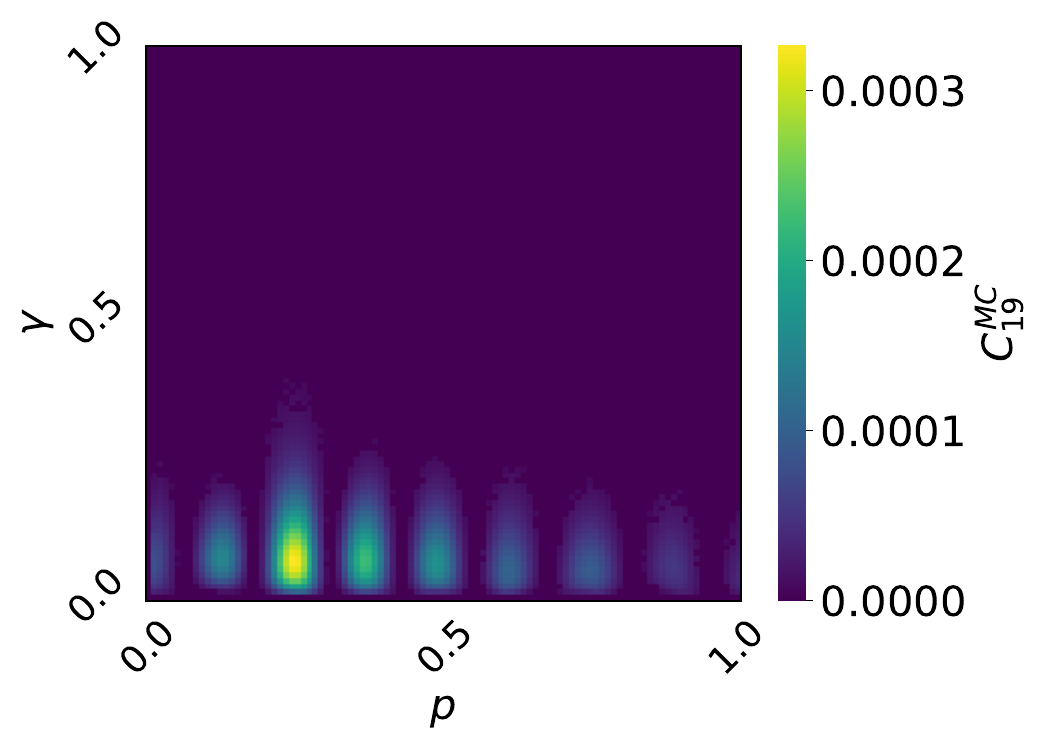}
}
\caption{
(a-t) Memory functions $C^{MC}_k$ of delay $0\leq k\leq 19$ by ascending order of $k$, using XZ basis measurement results. The yellow (light gray) region has finite memory functions. Please be warned that each heatmap has a different scale. Small values still have statistical meaning because each memory function is thresholded by a random shuffle surrogate \cite{Kubota_2021}.
}
\label{fig:xz_mc_delay}
\end{figure*}

Furthermore, by evaluating the memory functions of every Pauli string measurement (especially those of entangling basis) in the subset ESP experiment of Sec.~\ref{subsec:subset}, we observed parameter dependencies of delay of memory functions with finite value. An interesting example was the $XZ$ basis, in which parameter regions with finite memory had multiple islands in a longer delay. (See Fig.~\ref{fig:xz_mc_delay} for detail.)

\section{Conclusion}
This paper proposes non-stationary, subspace, and subset ESPs that are helpful in real-world RC scenarios. As a concrete application, we numerically analyzed a QRC with a well-known SK Hamiltonian, a reset-input encoding method for non-stationary ESP, and an adaptive entanglement system for a subset of non-stationary ESP. Our study revealed a partial correspondence between the non-stationary ESP, the subset and subspace ESP, and the information processing capabilities, as demonstrated using NARMA tasks and MC/IPC calculations.

Notably, using non-stationary ESP over the traditional ESP enables us to rule out dynamics that converge input independently to a fixed point. Furthermore, using subset non-stationary ESP enables us to predict the information processing capability of possibly disjoint systems, such as tensor product systems of qubits, as observed in subset non-stationary ESP experiments. Our theory provides novel perspectives for the practical design of QRC and other possibly non-stationary RC systems.
\section{Acknowledgments}
This work was supported by the MEXT Quantum Leap Flagship Program (MEXT Q-LEAP) Grant No. JPMXS0120319794.

\begin{appendix}
\section{Equivalence of ESP definitions}
\label{sec:esp_equiv}
Here, we prove the equivalence of two distinct ESP definitions introduced in \cite{Jaeger2001ESP} so that our discussion becomes general.

\begin{rem}
	The following ESP conditions are equivalent.

	Following the notations introduced in \cite{Jaeger_2001}, let us consider input sequences $\{\mathbf{u}_t\}_{t \in \mathcal{T}} \in U^\mathcal{T}$ where $U$ is compact. Let $U^{-\mathbb{N}}$ be a left-infinite sequence up to some $t \geq 0$, namely, $U^{-\mathbb{N}} = \{\cdots, \mathbf{u}_{t-1}, \mathbf{u}_{t}\}$.
	
	\begin{enumerate}
		\item Let a state sequence given inputs $\{\mathbf{u}_t\}$ be $\{\mathbf{x}_t\}$ such that $\mathbf{x}_t=T(\mathbf{x}_{t-1}, \mathbf{u}_t)$ for some $T: \mathcal{X} \times U \to \mathcal{X}$. Let $\{\mathbf{x}_t\}$ and $\{\mathbf{x}_t'\}$ be two different state sequences with the same input sequence and different initial states. Then,
		\begin{equation}
		\label{eqn:state-equiality}
			\forall \{\mathbf{u}_t\} \in U^{-\mathbb{N}},\ \underset{t \to \infty} \lim \|\mathbf{x}_t - \mathbf{x}_t'\| = 0
		\end{equation} 
		\item Let an \textit{echo function} be $\mathbf{E}: \left(U^{-\mathbb{N}} \to \mathbb{R}\right)^N$. Then,
		\begin{equation}
				\label{eqn:echo-function}
			\forall \{\mathbf{u}_t\} \in U^{-\mathbb{N}}, \mathbf{x}_t = \mathbf{E}(\{\mathbf{u}_t\}).
		\end{equation}
	\end{enumerate}
\end{rem}
\begin{proof}
	\begin{itemize}
		\item $2 \to 1$
			If $\mathbf{x}_t = \mathbf{E}(\{\mathbf{u}_t\})$, then $\mathbf{x}_t  = \mathbf{x}_t' = \mathbf{E}(\{\mathbf{u}_t\})$. This proves the necessity of the remark.
		\item $1 \to 2$\\
			We prove the contraposition of the fact. Namely, if $\mathbf{x}_t$ is not an echo function, then there exists an input sequence $\{\mathbf{u}_t\}$ such that $\mathbf{x}_t \neq \mathbf{x}_t'$.
			
			Indeed, if $\mathbf{x}_t$ is not an echo function, there exists at least one input-independent parameter $\theta$ such that $\mathbf{x}_t = \mathbf{E}'(\{\mathbf{u}_t\}, \theta)$. This form implies that the effect of $\theta$ remains finite after processing an infinite number of inputs for at least one input sequence because $\{\mathbf{u}_t\}$ is left-infinite. Therefore, for different states $\mathbf{x}_t$ and $\mathbf{x}_t'$ with different $\theta$, there exists at least one input sequence $\{\mathbf{u}_t\} \in U^{-\mathbb{N}}$ such that $\mathbf{x}_t  \neq \mathbf{x}_t'$. This proves the sufficiency of the remark.
	\end{itemize}
\end{proof}

The result above ensures that all known ESP definitions are equivalent because Eq.~\eqref{eqn:esp} is equivalent to Eq.~\eqref{eqn:state-equiality} and Eq.~(11) in \cite{Mart_nez_Pe_a_2023} is equivalent to Eq.~\eqref{eqn:echo-function}.

Furthermore, the following fact is implied.
\begin{rem}{Timestamp function}

If ESP holds and the state sequence $\{\mathbf{x}_t\}$ depends on input cycle $t$, there exists a function $\tilde{t}: U^{-\mathbb{N}} \to \mathbb{N}$ such that 
\begin{equation}
	\forall \{\mathbf{u}_t\} \in U^{-\mathbb{N}},\ \tilde{t}\left(\dots, \mathbf{u}_{t-1}, \mathbf{u}_t\right) = t.
\end{equation}

\end{rem}

		One of the confusions about the equality of the conditions comes when we let $\theta = t$, a time parameter. Since input sequence $\{\mathbf{u}_t\}$ also has time parameter $t$, we are keen to equate them. However, the time parameter $t$ of $\{\mathbf{u}_t\}$ is a constant between all dynamics with different initial states for each time step: a function of the input sequence itself. Namely, for states $\mathbf{x}$ and $\mathbf{x}'$ having the same input sequence $\{\mathbf{u}_t\}$ at every input cycle, $t$, the input cycle, does not differentiate their states.
		
		Therefore, if we set a time parameter $\theta = \tau$ as a parameter of the echo function, $\tau$ must not depend on the timestamp of the input sequence. It is an unrelated parameter to the cycle of the data input procedure and typically an initial time $t_0$.
		
		There is a work \cite{Kubota_2021} in which temporal information processing capacity (TIPC) analyzes the input cycle-dependent structure of state sequences. We argue that all input sequences are input cycle-dependent. TIPC examines how initial state dependency, which does not converge after washout, evolves through the input cycle. Namely, it analyzes the dynamics of initial state dependency.

\add{\section{ESP results for the other random Hamiltonians}
We calculated the ESP and non-stationary ESP indicators for QRCs with SK Hamiltonians with different parameters from the one we analyzed in Sec.~\ref{sec:ns_esp_qrc}. Each Hamiltonian parameter configuration is represented by: $H_2 (h\sim 0.013$ and $W\sim 1.05)$, $H_3 (h\sim 0.377$ and $W\sim 24.8)$, $H_4 (h\sim 57.2$ and $W\sim 47.5)$ and $H_5 (h\sim 48.3$ and $W\sim 0.0305)$.}

\add{As shown in Fig.~\ref{fig:esp_other_hamiltonians} below, after feeding the appropriate length of input signals, every setup reaches the point where the upper and bottom line is the only region in which the non-stationary ESP is considerably weaker compared to with other regions. The length input sequence required to reach that point is governed by the respective decay strength of fading memory in each Hamiltonian: for instance, $H_3$ and $H_4$ relate to the many-body localization phase as shown in \cite{Mart_nez_Pe_a_2021}, so it requires a longer input sequence to reach that point.}
\begin{figure*}
\centering
\subfloat[]{
\label{subfig:ns-esp-target2-200}
	\includegraphics[width=0.33\hsize]{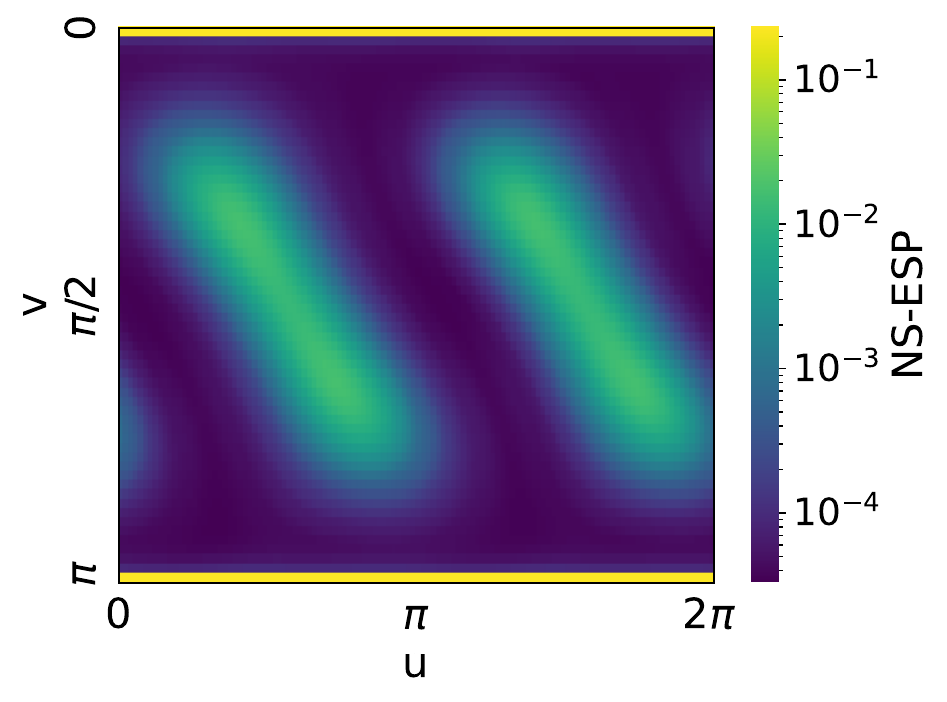}
}
\subfloat[]{
\label{subfig:ns-esp-target2-2000}
	\includegraphics[width=0.33\hsize]{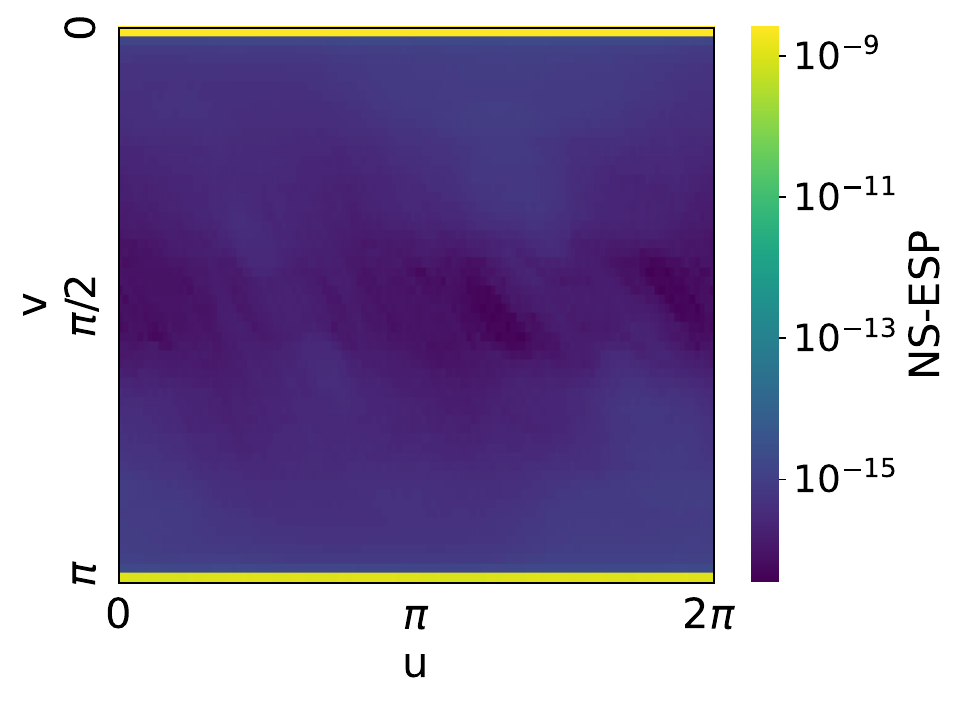}
}
\subfloat[]{
\label{subfig:ns-esp-target2-20000}
	\includegraphics[width=0.33\hsize]{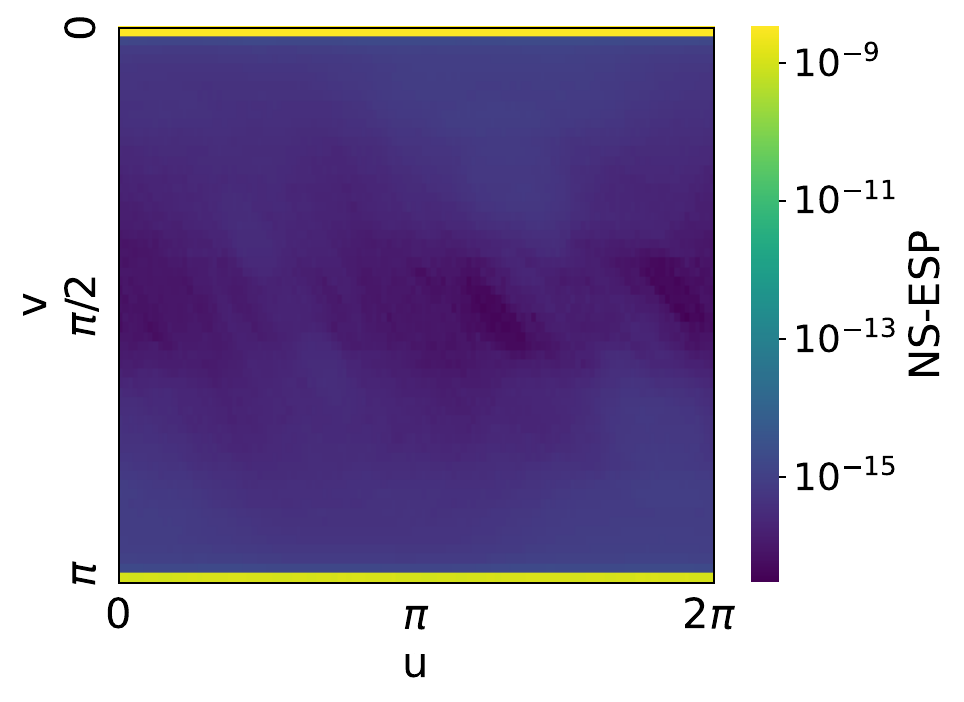}
}
\par
\subfloat[]{
\label{subfig:ns-esp-target3-200}
	\includegraphics[width=0.33\hsize]{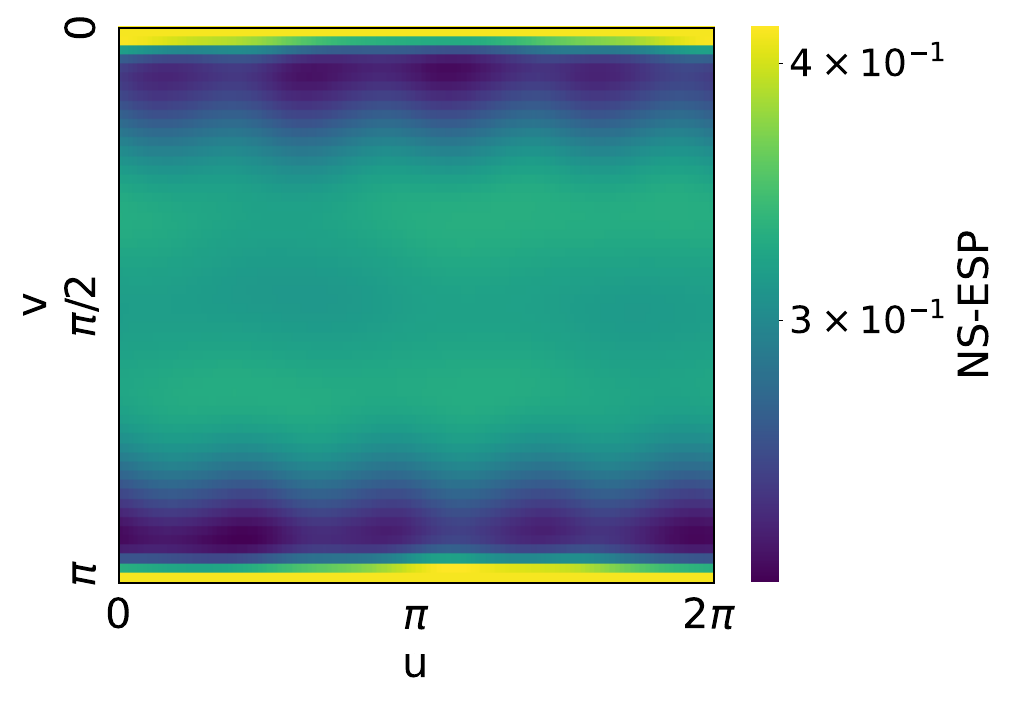}
}
\subfloat[]{
\label{subfig:ns-esp-target3-2000}
	\includegraphics[width=0.33\hsize]{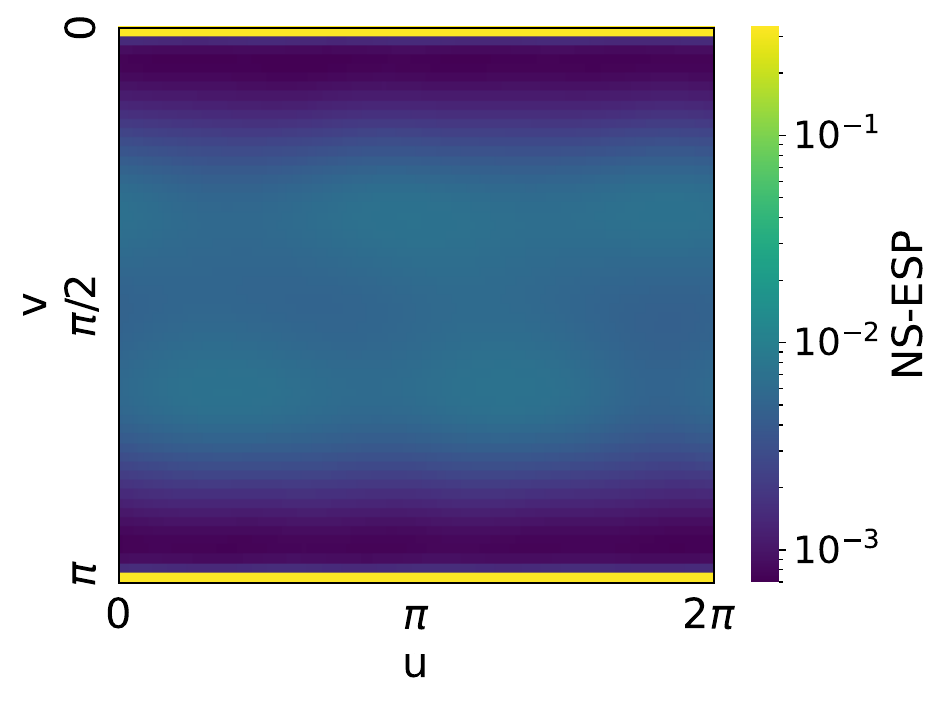}
}
\subfloat[]{
\label{subfig:ns-esp-target3-20000}
	\includegraphics[width=0.33\hsize]{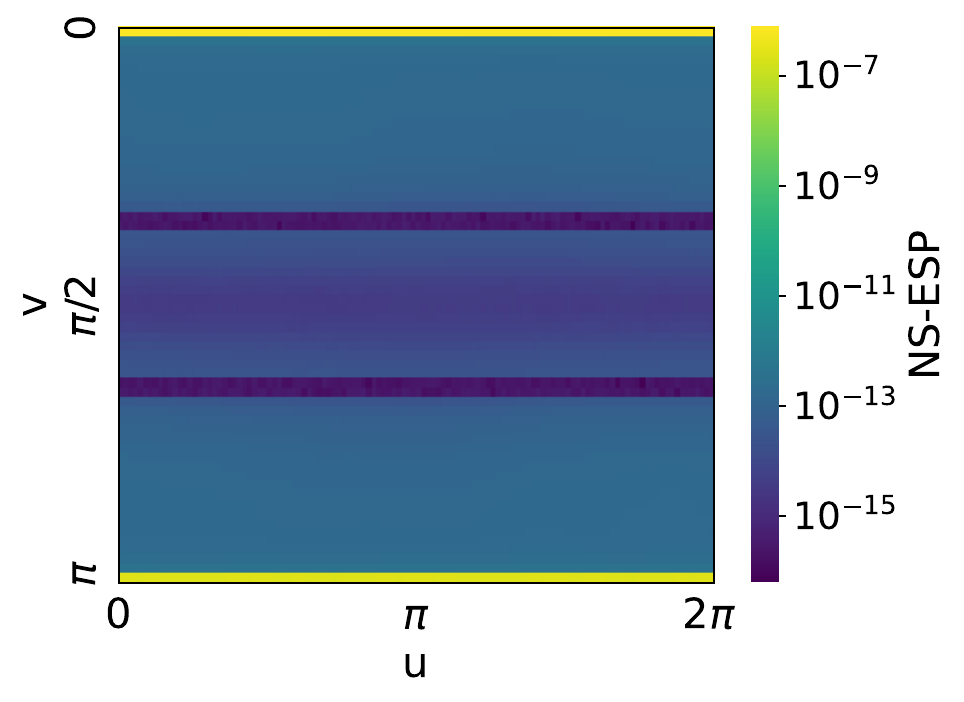}
}
\par
\subfloat[]{
\label{subfig:ns-esp-target4-200}
	\includegraphics[width=0.33\hsize]{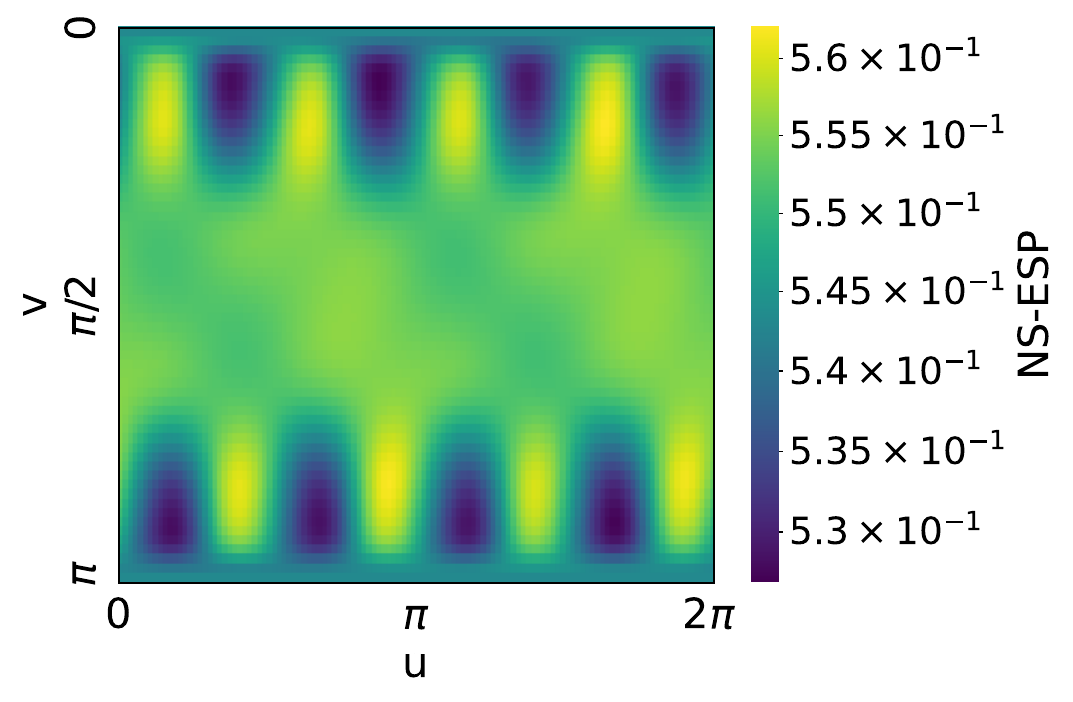}
}
\subfloat[]{
\label{subfig:ns-esp-target4-2000}
	\includegraphics[width=0.33\hsize]{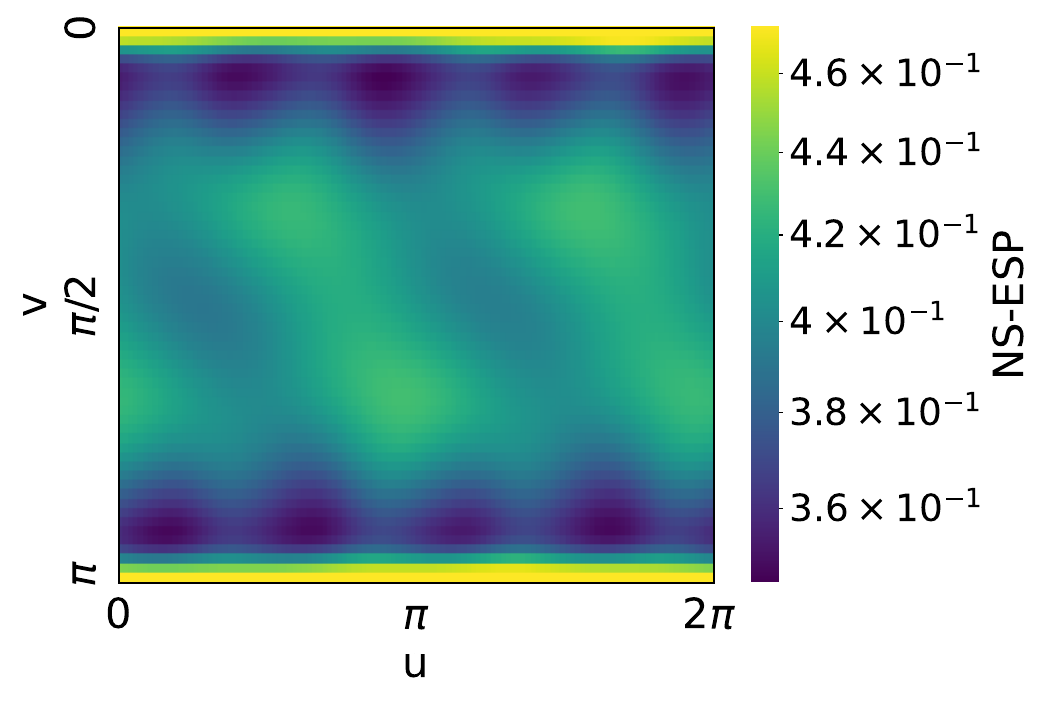}
}
\subfloat[]{
\label{subfig:ns-esp-target4-20000}
	\includegraphics[width=0.33\hsize]{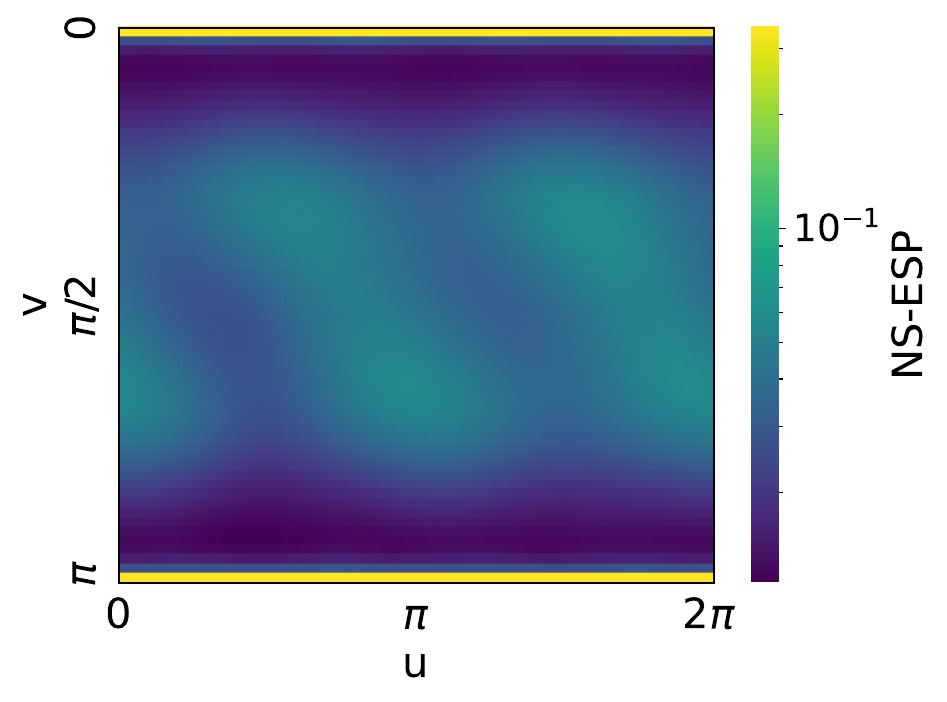}
}
\par
\subfloat[]{
\label{subfig:ns-esp-target5-200}
	\includegraphics[width=0.33\hsize]{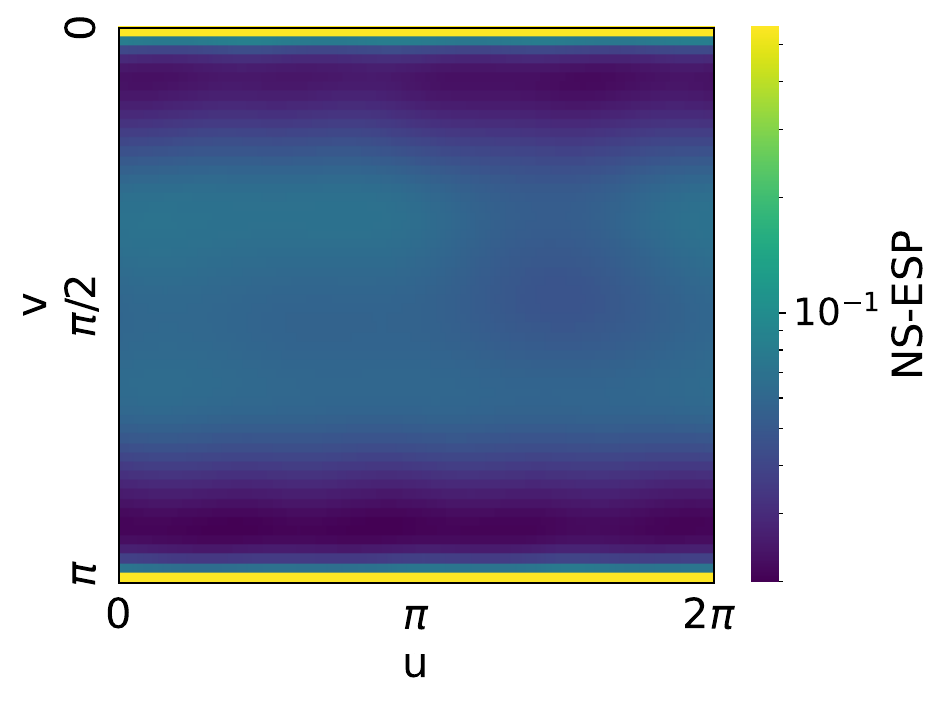}
}
\subfloat[]{
\label{subfig:ns-esp-target5-2000}
	\includegraphics[width=0.33\hsize]{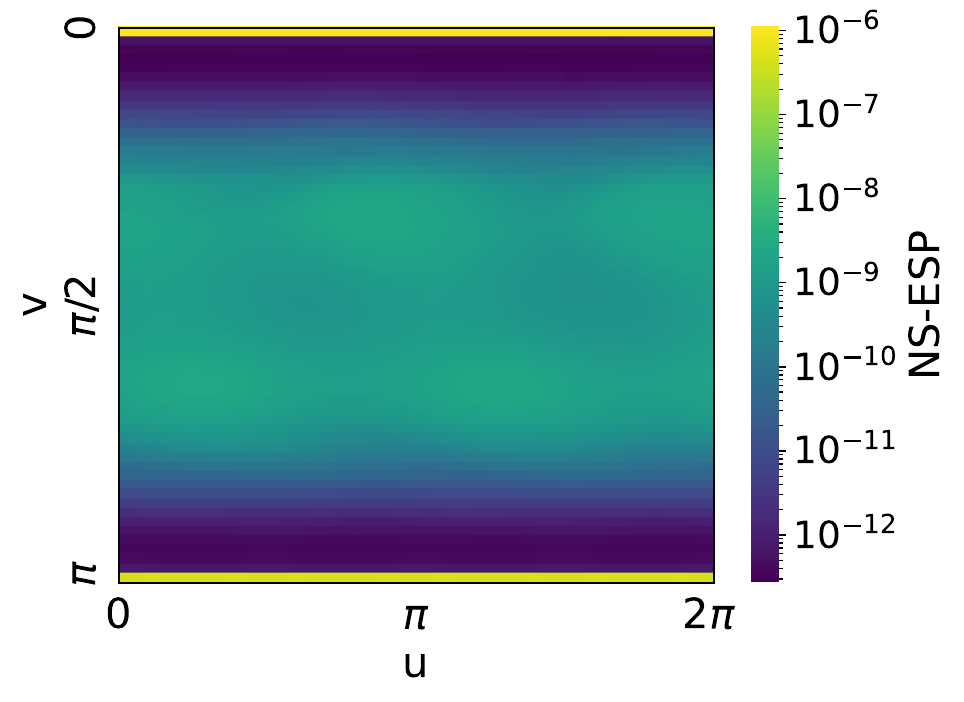}
}
\subfloat[]{
\label{subfig:ns-esp-target5-20000}
	\includegraphics[width=0.33\hsize]{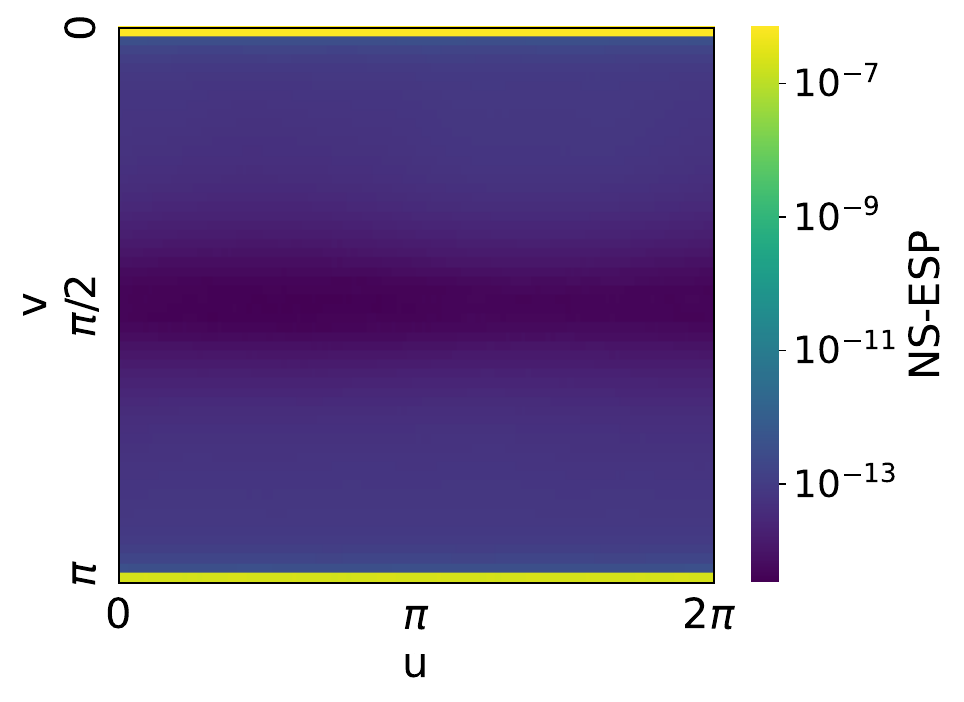}
}
\caption{
\add{Non-stationary ESP indicators for QRCs driven by the random Hamiltonians $H_2$ (a-c), $H_3$ (d-f), $H_4$ (g-i), and $H_5$ (j-l). $t=200$ for the left column (a, d, g, j), $t=2000$ for the middle column (b, e, h, k), and $t=20000$ for the right column (c, f, i, l), where $t$ represents the time steps used in the calculation of the non-stationary ESP indicator (Eq.~\ref{eqn:ns_esp_indicator}).}
}
\label{fig:esp_other_hamiltonians}
\end{figure*}

\end{appendix}
\normalem
\bibliography{main.bib}
\end{document}